\documentclass[11pt]{article}
\usepackage{epsfig} 
\usepackage{amssymb}
\usepackage{graphics}
\newcommand{\sect}[1]{ \section{#1} \setcounter{equation}{0} } 

\newcommand{\half}{\mbox{\small{$\frac{1}{2}$}}} 
\newcommand{\threehalves}{\mbox{\small{$\frac{3}{2}$}}}
\newcommand{\third}{\mbox{\small{$\frac{1}{3}$}}} 
 
\newcommand{\MSbar}{\overline{\mbox{MS}}} 
\newcommand{\MSbars}{\overline{\mbox{\footnotesize{MS}}}} 

\newcommand{\MOMs}{\mbox{\footnotesize{MOM}}}
\newcommand{\NA}{N_{\!A}}
\newcommand{\Nc}{N_{\!c}}
\newcommand{\Nf}{N_{\!f}}
\newcommand{\partialslash}{\partial \! \! \! /}
\newcommand{\partialline}{\partial \! \! \! \backslash}
\newcommand{\pline}{p \! \! \! \backslash}
\newcommand{\kline}{k \! \! \! \backslash}

\newcommand{\Cc}{\mathbb{C}}
\newcommand{\Pp}{\mathbb{P}}
\newcommand{\Zz}{\mathbb{Z}}

\begin{document}
\title{Five loop renormalization of the Wess-Zumino model}
\author{J.A. Gracey, \\ Theoretical Physics Division, \\ 
Department of Mathematical Sciences, \\ University of Liverpool, \\ P.O. Box 
147, \\ Liverpool, \\ L69 3BX, \\ United Kingdom.} 
\date{}
\maketitle 

\vspace{5cm} 
\noindent 
{\bf Abstract.} We renormalize the Wess-Zumino model at five loops in both the
minimal subtraction ($\MSbar$) and momentum subtraction (MOM) schemes. The
calculation is carried out automatically using a routine that performs the
$D$-algebra. Generalizations of the model to include $O(N)$ symmetry as well as
the case with real and complex tensor couplings are also considered. We confirm
that the emergent $SU(3)$ symmetry of six dimensional $O(N)$ $\phi^3$ theory is
also a property of the tensor $O(N)$ model. With the new loop order precision 
we compute critical exponents in the $\epsilon$ expansion for several of these 
generalizations as well as the XYZ model in order to compare with conformal 
bootstrap estimates in three dimensions. For example at five loops our estimate
for the correction to scaling exponent is in very good agreement for the 
Wess-Zumino model which equates to the emergent supersymmetric fixed point of 
the Gross-Neveu-Yukawa model. We also compute the rational number that is part 
of the six loop $\MSbar$ $\beta$-function.

\vspace{-18.5cm}
\hspace{13.2cm}
{\bf LTH 1266}

\newpage 

\sect{Introduction.}

The Wess-Zumino model constructed in \cite{1} is the simplest scalar 
supersymmetric quantum field theory in four dimensions with chiral symmetry 
that is renormalizable. It comprises two scalar fields and a Dirac fermion to 
have equal boson and fermion degrees of freedom. There are two interactions one
of which is a quartic scalar whereas the other is a scalar-Yukawa one. In this
respect it has the basic structure of the Standard Model in the absence of 
gauge fields and flavour symmetry groups. Consequently the Wess-Zumino model 
forms a sector of the extension of the Standard Model to the Minimal 
Supersymmetric Standard Model (MSSSM) and as such has been used as a simple
laboratory to explore aspects of that potential theory for new physics beyond 
the Standard Model. This property of the Wess-Zumino model has been one of the 
motivations for its study since its construction in 1974. While the original 
article considered the component field Lagrangian it has been reformulated in 
superspace \cite{2} where it involves two scalar superfields, one of which is 
chiral and the other anti-chiral. These separately have cubic self-interactions
in the superspace action. Several years after its inception the renormalization
group functions were determined beyond the one loop ones recorded in \cite{1}. 
Indeed the four loop expressions in the modified minimal subtraction ($\MSbar$)
scheme were determined in a very short time span from 1979 to 1982, 
\cite{3,4,5,6}. The three loop $\beta$-function in the momentum subtraction 
(MOM) scheme was also given in \cite{4}. One reason for the rapid progress was 
the calculational shortcut available from the supersymmetry Ward identity, 
\cite{1,2}. This ensures that there is only one independent renormalization 
constant in the massless theory which is either that of the wave function or 
the coupling constant. As the former is deduced from the $2$-point function 
this means that a relatively small number of Feynman graphs have to be 
evaluated even to four loops in order to deduced the $\beta$-function. While 
this was manageable at very low loop order, progress with the three and four 
loop renormalization was further advanced with the use of superspace 
techniques, \cite{2,4,6}. In addition to having a small number of supergraphs 
to consider the superspace approach circumvents the issue of $\gamma^5$ if a 
regularization involving analytically continuing the space-time dimension is 
employed, \cite{4}.

Aside from the main connection to a sector of the MSSSM the Wess-Zumino model
has enjoyed a renaissance of interest in recent years due, for example, to an
observation in condensed matter physics. In \cite{7,8,9,10} it was shown that 
supersymmetry was present on the boundary of a three dimensional topological 
insulator. This emergent supersymmetry is believed to be described by the 
Wess-Zumino model. Another instance where the Wess-Zumino model can emerge is 
in a two dimensional optical lattice with cold atom-molecule mixtures
\cite{11}. Equally there is a connection with the four dimensional 
Gross-Neveu-Yukawa model \cite{12} or XY Gross-Neveu model \cite{13,14,15}. 
This is a theory with a scalar-Yukawa and a quartic scalar interaction. Both 
interactions have independent coupling constants. However, it has been 
established \cite{8,13,14,15,16,17} that there is a Wilson-Fisher fixed point 
in $d$~$=$~$4$~$-$~$2\epsilon$ dimensions where the critical couplings are 
equal. Moreover the anomalous dimensions of all the fields are equal at 
criticality revealing the emergent supersymmetry. This has been established at
four loops in the $\epsilon$ expansion, \cite{15} and the exponents have been 
shown to be equal to those of the Wess-Zumino model, \cite{18}. The 
extrapolation to three dimensions is believed to be in the same universality 
class of the supersymmetry associated with the topological insulator.

Given this renewed interest in the Wess-Zumino model and the potential for
supersymmetry to be realized in Nature, albeit not through observations using a 
particle collider, the main aim of this article is to compute the five loop
$\beta$-function of the Wess-Zumino model. While this is around 40 years since 
the previous loop order appeared such a computation is possible now given the
revolution in automatically evaluating Feynman diagrams that has advanced the
field in the last decade. The main techniques that have been instrumental in 
this are the Laporta algorithm \cite{19} and the {\sc Forcer} package 
\cite{20,21}. The former is a routine that systematically uses integration by 
parts to relate specific classes of Feynman graphs to a small set of master 
integrals whose Laurent expansion in $\epsilon$ is known. The latter method is 
a four loop algorithm for the evaluation of $2$-point functions in 
$d$-dimensions and is the natural successor to the {\sc Mincer} package
\cite{22,23} that has been the workhorse of four dimensional massless multiloop 
calculations for a generation. For instance, both approaches have led to the 
five loop $\MSbar$ renormalization of Quantum Chromodynamics (QCD)
\cite{24,25,26,27}. Also the four loop $\beta$-function of six dimensional 
$\phi^3$ theory has been given in \cite{28}. More recently this has been 
superseded by the five loop result \cite{29,30}. The latter computation,
\cite{30}, was effected by a technique that successfully extended our loop 
knowledge of scalar theories to much higher orders. The particular method is 
known as graphical functions \cite{31,32,33}. Prior to \cite{29,30} the six and
seven loop $\phi^4$ $\MSbar$ $\beta$-functions were computed using algebraic 
geometry as well as graphical functions, \cite{32,34}. Indeed it was mentioned 
in \cite{31} that it may be possible to extend the field anomalous dimension to
{\em eight} loops in $\MSbar$.

We will use both the Laporta and {\sc Forcer} techniques in this article
together with a routine developed here to automatically carry out the 
$D$-algebra associated with superspace calculations specifically for the 
Wess-Zumino model. Another motivation for extending the renormalization to five
loops is that in recent years the conformal bootstrap and functional 
renormalization group techniques have been successful in determining critical 
exponents at very high numerical precision. These methods have also been used 
to study the Wess-Zumino model in three dimensions partly for the emergent 
supersymmetry reasons but also for other more mathematical physics problems, 
\cite{17,35,36,37,38,39}. Therefore we will carry out the analogous 
renormalization of these theories to have five loop precision for the exponents
of various operators as well as the correction to scaling exponent by using the
$\epsilon$ expansion and extracting estimates in three dimensions. For 
instance, in \cite{40} the complex one dimensional conformal manifold that 
underlies the infrared behaviour of a class of ${\cal N}$~$=$~$2$
supersymmetric theories in three dimensions was studied in depth using the
conformal bootstrap. One aspect of the study of these more mathematical three 
dimensional theories is that certain dualities have been found to exist. For 
instance, there is believed to be a dual connection between supersymmetric 
Quantum Electrodynamics and an $SU(3)$ Wess-Zumino model, 
\cite{41,42,43,44,45,46}. In this context we will also examine the five loop 
structure of the $O(N)$ model in two formulations. One is the standard one of 
the Hubbard-Stratonovich decomposition used for $\phi^4$ theory. Indeed this 
case has already been examined in the large $N$ expansion \cite{47,48,49} and 
we will use the information contained in the $O(1/N^3)$ $d$-dimensional 
critical exponents of \cite{48,49} as a non-trivial check on our five loop 
renormalization group functions. However, there is an alternative formulation 
of the $O(N)$ Wess-Zumino model based on a tensor decomposition of the $O(N)$ 
quartic interaction. This was studied in non-supersymmetric $\phi^3$ theory in 
six dimensions in \cite{50,51} at low loop order before being extended to four 
loops in \cite{52}. For the $O(3)$ tensor model an emergent $SU(3)$ symmetric 
fixed point was found \cite{50,52}. The exponents of the constituent scalar 
fields are equal as are the critical couplings thereby admitting the larger 
symmetry. This is in complete analogy with the emergent supersymmetry in the 
chiral XY Gross-Neveu model. As the tensor $O(N)$ Wess-Zumino model has the 
same formal cubic interaction we will confirm that the tensor $O(3)$ 
Wess-Zumino model too has an emergent $SU(3)$ fixed point which potentially 
adds to the set of theories connected to the dual behaviour in three 
dimensions. In light of this it is not inconceivable that the chiral XY 
Gross-Neveu theory can be extended to have a parallel tensor symmetry. In that 
case the emergent supersymmetry and $SU(3)$ symmetry should occur together at 
one of the fixed points of that tensor theory.

The paper is organized as follows. The basic properties of the Wess-Zumino
model that are necessary for the five loop renormalization are introduced in 
Section $2$. The computational strategy for this is reviewed in Section $3$ in
the context of the four loop renormalization while the details of the five loop
algorithm that we used are given in Section $4$. The main results for the 
original Wess-Zumino model are given in Section $5$ where the $\MSbar$ and MOM 
renormalization group functions are recorded. The next few sections are devoted
to the extension of the theory to include various symmetries. For instance, a 
group valued coupling is considered in Section $6$ where the $\epsilon$ 
expansion is used to compare exponents with estimates of the same quantities 
from the functional renormalization group and conformal bootstrap techniques. 
Endowing the Wess-Zumino model with an $O(N)$ symmetry is the subject of 
Sections $7$ and $8$ with the latter concentrating on the tensor $O(N)$ version
of the model. Section $9$ is devoted to the case where the basic coupling 
constant is replaced by a rank three symmetric tensor coupling. This forms the 
groundwork for studying the exponents connected with the three dimensional 
conformal manifold which is discussed in Section $10$. While the focus will 
have been on five loops to this point, Section $11$ explores some of the issues
that would arise if the six loop renormalization were to be computed. In fact 
we will provide the rational part of the six loop $\beta$-function in the 
$\MSbar$ scheme from the MOM scheme expression that was deduced from a Hopf 
algebra argument. Concluding renmarks are provided in Section $12$ and two 
appendices contain definitions and details of the tensor coupling 
renormalization.

\sect{Background.}

In this section we review the Wess-Zumino model \cite{1} and its properties 
that are relevant for the renormalization. The superspace bare action is given 
by
\begin{equation}
S ~=~ \int d^4 x \left[ \int d^2 \theta d^2 \bar{\theta} \,
\bar{\Phi}_0 (x,\bar{\theta}) e^{-2 \theta {\partialline} \bar{\theta}} 
\Phi_0 (x,\theta) ~+~ 
\frac{g_0}{3!} \int d^2 \theta \, \Phi_0^3(x,\theta) ~+~ 
\frac{g_0}{3!} \int d^2 \bar{\theta} \, \bar{\Phi}_0^3(x,\bar{\theta}) \right] 
\label{lagwz}
\end{equation}
where we use type I chiral bare superfields $\Phi_0(x,\theta)$ and
$\bar{\Phi}_0(x,\bar{\theta})$ and $g_0$ is the bare real coupling constant. 
The superspace coordinates $\theta$ and $\bar{\theta}$ are anticommuting and
represented by $2$ component spinors. In light of this the $2$~$\times$~$2$ 
covariant Pauli spin matrices $\sigma^\mu$ are used in spinor space leading to
the shorthand notation $\partialline$~$=$~$\sigma^\mu \partial_\mu$. The
$\sigma^\mu$ matrices satisfy the same Clifford algebra as the usual Dirac
$\gamma$ matrices. This version of the action, (\ref{lagwz}), was used for the
four loop calculation of \cite{6}. When the model was renormalized at lower
loop order, the component Lagrangian was employed, \cite{1,3}, and for 
completeness we note that the bare Lagrangian in that case is
\begin{equation}
L^{\mbox{\footnotesize{WZ}}} ~=~ i \bar{\psi}_0 \partialslash \psi_0 ~+~
\frac{1}{2} \left( \partial_\mu \sigma_0 \right)^2 ~+~
\frac{1}{2} \left( \partial_\mu \pi_0 \right)^2 ~+~
g_0 \bar{\psi} \left( \sigma_0 + i \pi_0 \gamma^5 \right) \psi ~+~ 
\frac{1}{24} g_0^2 \left( \sigma_0^2 + \pi_0^2 \right)^2 ~.
\label{lagwzc}
\end{equation}
It is this form of the Wess-Zumino Lagrangian that demonstrates the connection
with the emergent supersymmetry at one of the fixed points of the chiral XY 
Gross-Neveu-Yukawa theory, \cite{8,13,14,15,16,17}. The only difference between 
(\ref{lagwzc}) and that of the Gross-Neveu-Yukawa Lagrangian is that there are 
two coupling constants $g_1$ and $g_2$ respectively for the cubic and quartic 
interactions. At the emergent supersymmetry fixed point both $g_1$ and $g_2$ 
are equivalent, \cite{8,13,14,15,16,17}. Moreover the anomalous dimensions of 
all the fields are equivalent at the fixed point.

One useful property of (\ref{lagwz}) that we used in the renormalization is
that of the supersymmetry Ward identity, \cite{1,3}. If we define renormalized 
entities via the renormalization constants $Z_\phi$ and $Z_g$ with
\begin{equation}
\Phi_0 ~=~ \sqrt{Z_\Phi} \Phi ~~,~~
\bar{\Phi}_0 ~=~ \sqrt{Z_\Phi} \bar{\Phi} ~~,~~
g_0 ~=~ \mu^\epsilon Z_g g
\label{defrencon}
\end{equation}
where $\mu$ is a mass dimension $1$ object in $d$~$=$~$4$~$-$~$2\epsilon$
dimensions, then there is only one independent renormalization since it has 
been shown that the vertex function is finite \cite{1,3}. As a consequence we 
have
\begin{equation}
Z_g Z_\Phi^{\threehalves} ~=~ 1 
\label{susywi}
\end{equation}
which implies
\begin{equation}
\beta(a) ~=~ 3 a \gamma_\Phi(a)
\end{equation}
where
\begin{equation}
a ~=~ \frac{g^2}{16\pi^2}
\end{equation}
and $\gamma_\Phi(a)$ is the anomalous dimension of $\Phi$ and $\bar{\Phi}$. 

{\begin{table}[ht]
\begin{center}
\begin{tabular}{|c||r|r|r|}
\hline
$L$ & Real field & Complex field & Superfield $k_L$ \\
\hline
$1$ & $1$ & $1$ & $1$ \\
$2$ & $8$ & $7$ & $1$ \\
$3$ & $96$ & $90$ & $4$ \\
$4$ & $1942$ & $1797$ & $13$ \\
$5$ & $49710$ & $45183$ & $63$ \\
\hline
Total & $51757$ & $47078$ & $82$ \\
\hline
\end{tabular}
\end{center}
\begin{center}
\caption{Number of graphs at each loop order $L$ for $2$-point functions using
real component, complex component and superfield Lagrangians.}
\label{feynnum}
\end{center}
\end{table}}

{\begin{figure}[ht]
\begin{center}
\includegraphics[width=12cm,height=1.9cm]{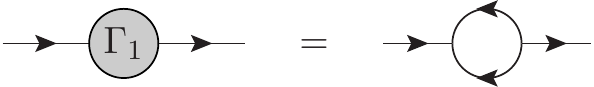}
\end{center}
\caption{One loop 1PI $2$-point function.}
\label{figtw1}
\end{figure}}

Having discussed the formulation of the superspace action we now outline the
strategy taken to carry out the five loop renormalization. One way to gauge the 
magnitude of a high loop order computation is to tally up the number of Feynman
graphs that have to be computed. This has been recorded in Table \ref{feynnum} 
where the data for the $2$-point function are given. These were compiled using 
the {\sc Qgraf} package, \cite{53}. Due to the supersymmetry Ward identity the 
vertex function is completely finite and so those graphs do not have to be 
calculated. There are several ways of counting the diagrams for (\ref{lagwz}) 
which will determine the strategy we will follow. Aside from a superspace 
approach, where the graph count is given in the final column of Table 
\ref{feynnum}, the theory can be formulated in terms of component fields. For 
(\ref{lagwz}) one can have real bosonic fields, as in (\ref{lagwzc}), or 
complex ones. The numbers of graphs for the bosonic field $2$-point functions 
are provided in the table too. Clearly there is a significantly larger number
of graphs for both component field calculations. We have chosen not to effect a
calculation for either component Lagrangian. This is not merely due to the 
number of graphs but also because in that case one would have to use 
dimensional reduction \cite{54} rather than dimensional regularization as the
latter does not preserve supersymmetry. The former regularization needs to be
implemented with care since additional evanescent fields have to be included in
the dimensionally regularized Lagrangian, \cite{55,56,57}. By contrast, 
although the superfield formalism has less than a total of $100$ graphs to 
compute, the superspace propagator in momentum space for (\ref{lagwz}) is
\begin{equation}
\langle \Phi(p,\theta) \bar{\Phi}(-p,\bar{\theta}) \rangle ~=~ 
\frac{\exp{(2 \theta \pline \bar{\theta}})}{p^2}
\label{propss}
\end{equation}
where $p$ is the momentum. Not only is the loop momentum integrated over in
superspace Feynman integrals but also the internal $\theta$ coordinates that
arise at each vertex of a supergraph. In \cite{4} a different form of the
superpropagator was used which involved the supercovariant derivatives
$D_\alpha$ and $\bar{D}^{\dot{\alpha}}$. These satisfy an algebra, known as
the $D$-algebra, which is used to simplify each superspace integral before the
integration over the loop momenta can be carried out. Ordinarily the
$D$-algebra is implemented by hand, which is straightforward to three loops for
(\ref{lagwz}), but this is not a practical approach for higher order 
calculations. As the superpropagator takes the form (\ref{propss}) in 
(\ref{lagwz}) it is possible to implement the corresponding $D$-algebra in an 
automatic Feynman diagram calculation. To do so we have written a module in the
symbolic manipulation language {\sc Form} and its threaded version {\sc Tform},
\cite{58,59}, to achieve this. Indeed the full computation could only be 
carried out with several key features of the language. For instance, the 
non-commuting function facility of {\sc Form} was essential for handling the 
$D$-algebra. Moreover, once it has been applied to each Feynman graph they can 
each be evaluated in dimensional regularization which is what we use 
throughout.

{\begin{figure}[ht]
\begin{center}
\includegraphics[width=12cm,height=3.3cm]{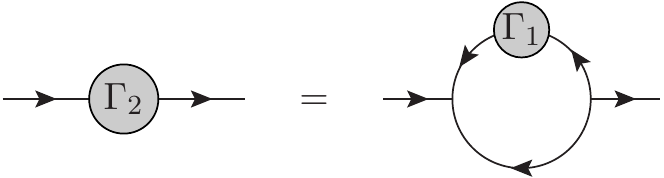}
\end{center}
\caption{Two loop 1PI $2$-point function.}
\label{figtw2}
\end{figure}}

\sect{Computational details.}

We now discuss the technical aspects behind the five loop calculation which 
will involve explaining the algorithm for constructing an automatic five loop
evaluation. In order to provide the necessary introduction to all the
ingredients required for this we focus on the lower loop Feynman graphs for the
moment and outline the first step of the process which is to reduce the 
superspace integrals to momentum space ones. For instance the one and two loop 
graphs contributing to the $1$-particle irreducible $\Phi$ $2$-point function 
are illustrated in Figures \ref{figtw1} and \ref{figtw2}. Our notation 
throughout will be that Feynman graphs in superspace will have directed lines 
as in these two figures. In this respect we note that from (\ref{lagwz}) the 
arrows on a propagator will all be directed towards the vertex or away. The 
immediate consequence for this is that there are no Feynman diagrams with 
subgraphs with an odd number of propagators. This is evident in Figures 
\ref{figtw1} and \ref{figtw2} as well as ones that appear later. Though where 
some figures have undirected propagators these represent Feynman integrals in 
ordinary momentum space and not superspace. We will also use $\Gamma_n$ to 
denote the $1$-particle irreducible graphs at $n$ loops and $C_n$ to indicate 
the connected $2$-point Green's function at the same order. This will simplify 
our illustration of the higher loop contributions to the $2$-point function.

{\begin{figure}[ht]
\begin{center}
\includegraphics[width=4.4cm,height=2.4cm]{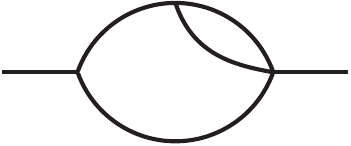}
\end{center}
\caption{Momentum space representation of $\Gamma_2$.} 
\label{figsc2}
\end{figure}}

For $\Gamma_1$ and $\Gamma_2$ the $D$-algebra is simple to implement. Since the
$\theta$ and $\bar{\theta}$ dependence in (\ref{propss}) is in the exponential
of each propagator then each graph will have one exponential that depends on 
all the anticommuting variables of each vertex of a Feynman diagram. So, for 
example, since $\Gamma_1$ has only two external vertices the overall 
exponential only depends on the external vertex variables and factors off 
consistent with renormalizability in superspace. In fact this is a feature of 
all higher loop graphs where the same factor emerges overall, \cite{6}. 
Moreover when $\Gamma_1$ appears embedded in a higher loop graph this factor 
that was external contributes to the $D$-algebra calculation of the remaining 
part of the higher loop graph. So for $\Gamma_2$ the only anticommuting 
variable dependence that remains is a factor 
$\exp{(2 \theta_1 \kline \bar{\theta}_1})$ where $k$ is the loop momentum and 
$\theta_1$ and $\bar{\theta}_1$ are to be integrated over, \cite{6}. This is 
after a change of variables on the original internal anticommuting variables. 
Expanding the exponential then only the quadratic terms are relevant for the 
$\theta_1$ and $\bar{\theta}_1$ integration after a trace is taken over the 
$\sigma^\mu$ matrices, \cite{6}. This is readily carried out by mapping the 
traces to the usual $\gamma$-matrix trace routine but adjusted so that the 
trace normalization is $2$ and not $4$. The resulting momentum space Feynman 
integral is represented by the graph of Figure \ref{figsc2}. We have detailed 
this relatively simple calculation as it is an example of a deeper observation 
for the $D$-algebra of $2$-point subgraphs in higher loop graphs. It turns out 
that in the resulting momentum space integral one of the propagators connecting
any $\Gamma_n$ subgraph is deleted in the same way as in Figure \ref{figsc2}. 
This lemma was useful in the five loop calculation. 

{\begin{figure}[hb]
\begin{center}
\includegraphics[width=15cm,height=6.9cm]{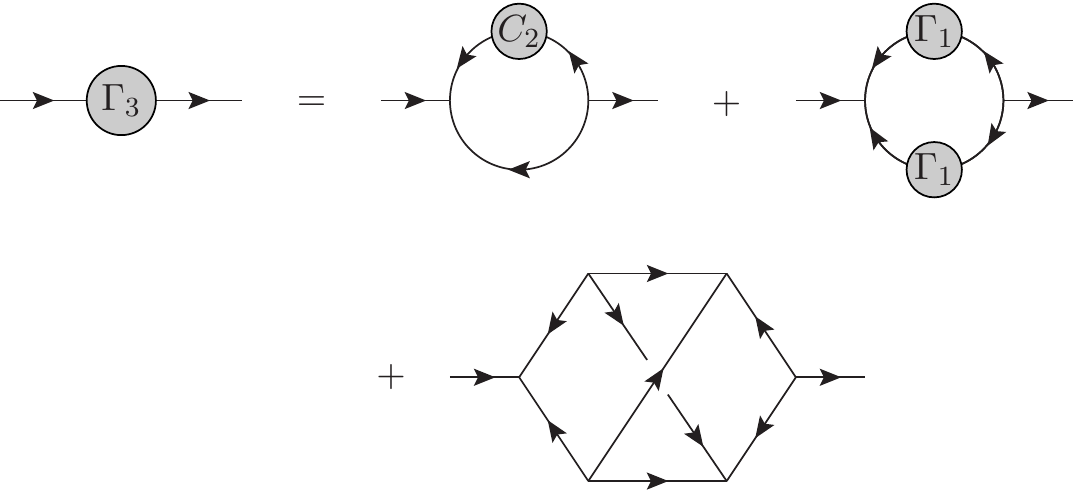}
\end{center}
\caption{Three loop 1PI $2$-point function.}
\label{figtw3}
\end{figure}}

At next order the $4$ three loop graphs are summarized in Figure \ref{figtw3}
where $C_2$ contains two diagrams. The non-planar graph is primitive and is
divergent. This is in contrast to the identical momentum space non-planar 
integral with undirected edges which is finite being equal to $20 \zeta_5$
where $\zeta_n$ is the Riemann zeta function. See, for example, the articles 
\cite{60,61,62,63} for the early discussion on the connection of the Riemann 
zeta series with the topology of high loop Feynman graph. To evaluate the 
primitive graph the $D$-algebra needs to be applied. This results in a set of 
momentum space integrals that are given in Figure \ref{figth1}. In displaying 
these we note that in total there are $14$ integrals but we have used 
left-right and up-down symmetry to reduce these to the four independent 
topologies. The non-planar graph contains the irreducible numerator which 
becomes apparent when the trace is taken over the fermion propagators which are
represented by the dotted lines. It is important to note that these integrals 
result from the $D$-algebra and have no connection with the Feynman integrals 
that one would have to compute using the component Lagrangian. We have detailed
the reduction for this graph as it differs from the way it was evaluated in the
four loop calculation of \cite{6}. There the external momentum was nullified in
the numerator of the integral after carrying out the integration over the
anticommuting superspace coordinates. For the five loop renormalization we have
to determine the integral to the $O(\epsilon)$ term rather than just isolate
the divergence. We note that comment was also made in \cite{64} as to how to 
effect the $D$-algebra for this topology.

{\begin{figure}[hb]
\begin{center}
\includegraphics[width=12cm,height=7.0cm]{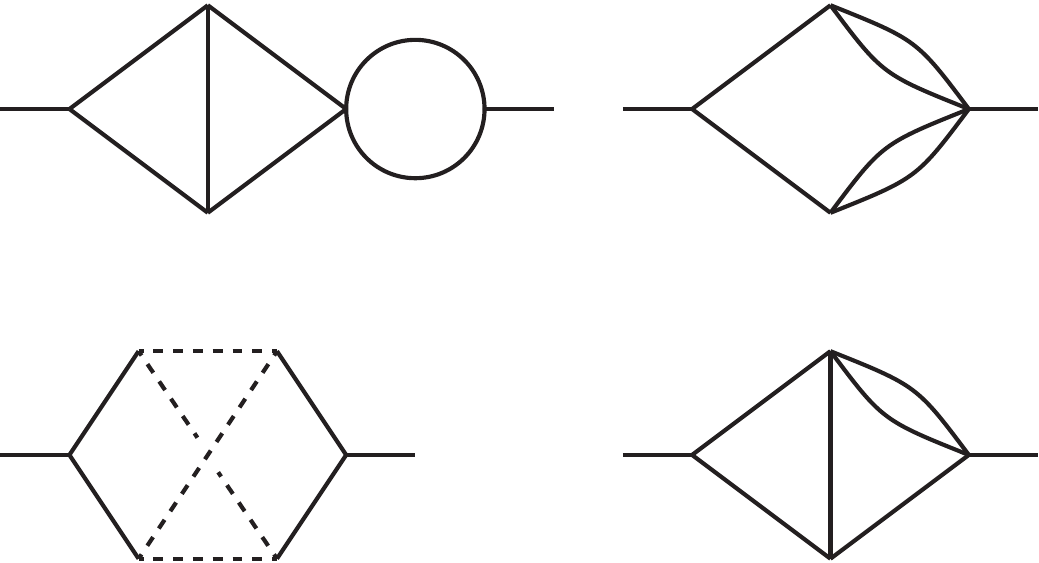}
\end{center}
\caption{Momentum space integrals after applying the $D$-algebra to the three
loop non-planar graph.}
\label{figth1}
\end{figure}}

At the next loop order the $13$ $2$-point function graphs are given in Figure
\ref{figtw4} where we have introduced a shorthand definition of the two loop
non-planar vertex which will be denoted by $V_2$ and is defined in Figure
\ref{figvnp}. The subgraph $\bar{V}_2$ of Figure \ref{figtw4} corresponds to
the graph of Figure \ref{figvnp} but with the direction of the external legs
reversed which is the origin of the conjugate notation. In Figure \ref{figtw4} 
and later figures we do not display all the subgraph mirror images. To 
illustrate what we mean by subgraph mirror image there is another graph similar
to the final graph on the first row of Figure \ref{figtw4} where the 
$\bar{V}_2$ subgraph is translated to the other external vertex whence it would
become $V_2$. However in performing this translation there is {\em no} 
reflection of the direction of any of the propagators which remains unchanged. 
The graphs of Figure \ref{figtw4} follow a similar pattern to those at three 
loops in that the majority are decorations of the previous loop order. This 
includes the three cases where there are propagator corrections on the three 
loop primitive. The remaining undecorated planar four loop graph is a primitive
at this order. It will have to be evaluated without the re-routing 
simplification that was used in \cite{6} since we will need the finite part. 
Moreover it transpires that there are a significantly larger number of 
momentum space integrals that result from the $D$-algebra compared to those of 
the three loop primitive. 

{\begin{figure}[ht]
\begin{center}
\includegraphics[width=16.0cm,height=5.8cm]{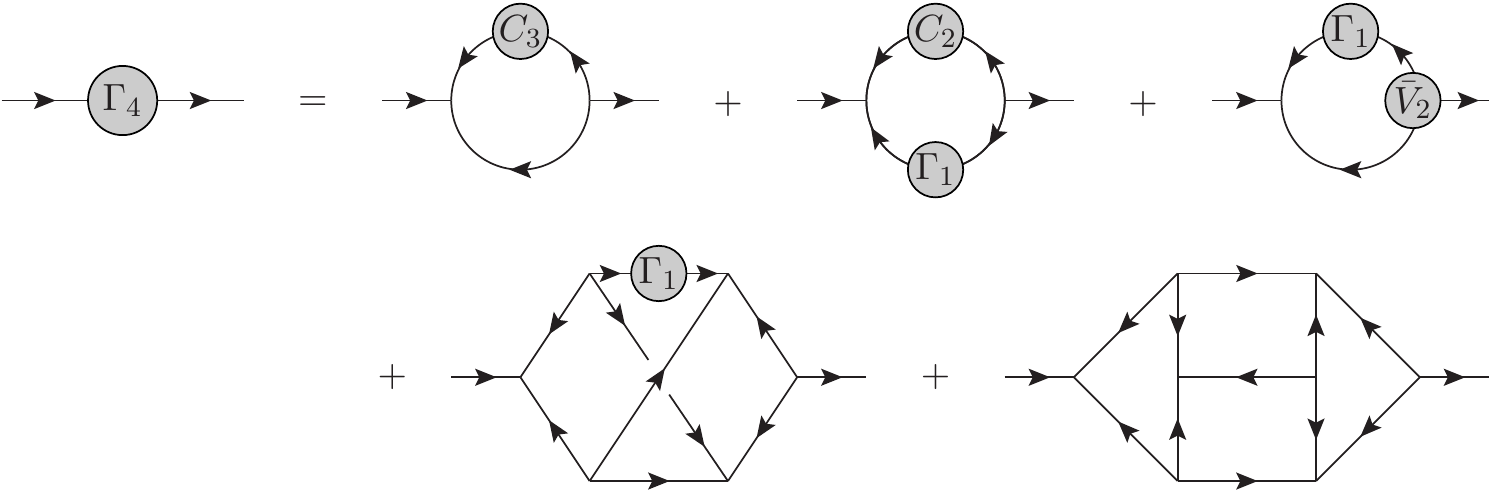}
\end{center}
\caption{Four loop 1PI $2$-point function.}
\label{figtw4}
\end{figure}}

{\begin{figure}[ht]
\begin{center}
\includegraphics[width=9.95cm,height=4.0cm]{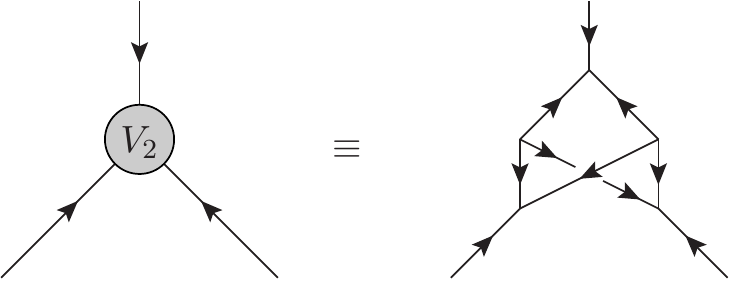}
\end{center}
\caption{Two loop non-planar vertex correction.}
\label{figvnp}
\end{figure}}

Although our aim is to renormalize (\ref{lagwz}) to five loops we pause at this
point to discuss the techniques we used to evaluate the momentum space 
integrals. To four loops the main tools we employed were the three and four 
loop packages {\sc Mincer}, \cite{22,23}, and {\sc Forcer}, \cite{20,21}, 
respectively. These are {\sc Form} encoded packages that evaluate dimensionally
regularized $2$-point functions up to various orders in $\epsilon$. While 
{\sc Mincer} is tied to theories in four dimensions {\sc Forcer} has the 
capacity to determine the $\epsilon$ expansion of momentum space integrals in 
theories with even critical dimensions. The usefulness of {\sc Mincer} for 
example in its application to the Wess-Zumino model is that it can determine 
the part of the $\beta$-function that solely involves rational numbers to five 
loops. While it can equally be applied to the evaluation of most of the four 
loop graphs we had to use {\sc Forcer} to find the primitive of Figure 
\ref{figtw4} to the finite part. Another technique we used, that is not limited
to the computation of $2$-point functions, was the Laporta algorithm \cite{19} 
encoded in the {\sc Reduze} package, \cite{65,66}. This was primarily required 
to check the four loop primitive graphs but was also used more extensively at 
five loops to verify the simple pole of certain difficult primitives. In 
applying both {\sc Mincer} and {\sc Forcer} to all the momentum space integrals
that result from the $D$-algebra we have verified the four loop 
$\beta$-function of \cite{6}. As far as we are aware this is the first direct 
evaluation of the graphs where there has been simplification involving the
external momenta to extract the divergences.

{\begin{figure}[hb]
\begin{center}
\includegraphics[width=16.0cm,height=3.0cm]{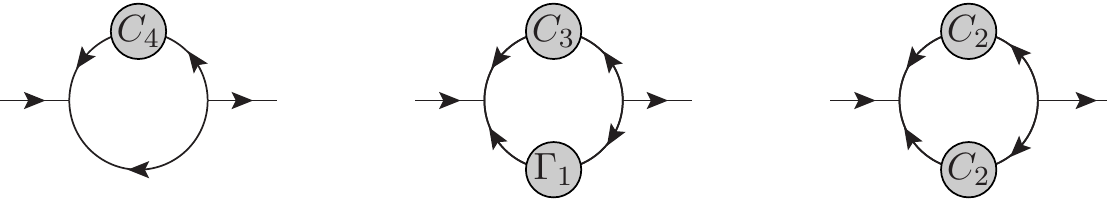}
\end{center}
\caption{Five loop graphs based on the decoration of $\Gamma_1$.}
\label{figtw51}
\end{figure}}

{\begin{figure}[ht]
\begin{center}
\includegraphics[width=16.0cm,height=11.0cm]{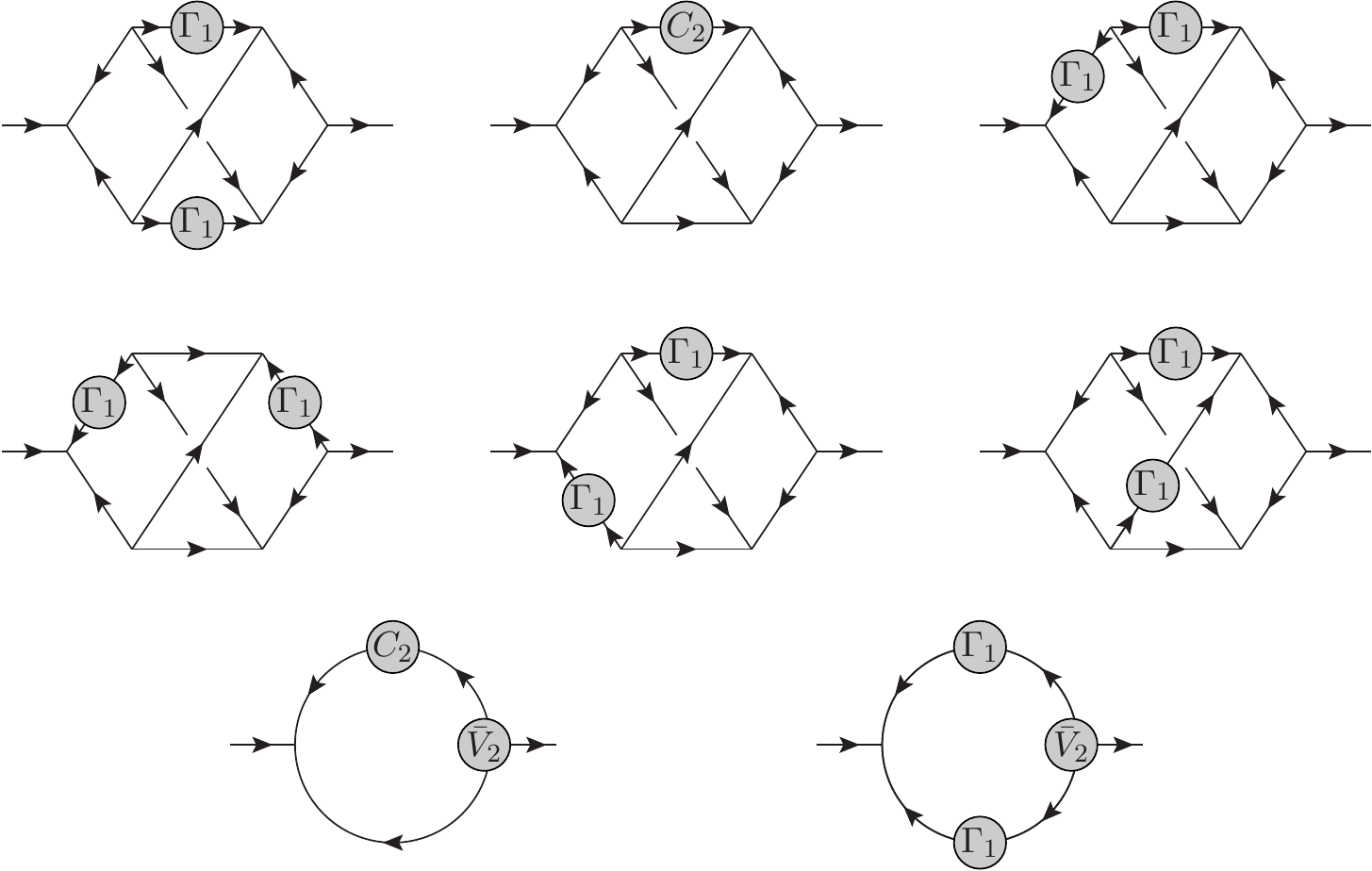}
\end{center}
\caption{Five loop graphs based on decoration of three loop primitive graph.}
\label{figtw53}
\end{figure}}

\sect{Five loop calculation.}

We turn now to the details of the five loop renormalization which first 
requires the evaluation of the $63$ graphs. We have chosen to illustrate these
in a sequence of Figures and classify the graphs by the underlying skeleton
topology. Those given by propagator dressings of $\Gamma_1$ are shown in Figure 
\ref{figtw51} where we note that $C_3$ and $C_4$ include the respective three 
and four loop primitives. As all the subgraphs within $C_n$ and $\Gamma_n$ in 
the figure are available to the finite part from lower loop computations their
contributions to $\gamma_\Phi(a)$ are straightforward to determine. However
this is not the case for the decoration of the three loop primitive where the
graphs are illustrated in Figure \ref{figtw3}. The reason for this is that
after performing the superspace integration over the internal anticommuting
coordinates the set of momentum space integrals do not have a direct
correspondence with the decoration of the topologies of Figure \ref{figth1} in 
all possible ways. This is not unrelated to the irreducible scalar products
that arise. For an $L$ loop $2$-point Feynman graph there are 
$\half (L-1)(L-2)$ irreducible scalar products. So to address this issue using
a Laporta algorithm approach would require an integral reduction of significant
size. Instead as the four loop {\sc Forcer} package has no direct applicability
we have followed a different tactic and that is to apply the method outlined in
the five loop renormalization of QCD in \cite{25}. There the divergent part of
similar five loop integrals was determined by a combination of infrared 
rearrangement and the method of subtractions. The external momentum is 
re-routed through the graph such that it enters through one current external 
vertex but exits via the first vertex adjacent to that one. For some of the 
graphs of Figure \ref{figtw53} there are several ways of achieving this which 
gives a check on the procedure. As noted in \cite{25} this produces an integral
containing a four loop $2$-point subgraph that can then be evaluated using the 
{\sc Forcer} algorithm, \cite{20,21}. In other words this package is used 
{\em indirectly} to extract the five loop divergences. For the Wess-Zumino 
model there are several additional simplifications compared to the QCD case. 
Aside from the fact that the superspace graphs are zero dimensional, there are 
fewer graphs and within these there are a small set of irreducible scalar 
products. Therefore we have constructed a procedure to effect the subtraction 
approach for the subset of graphs of Figure \ref{figtw53}. As a check on our 
method we have applied it to the similar decorations of the three loop 
primitive shown in Figure \ref{figtw4} since we know the correct answer from 
their direct evaluation in {\sc Forcer}. 

{\begin{figure}[ht]
\begin{center}
\includegraphics[width=16.0cm,height=10.0cm]{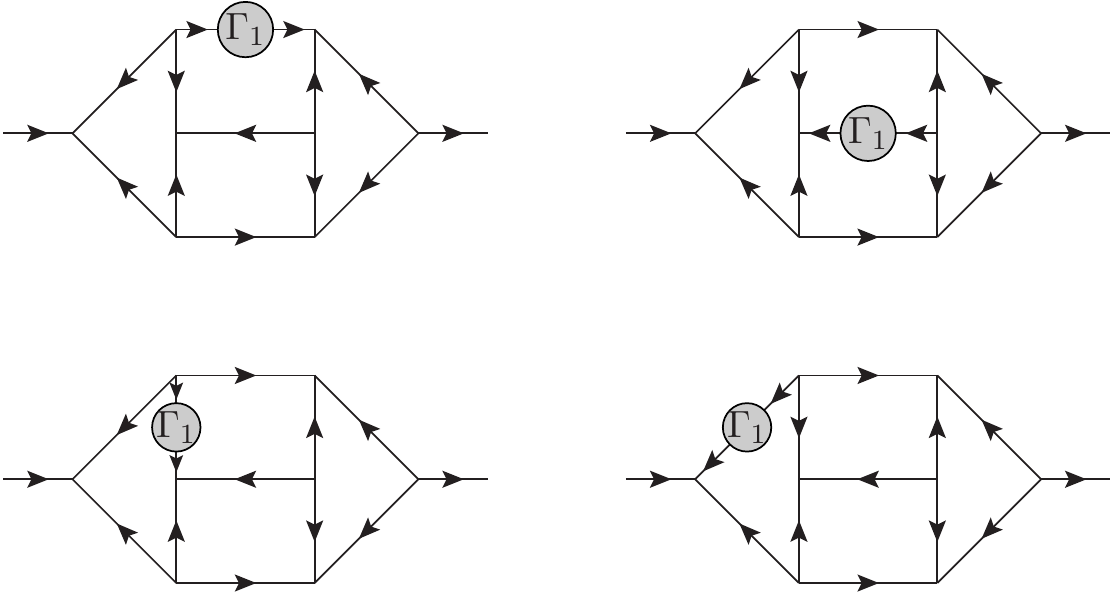}
\end{center}
\caption{Five loop graphs based on decoration of four loop primitive graph.}
\label{figtw54}
\end{figure}}

In applying that check we thereby verify that it is a valid procedure for 
evaluating the decoration of the four loop primitive graph of Figure
\ref{figtw4}. The corresponding representative five loop graphs are shown in 
Figure \ref{figtw54} and it is clear that the re-routing approach that exploits
{\sc Forcer} is one of the few strategies we have. However for this skeleton 
topology we were also able to check both poles in $\epsilon$ of the four graphs
of Figure \ref{figtw54} by following the algorithm given in \cite{6} for the 
underlying four loop graph. That method did not re-route the external momentum 
but set the external momentum to zero where it appeared in the numerator of the
integral after the $D$-algebra had been applied. At five loops this produced a 
topology with a four loop $2$-point subgraph which had a different structure to
that of the external momentum re-routing but which could equally well be 
evaluated using {\sc Forcer}. For each of the four cases we obtained consistent
expressions for the divergences. 

{\begin{figure}[ht]
\begin{center}
\includegraphics[width=14.0cm,height=9.0cm]{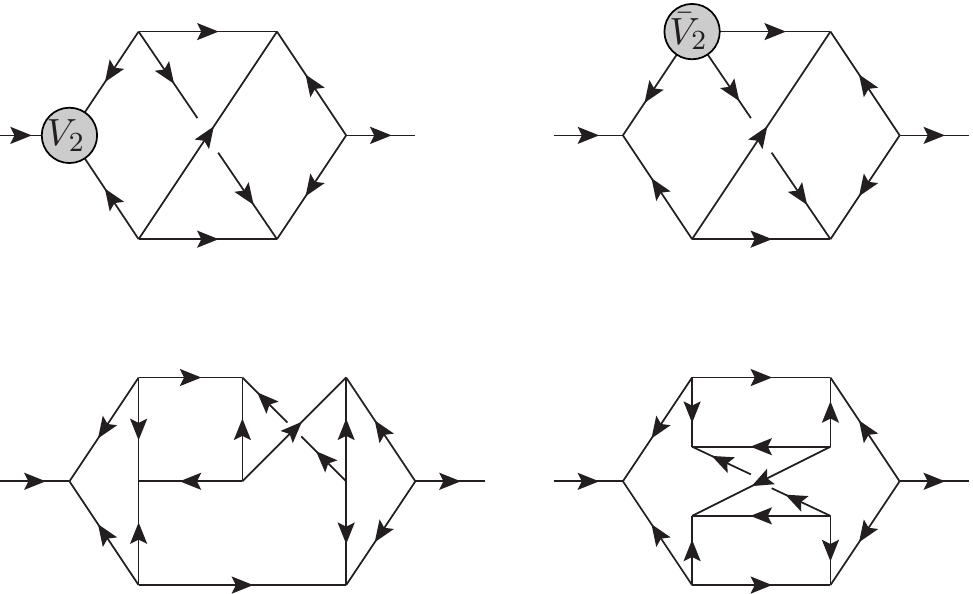}
\end{center}
\caption{Five loop primitive graphs.}
\label{figpr5}
\end{figure}}

The final subset of graphs for the five loop renormalization are provided in
Figure \ref{figpr5} and are the primitives. These can be divided into two
classes. One class involves the decoration of the three loop primitives by
non-planar vertex corrections. In fact the first graph on the top row is
$\Gamma_1$ where both external vertices are dressed with $V_2$ and $\bar{V}_2$.
For both these graphs we have evaluated them in several different ways. For the
double dressing of $\Gamma_1$, for instance, we can merely multiply the pole of
$\Gamma_1$ by the finite value of $V_2$. We have determined this by computing
the two loop vertex function using either {\sc Mincer} or {\sc Forcer} with one
external momentum nullified. As an alternative we have also computed the
underlying integral without any restriction on the external momentum. In other 
words the integral is evaluated at a non-exceptional subtraction point. More 
specifically we considered the fully symmetric point where the squares of the 
external momenta are all equal. After applying the {\sc Form} $D$-algebra 
module we used the {\sc Reduze} encoding of the Laporta algorithm to express 
the diagram in terms of the various two loop master integrals which are 
available in \cite{67,68,69,70}. Either method produces the value of $3\zeta_3$
for the finite part of $V_2$ and its conjugate. With this value it transpires 
that both graphs in the top row of Figure \ref{figpr5} are proportional to 
$\zeta_3^2$. In each case we have checked this argument by re-routing the 
external momentum. As the graphs are primitive where the momentum enters the 
graph and leaves is not important as long as it is at two separate vertices. 
This includes the case where only one external momentum is re-routed which we 
used on the lower loop decorated primitives. The divergence was extracted using
{\sc Forcer}. Whichever approach we used the same simple pole resulted for both 
these graphs. It also tallies with the method used in \cite{6} for the 
underlying skeleton topology. What is worth noting about this primitive is that
in non-supersymmetric models graphs with a non-planar vertex subgraph 
correction would not ordinarily be regarded as a primitive. Indeed in the 
conventional understanding of the appearance of $\zeta_n$ to five loops in 
$2$-point function calculations the primitives are associated with $\zeta_3$, 
$\zeta_5$ and $\zeta_7$. This product of $\zeta_n$ values in a primitive 
appears to be solely peculiar to the Wess-Zumino model. This leaves the graphs 
of the lower row of Figure \ref{figpr5} to evaluate. These do not have any 
vertex subgraphs and so we do not have the same guidance into the final residue
of the simple pole. However we have applied the same techniques to extract the 
divergence and find that both involve the underlying number which is 
$\frac{441}{8} \zeta_7$ if one omits the symmetry factor. That this combination
appears is not surprising since it is not unrelated to a parallel primitive 
Feynman graph in scalar $\phi^4$ theory. In \cite{61,62,63,71} the primitive 
graph was evaluated by the use of conformal integration or the uniqueness 
method, \cite{72,73,74}, after an initial numerical evaluation \cite{61,62,63}.
In fact the residue was also recorded for what is termed the zigzag graph in 
the prescient work of Broadhurst in \cite{60}. In particular it is recorded in 
Table 3 of that article where it corresponds to diagram c of Figure 6 there. 
The residue of the other five loop primitive shown in the first row of Figure 
\ref{figpr5} is also apparent in Table 3 of \cite{60} via diagrams d and e of 
Figure 6. The fact that the zigzag topology arises in the seemingly 
topologically unconnected lower row graphs of Figure \ref{figpr5} is as a 
consequence of the $D$-algebra. In the simplification of the numerator scalar 
products after using the method of \cite{6} several propagators are deleted to 
leave the zigzag graph. 

Having outlined in detail in this and the previous section how we have 
evaluated all the diagrams to five loops to the requisite order in $\epsilon$
to carry out the full renormalization we now note some of the practical aspects
of the automatic routine we have constructed. First all the superspace graphs
are generated electronically using the {\sc Fortran} based {\sc Qgraf}
package, \cite{53}. To ease the implementation of the $D$-algebra routine that 
we have written we use the {\sc Qgraf} setting that equates to the {\sc Mincer}
or {\sc Forcer} setup where each propagator is allocated a momentum $p_i$. 
After the $D$-algebra has been carried out either the energy-momentum 
conservation is implemented at each vertex to reduce the number of $p_i$ to the
number of loops or values of each $p_i$ are substituted explicitly. The latter 
is used for the cases where the {\sc Reduze} package was required since the 
integral families are defined by the explicit values of the internal loop 
momenta. This represents the core of the integration routine. Though for those 
five loop graphs where a re-routing was necessary to find the divergence the 
value was constructed in a separate routine and the result included in the 
automatic calculation which reduces the run time. This is particularly 
important since although the focus thus far has been on the renormalization of 
(\ref{lagwz}) we have also considered extensions of this action such as that
with $O(N)$ symmetry which have a significantly larger number of graphs to be 
determined. Once all the graphs have been computed they are summed before the 
renormalization is carried out. This follows the established routine of
\cite{75} where the calculation is carried out for bare parameters which in the
Wess-Zumino case is the coupling constant. Its renormalized partner is 
introduced through (\ref{defrencon}). As there is one independent 
renormalization constant the coupling constant counterterms are formally
deduced by iteratively solving (\ref{susywi}) and expressing them in terms of 
the $Z_\Phi$ counterterms. These relations are then included in the routine 
that ultimately determines the values of the $Z_\Phi$ counterterms. We close 
with a final remark on the evaluation of the diagrams. Although early loop 
computations of the $\beta$-function primarily concentrated on extracting the 
result in the $\MSbar$ scheme, in \cite{4} the $\beta$-function in the momentum
(MOM) subtraction scheme was also determined at three loops. This required 
knowledge of the higher order terms in the $\epsilon$ expansion of each Feynman
graph to two loops. Those at three loop were not necessary, \cite{4}, as they 
would contribute to the four loop MOM $\beta$-function. Therefore, as we have 
used {\sc Forcer} to compute the four loop graphs we have also found the finite
part of those diagrams as well as the $O(\epsilon)$ terms. So we will also be 
able to determine the five loop MOM scheme $\beta$-function for (\ref{lagwz}) 
and its extensions.

\sect{Results.}

After discussing the technical details of how we evaluated all $63$ five loop 
graphs we now provide the results together with comments on internal checks on
the final renormalization group functions. We find in the $\MSbar$ scheme that 
the field anomalous dimension is
\begin{eqnarray}
\gamma_\Phi(a) &=& \frac{1}{2} a ~-~ \frac{1}{2} a^2 ~+~ 
\left[ 12 \zeta_3 + 5 \right] \frac{a^3}{8} ~+~
\left[ 18 \zeta_4 - 60 \zeta_3 - 80 \zeta_5 - 9 \right] \frac{a^4}{8}
\nonumber \\
&& +~ \left[ 504 \zeta_3^2 + 858 \zeta_3 - 441 \zeta_4 + 1828 \zeta_5
- 900 \zeta_6 + 2646 \zeta_7 + 79 \right] \frac{a^5}{32} ~+~ O(a^7)
\label{gam5}
\end{eqnarray}
implying
\begin{eqnarray}
\beta(a) &=& \frac{3}{2} a^2 ~-~ \frac{3}{2} a^3 ~+~ 
\left[ 36 \zeta_3 + 15 \right] \frac{a^4}{8} ~+~
\left[ 54 \zeta_4 - 180 \zeta_3 - 240 \zeta_5 - 27 \right] \frac{a^5}{8}
\nonumber \\
&& +~ \left[ 1512 \zeta_3^2 + 2574 \zeta_3 - 1323 \zeta_4 + 5484 \zeta_5
- 2700 \zeta_6 + 7938 \zeta_7 + 237 \right] \frac{a^6}{32} \nonumber \\
&& +~ O(a^7)
\label{bet5}
\end{eqnarray}
for the $\beta$-function which are some of the main results of the article. In 
arriving at (\ref{gam5}) the non-simple poles of $Z_\Phi$ are not independent 
from the property of the renormalization group and are related to the residues 
of the lower loop order poles. That this is consistent validates that aspect of
the calculation. Another non-trivial check on the result will be discussed in a
later section. Also structurally the five loop $\beta$-function is formally the
same as its scalar $\phi^4$ counterpart, \cite{61,62,63,75}, in terms of the 
rational and irrational dependence. 

As the MOM scheme was considered in \cite{4} we can also provide the
renormalization group functions to five loops for that case. For (\ref{lagwz})
the MOM scheme is defined such that at the subtraction point there are no
$O(a)$ corrections to the $2$-point function. In other words after 
renormalization in that scheme the $2$-point function is unity in superspace at
the subtraction point. This will determine the MOM expression for $Z_\Phi$.
However in extracting it from the $2$-point function the coupling constant has
also to be renormalized in the same scheme. This is effected by ensuring that
the supersymmetry Ward identity (\ref{susywi}) is preserved as otherwise the
scheme would not be consistent with this symmetry. Applying this procedure to 
the $2$-point function and retaining the necessary terms depending on 
$\epsilon$ at each loop order we arrive at the results
\begin{eqnarray}
\gamma_\Phi^{\MOMs}(a) &=& \frac{1}{2} a ~-~ \frac{1}{2} a^2 ~+~ 
\left[ 6 \zeta_3 + 7 \right] \frac{a^3}{4} ~-~
\left[ 13 \zeta_3 + 20 \zeta_5 + 20 \right] \frac{a^4}{2}
\nonumber \\
&& +~ \left[ 216 \zeta_3^2 + 772 \zeta_3 + 230 \zeta_5 + 1323 \zeta_7 
+ 1222 \right] \frac{a^5}{16} ~+~ O(a^6)
\label{gam5mom}
\end{eqnarray}
and
\begin{eqnarray}
\beta^{\MOMs}(a) &=& \frac{3}{2} a^2 ~-~ \frac{3}{2} a^3 ~+~ 
3 \left[ 6 \zeta_3 + 7 \right] \frac{a^4}{4} ~-~
3 \left[ 13 \zeta_3 + 20 \zeta_5 + 20 \right] \frac{a^5}{2}
\nonumber \\
&& +~ 3 \left[ 216 \zeta_3^2 + 772 \zeta_3 + 230 \zeta_5 + 1323 \zeta_7 
+ 1222 \right] \frac{a^6}{16} ~+~ O(a^7)
\label{bet5mom}
\end{eqnarray}
where both are provided for later purposes. Our convention is that when a 
renormalization group function is labelled with MOM then the coupling constant
$a$ is the MOM coupling constant rather than the $\MSbar$ one. For cases where
there is potential ambiguity we denote the MOM coupling constant by
$a^{\MOMs}$. Where this no ambiguity $a$ will be regarded as the $\MSbar$
variable. There are several interesting features of (\ref{gam5mom}) and 
(\ref{bet5mom}). First the coefficients of the one and two loop terms of
$\gamma_\Phi^{\MOMs}(a)$ are the same as the $\MSbar$ $\gamma_\Phi(a)$. This is
a consequence of the supersymmetry Ward identity ensuring the $\beta$-function
and $\gamma_\Phi(a)$ are proportional. It appears to contradict the accepted
position that only the $\beta$-function in a single coupling theory is scheme 
independent at two loops. In scalar $\phi^4$ theory the two loop term of the
field anomalous dimension is independent of the renormalization scheme but this
is for a trivial reason since it is the first non-zero term. The other peculiar
feature of (\ref{gam5mom}) for example is that there are no terms involving
$\zeta_{2n}$. In other words only the odd integer argument Riemann zeta
function numbers are present. Hence there are no terms which involve even
powers of $\pi$ at least to five loops.

While we have found the five loop result for $\gamma_\Phi^{\MOMs}(a)$ by direct
evaluation it is possible to determine it by another method. This was discussed
in \cite{4} and involves constructing the map between the coupling constant in 
one scheme with that in the other. It only requires the four loop calculation 
of $Z_\Phi$ is each scheme to achieve this. First, we define the two conversion
functions
\begin{equation}
C_g(a) ~=~ \left( \frac{Z_g^{\MSbars}}{Z_g^{\MOMs}} \right)^2 ~~~~,~~~~
C_\Phi(a) ~=~ \frac{Z_\Phi^{\MOMs}}{Z_\Phi^{\MSbars}}
\end{equation}
where each renormalization constant depends on the coupling constant in the
indicated scheme. Although each renormalization constant has poles in 
$\epsilon$ the conversion function is finite as $\epsilon$~$\to$~$0$. This is
because the variables $a$ and $a^{\mbox{$\MOMs$}}$ are not independent and in
fact ensuring $C_g(a)$ is finite order by order determines the relation between
the two. Thus we find 
\begin{eqnarray}
a^{\MOMs} &=& a \left[ 1 ~-~ 3 a ~+~ \frac{57}{4} a^2 ~-~ 
[ 64 \zeta_3 + 18 \zeta_4 + 659 ] \frac{a^3}{8} 
\right. \nonumber \\
&& \left. ~~~+~ 
[ 2094 \zeta_3 - 24 \zeta_3^2 + 351 \zeta_4 + 504 \zeta_5 + 300 \zeta_6 
+ 8895 ] \frac{a^4}{16} \right] ~+~ O(a^6)
\label{ccmommap}
\end{eqnarray}
where $a$ on the right side is in the $\MSbar$ scheme. Equally once 
(\ref{ccmommap}) has been established the wave function scheme conversion 
function $C_\Phi(a)$ can be deduced as
\begin{eqnarray}
C_\Phi(a) &=& 1 ~-~ a ~+~ \frac{15}{4} a^2 ~-~ 
[ 64 \zeta_3 + 18 \zeta_4 + 471 ] \frac{a^3}{24} \nonumber \\
&& +~ [ 1838 \zeta_3 - 24 \zeta_3^2 + 279 \zeta_4 + 504 \zeta_5 + 300 \zeta_6
+ 6156 ] \frac{a^4}{48} ~+~ O(a^5) ~.
\end{eqnarray}
Equipped with these relations and using the renormalization group formalism the
MOM renormalization group functions can be calculated using
\begin{equation}
\beta^{\mbox{$\MOMs$}} (a^{\mbox{$\MOMs$}}) ~=~
\left[ \beta(a) \frac{\partial a_{\mbox{$\MOMs$}}}{\partial a} 
\right]_{ \MSbars \rightarrow \MOMs }
\label{betamomms}
\end{equation}
and
\begin{equation}
\gamma_\Phi^{\MOMs} ( a^{\MOMs} ) ~=~ \left[ \gamma_\Phi^{\MSbars}(a) ~+~ 
\beta^{\MSbars}(a) \frac{\partial ~}{\partial a} \ln C_\Phi^{\MOMs}(a) 
\right]_{ \MSbars \rightarrow \MOMs } 
\label{gammamomms}
\end{equation}
where the restriction indicates that because the quantity inside the square 
brackets is a function of $a$ it has to be mapped to the $a^{\mbox{$\MOMs$}}$
variable. This is achieved by the mapping which is the inverse of 
(\ref{ccmommap}). Following this we reproduce the {\em five} loop MOM results
(\ref{gam5mom}) and (\ref{bet5mom}). Only four loop information is required for
this exercise which is also the reason why the finite parts of the five loop 
Feynman graphs are not required to determine the five loop MOM renormalization 
group functions.

\sect{Group valued Wess-Zumino model.}

We now turn to a variation on (\ref{lagwz}) which is to have a multiplet of $N$
superfields where the interaction contains a real tensor denoted by $d^{ijk}$ 
where $1$~$\leq$~$i$~$\leq$~$N$. The bare action is
\begin{equation}
S ~=~ \int d^4 x \! \left[ \int d^2 \theta d^2 \bar{\theta} \,
\bar{\Phi}_0^i (x,\bar{\theta}) e^{-2 \theta {\partialline} \bar{\theta}} 
\Phi_0^i (x,\theta) \,+\,
g_0 \frac{d^{ijk}}{3!} \! \int d^2 \theta \, \Phi_0^i \Phi_0^j \Phi_0^k \,+\,
g_0 \frac{d^{ijk}}{3!} \! \int d^2 \bar{\theta} \, 
\bar{\Phi}_0^i \bar{\Phi}_0^j \bar{\Phi}_0^k
\right] 
\label{lagwzt}
\end{equation}
where the aim is to determine the coupling constant renormalization. The
notation for the tensor derives from that of six dimensional scalar $\phi^3$
theory \cite{77,78}. To accommodate the different combinations of tensors that
appear in loop calculations a useful notation was also provuided in
\cite{77,78} and extended to the four loop renormalization in \cite{79}. This
will introduce scalar objects $T_i$ that play a similar role as the group
Casimirs of a non-abelian gauge theory. As the diagrams comprising the 
$2$-point function of (\ref{lagwzt}) only have subgraphs with an even number of
propagators, we only need to recall the relevant tensor combinations that will 
appear to five loops. These are 
\begin{eqnarray}
&& T_2 \delta^{i j} ~=~ d^{i i_1 i_2} d^{j i_1 i_2} \nonumber \\
&& T_5 d^{i j k} ~=~ 
d^{i i_1 i_2} d^{j i_3 i_4} d^{k i_5 i_6} d^{i_1 i_3 i_5} d^{i_2 i_4 i_6} 
\nonumber \\
&& T_{71} d^{i j k} ~=~ 
d^{i i_1 i_2} d^{j i_3 i_4} d^{k i_5 i_6} d^{i_1 i_3 i_7} d^{i_2 i_5 i_8}
d^{i_4 i_6 i_9} d^{i_7 i_8 i_9} \nonumber \\
&& T_{94} d^{i j k} ~=~
d^{i i_1 i_2} d^{j i_3 i_4} d^{k i_5 i_{12}} d^{i_1 i_5 i_6} d^{i_2 i_7 i_8} 
d^{i_3 i_9 i_{12}} d^{i_4 i_{10} i_{11}} d^{i_6 i_7 i_{10}}
d^{i_8 i_9 i_{11}} ~. 
\label{Tdef}
\end{eqnarray}
The first digit of the subscript of any $T_{i}$ indicates the number of
$d^{ijk}$ tensors comprising the underlying graph or equivalently the number of
propagators. So $T_2$ denotes the one loop $2$-point bubble. The others 
correspond to vertex functions at two, three and four loops respectively. 
Contracting these tensors with another tensor produces a $2$-point function 
topology. These then isolate the respective three and four loop primitive 
graphs of Figures \ref{figtw3} and \ref{figtw4}. At five loops the graphs that 
involve $T_{94}$ are those of the lower row of Figure \ref{figpr5}. Those in 
the top row involve $T_5^2$. One advantage of this notation is that the 
contribution to the renormalization group functions from the primitive at each 
loop order can be identified and followed within a calculation. Such an 
analysis was performed for scalar $\phi^4$ theory in \cite{34} and suggested 
that the percentage contribution from the primitive graphs at each loop order 
increases with the number of loops. 
 
Therefore we have computed the renormalization group functions for 
(\ref{lagwzt}) and find  
\begin{eqnarray}
\gamma_T(a) &=& \frac{1}{2} T_2 a ~-~ \frac{1}{2} T_2^2 a^2 ~+~ 
T_2 \left[ 12 \zeta_3 T_5 + 5 T_2^2 \right] \frac{a^3}{8} 
\nonumber \\
&& +~ T_2 \left[ 18 \zeta_4 T_2 T_5 - 60 \zeta_3  T_2 T_5 - 80 \zeta_5 T_{71} 
- 9 T_2^3 \right] \frac{a^4}{8}
\nonumber \\
&& +~ T_2 \left[ 12 \zeta_3 T_2^4 + 79 T_2^4 + 846 \zeta_3 T_2^2 T_5 
- 441 \zeta_4 T_2^2 T_5 - 612 \zeta_5 T_2^2 T_5
- 216 \zeta_3^2 T_2 T_{71} 
\right. \nonumber \\
&& \left. ~~~~~~~
+ 2440 \zeta_5 T_2 T_{71} - 900 \zeta_6 T_2 T_{71} + 720 \zeta_3^2 T_5^2 
+ 2646 \zeta_7 T_{94} \right] \frac{a^6}{32} ~+~ O(a^7)
\label{gam5T}
\end{eqnarray}
for the anomalous dimension in the $\MSbar$ scheme. As there is only one 
coupling and chiral field in (\ref{lagwzt}) the original supersymmetry Ward 
identity (\ref{susywi}) is satisfied. At the same time it is a simple matter to
determine the MOM scheme version of (\ref{gam5T}) giving
\begin{eqnarray}
\gamma_T^{\MOMs}(a) &=& \frac{1}{2} T_2 a ~-~ \frac{1}{2} T_2^2 a^2 
\nonumber \\
&& +~ T_2 \left[ 6 \zeta_3 T_5 + 7 T_2^2 \right] \frac{a^3}{4} ~-~
T_2 \left[ 15 \zeta_3 T_2 T_5 - 2 \zeta_3 T_2^3 + 20 \zeta_5 T_{71}
+ 20 T_2^3 \right] \frac{a^4}{2} \nonumber \\
&& +~ T_2 \left[ 1222 T_2^4 - 164 \zeta_3 T_2^4 + 936 \zeta_3 T_2^2 T_5 
- 810 \zeta_5 T_2^2 T_5 - 144 \zeta_3^2 T_2 T_{71}
\right. \nonumber \\
&& \left. ~~~~~~~~
+ 1040 \zeta_5 T_2 T_{71} + 360 \zeta_3^2 T_5^2 + 1323 \zeta_7 T_{94}
\right] \frac{a^5}{16} ~+~ O(a^6) 
\label{gam5Tmom}
\end{eqnarray}
where like (\ref{bet5mom}) there are no even zetas. Formally setting 
$T_i$~$=$~$1$ for all $i$ recovers the analogous equations of the previous
section. It is clear from both expressions that the coefficients of the 
primitives are unchanged at the loop order where they first appear. We note 
that the coupling constant map is
\begin{eqnarray}
a_T^{\MOMs} &=& \left[ 1 ~-~ 3 T_2 a ~+~ \frac{57}{4} T_2^2 a^2 ~-~ 
T_2 [ 72 \zeta_3 T_2^2 - 8 \zeta_3 T_2 T_5 + 18 \zeta_4 T_2 T_5 + 659 T_2^2 ] 
\frac{a^3}{8} 
\right. \nonumber \\
&& \left. ~+
T_2 [ 8895 T_2^3 - 300 \zeta_3 T_2^3 + 2394 \zeta_3 T_2 T_5
+ 351 \zeta_4 T_2 T_5 - 336 \zeta_5 T_2 T_5 - 24 \zeta_3^2 T_{71}
\right. \nonumber \\
&& \left. ~~~~~~~~
+ 840 \zeta_5 T_{71} + 300 \zeta_6 T_{71} ] \frac{a^4}{16} \right] a ~+~ 
O(a^6) ~.
\end{eqnarray}
To gauge the primitive contribution the numerical evaluations of (\ref{gam5T})
and (\ref{gam5Tmom}) are
\begin{eqnarray}
\gamma_T(a) &=& \frac{1}{2} T_2 a - \frac{1}{2} T_2^2 a^2 
+ T_2 \left[ 0.625 T_2^2 + 1.803085 T_5 \right] a^3 \nonumber \\ 
&& -~ T_2 \left[ 1.125 T_2^3 + 6.580199 T_2 T_5 + 10.369277 T_{71} \right] a^4
\nonumber \\
&& +~ T_2 \left[ 2.919521 T_2^4 - 2.967631 T_2^2 T_5 + 38.872050 T_2 T_{71} +
32.511168 T_5^2 \right. \nonumber \\
&& \left. ~~~~~~~ + 83.377881 T_{94} \right] a^5 ~+~ O(a^6)
\end{eqnarray}
and
\begin{eqnarray}
\gamma_T^{\MOMs}(a) &=& \frac{1}{2} T_2 a - \frac{1}{2} T_2^2 a^2 ~+~ 
T_2  \left[ 1.75 T_2^2 + 1.803085 T_5 \right] a^3 \nonumber \\
&& -~ T_2 \left[ 8.797943 T_2^3 + 9.015427 T_2 T_5 
+ 10.369277 T_{71} \right] a^4 \nonumber \\
&& +~ T_2 \left[ 64.053917 T_2^4 + 17.825861 T_2^2 T_5 + 54.395837 T_2 T_{71} 
+ 32.511168 T_5^2 \right. \nonumber \\
&& \left. ~~~~~~~ + 83.377881 T_{94} \right] a^5 ~+~ O(a^6)
\end{eqnarray}
respectively. If we recall that at five loops the graphs of the upper row of
Figure \ref{figpr5} are what we termed product primitives we can identity their
contributions from the coefficient of $T_2 T_5^2$. This is because $T_5$ is 
associated with the graph $V_2$. If we compute the contribution from the 
primitives at three, four and five loop order we find that respectively they
contribute $74.26\%$, $57.37\%$ and $74.91\%$. At lower orders it is not
meaningful to quote values as it would be $100\%$ at one loop and there are no
two loop primitives. For the MOM scheme the analogous numbers are $50.75\%$,
$36.79\%$ and $45.96\%$. The smaller relative contribution for the MOM scheme
is due primarily to the increase in the coefficient of the $T_2^L$ terms at
each loop order $L$. However for the $\MSbar$ scheme the observation of 
\cite{34} that the primitives make an increasing contribution at higher orders 
for $\phi^4$ theory seems to hold here too for the $\MSbar$ scheme albeit at 
one loop order fewer than \cite{34}. It would be interesting if another scheme 
could be studied for the non-supersymmetric theory. 

An additional motivation for examining the $\beta$-function of (\ref{lagwzt}) 
is that it provides another relatively trivial check on our five loop
computation. It transpires that the coefficients of the terms of $T_2^L$ in
(\ref{gam5Tmom}) have already been computed before. More specifically we mean 
the three loop and higher coefficients since the one and two loop terms are 
scheme independent. We stress that we are indeed referring to the MOM result 
rather than the $\MSbar$ one. In \cite{64,80,81} $\gamma_\Phi(a)$ was studied 
using the Hopf algebra construction of Broadhurst and Kreimer, \cite{82,83}. 
Specifically it was used to determine the scalar field anomalous dimension in 
scalar $\phi^3$ and scalar Yukawa theories for a specific class of Feynman 
diagrams. In particular the Dyson-Schwinger equation for embedding of basic one
loop propagator correction within the skeleton one loop graph itself was 
constructed and solved for the anomalous dimension. This was extended in 
\cite{81} to the Wess-Zumino model where the supersymmetry Ward identity was 
important in constructing and solving the corresponding Dyson-Schwinger
equation. Moreover, it is the first case we believe where the $\beta$-function
of any theory was accessed this way in the Hopf approach. Consequently the 
first $200$ coefficients of $\gamma_\Phi(a)$ were determined for (\ref{lagwz})
with the analytic form given for the first $12$ terms for the class of diagrams
considered. While the analysis of \cite{81} centred on the theory with action 
(\ref{lagwz}) a subset of the graphs making up the coefficients of (\ref{gam5})
were found.  These are straightforward to isolate with the labelling used for 
(\ref{lagwzt}). As \cite{81} used the iteration of the one loop bubble the 
$T_2^L$ terms of our five loop $\beta$-function should tally with the Hopf 
algebra case. The question of which scheme was used can be established by the 
renormalization condition used in \cite{81} and it is clear it corresponds to 
the MOM one of \cite{4}. This therefore represents a specific check on the 
$T_2^L$ coefficients of (\ref{gam5Tmom}).

Having established the five loop renormalization group functions we can now
extract estimates for several critical exponents in the $\epsilon$ expansion at
the Wilson-Fisher fixed point where again we take $d$~$=$~$4$~$-$~$2\epsilon$.
The specific exponents we will compute are $\eta$~$=$~$\gamma_\Phi(a^\ast)$ and
the correction to scaling exponent $2\beta^\prime(a^\ast)$ where $a^\ast$ is 
the critical coupling constant. We will denote this combination here and later 
by $\hat{\omega}$ rather than the more usual unhatted version to avoid conflict
with notation in a later section. From (\ref{bet5}) we find
\begin{eqnarray}
\hat{\omega} &=& 2 \epsilon ~-~ \frac{4}{3} \epsilon^2 ~+~ 
\frac{4}{9} [12 \zeta_3 + 1 ] \epsilon^3 ~+~ 
\frac{4}{27} [ 54 \zeta_4 - 84 \zeta_3 - 240 \zeta_5 - 7 ] \epsilon^4 
\nonumber \\
&& +~ \frac{4}{81} [ 576 \zeta_3^2 + 396 \zeta_3 - 378 \zeta_4 + 1416 \zeta_5
- 1800 \zeta_6 + 5292 \zeta_7 + 19 ] \epsilon^5 ~+~ O(\epsilon^6) 
\label{omeps}
\end{eqnarray}
or
\begin{eqnarray}
\hat{\omega} &=& 2 \epsilon ~-~ 1.333333 \epsilon^2 ~+~ 6.855415 \epsilon^3 ~-~
44.205924 \epsilon^4 ~+~ 290.935250 \epsilon^5 ~+~ O(\epsilon^6)
\end{eqnarray}
numerically. The situation with $\eta$ is somewhat simpler in perturbation 
theory due to the supersymmetry Ward identity as has been noted in \cite{15,35}
for example. As the dimensionality of the coupling constant manifests itself in
the $O(a)$ term of $\beta(a)$ in $d$-dimensions then (\ref{gam5}) implies
\begin{equation}
\eta ~=~ \frac{1}{3} \epsilon
\end{equation}
{\em exactly}. For the more general group valued case (\ref{lagwzt}), and for 
later purposes, we note that the critical coupling is  
\begin{eqnarray}
a_T^\ast &=& \frac{2}{3 T_2} \epsilon + \frac{4}{9 T_2} \epsilon^2
+ 2 [ T_2^2 - 4 \zeta_3 T_5 ] \frac{\epsilon^3}{9 T_2^3} 
+ 8 [ 2 T_2^3 - 9 \zeta_4 T_2 T_5 + 40 \zeta_5 T_{71} ]
\frac{\epsilon^4}{81 T_2^4} 
\nonumber \\
&& +~ 2 [ 16 T_2^4 - 12 \zeta_3 T_2^4 - 54 \zeta_3 T_2^2 T_5 
+ 9 \zeta_4 T_2^2 T_5 + 612 \zeta_5 T_2^2 T_5 + 216 \zeta_3^2 T_2 T_{71} 
\nonumber \\
&& ~~~~~- 520 \zeta_5 T_2 T_{71} + 900 \zeta_6 T_2 T_{71} - 288 \zeta_3^2 T_5^2
- 2646 \zeta_7 T_{94} ] \frac{\epsilon^5}{243 T_2^5} ~+~ O(\epsilon^6)
\label{critccT2}
\end{eqnarray}
implying
\begin{eqnarray}
\hat{\omega}_T &=& 
2 \epsilon ~-~ \frac{4}{3} \epsilon^2 ~+~ \frac{4}{9} \epsilon^3 ~-~ 
\frac{28}{27} \epsilon^4 ~+~ 4 [24 \zeta_3 + 19 ] \frac{\epsilon^5}{81} 
\nonumber \\
&& +~ \left[ \frac{16}{3} \zeta_3 \epsilon^3 ~+~ 
\frac{8}{9} [ 9 \zeta_4 - 14 \zeta_3 ] \epsilon^4 ~+~ 
\frac{8}{27} [ 62 \zeta_3 - 63 \zeta_4 - 204 \zeta_5 ] \epsilon^5 \right]
\frac{T_5}{T_2^2} \nonumber \\
&&
+~ \left[ -~ \frac{320}{9} \zeta_5  \epsilon^4 ~+~ 
\frac{32}{27} [ 110 \zeta_5 - 18 \zeta_3^2 - 75 \zeta_6 ] \epsilon^5
\right] \frac{T_{71}}{T_2^3} \nonumber \\
&&
+~ \frac{448}{9} \zeta_3^2 \frac{T_5^2}{T_2^4} \epsilon^5 ~+~ 
\frac{784}{3} \zeta_7 \frac{T_{94}}{T_2^4} \epsilon^5 ~+~ O(\epsilon^6)
\label{omegaT2}
\end{eqnarray}
where we have ordered the expansion in terms of the group invariants. The power
of the leading term in $\epsilon$ of each of the invariants tallies with the
loop order of the $\beta$-function where the corresponding $T_i$ first appears.
The leading order $T_i$ independent terms correspond to the bubble insertions 
associated with $T_2$ with the primitive ranked by powers of $1/T_2$.

{\begin{table}[ht]
\begin{center}
\begin{tabular}{|c||r|r|r|}
\hline
$L$ & Pad\'{e} & Value & Average \\
\hline
$2$ & $[2,0]$ & $0.666667$ & $0.666667$ \\
\hline
$3$ & $[2,1]$ & $0.906650$ & $0.906650$ \\
\hline
$4$ & $[3,1]$ & $0.869530$ & $$ \\
    & $[2,2]$ & $0.872352$ & $0.870940$ \\
\hline
$5$ & $[4,1]$ & $0.879670$ & $$ \\
    & $[3,2]$ & $0.877593$ & $$ \\
    & $[2,3]$ & $0.878492$ & $0.878585$ \\
\hline
\end{tabular}
\end{center}
\begin{center}
\caption{Estimates for $\hat{\omega}$ in three dimensions from Pad\'{e} 
approximants.}
\label{omegad3}
\end{center}
\end{table}}

One reason for determining $\hat{\omega}$ in (\ref{omeps}) is that there has
been interest in estimating this exponent in three dimensions using various
methods, \cite{15,18,35,36,37,38,39,84}. Therefore with the five loop extension
of (\ref{bet5}) we can update the four loop $\epsilon$ expansion estimate noted
in \cite{38}. To do this we have evaluated Pad\'{e} approximants which are
recorded in Table \ref{omegad3}. In addition to the five loop estimates for
completeness we have provided lower loop approximants. In the table only
estimates in three dimensions are given where there were no singularities in 
the Pad\'{e} approximant between $4$ and $3$ dimensions. In other words the
approximant has to be continuously connected to the value in the critical
dimension. The final column gives the average of the approximants at each loop
order. If one focuses on the three and higher loop averages it would appear
that the approximants are converging but perhaps oscillating about the true
value. In order to place the five loop estimate in perspective we have gathered
results from earlier work on the exponent and recorded them chronologically in
Table \ref{expd3sum}. Aside from the $\epsilon$ expansion the two main 
techniques are the conformal bootstrap and the functional renormalization 
group. Some comments are in order. Errors on estimates are those given in the 
corresponding paper. In \cite{37} two sets of values were provided and 
distinguished by the parameter $n$. We have noted both sets but mention that 
the authors regarded the $n$~$=$~$2$ data as superior. Also the value we quote 
for $\hat{\omega}$ is that designated as supersymmetric in Table I of 
\cite{37}. The bracketed value for $1/\nu$ from \cite{36} was derived from the 
estimate of $\eta$ using the superscaling law of \cite{37,85,86}
\begin{equation}
\frac{1}{\nu} ~=~ \frac{1}{2} \left( d ~-~ \eta \right) ~. 
\label{supscal}
\end{equation}
We have also used this to extract the value recorded in the table from the
exact value of $\frac{1}{6}$ for $\eta$ which would imply that 
$\frac{1}{\nu}$~$=$~$\frac{17}{12}$. In \cite{35} the value of $\nu$ was 
determined but we have converted it to $\frac{1}{\nu}$ for consistency with the
other entries in the table. This was used to deduce $\eta$ from the 
superscaling law. While the values of the exponents from \cite{84} are noted as
$\epsilon$ expansion they are not deduced in the same way as those of this 
paper. Instead they represent the result of a matched Pad\'{e} approach where 
the $\epsilon$ expansion of two theories in the same universality class are 
used but one theory has a critical dimension of $2$ while the other is
renormalizable in $4$. Moreover the universality class is the 
Gross-Neveu-Yukawa one and the values in the table correspond to those for the
emergent supersymmetry. As we took a direct supersymmetric approach our values
for $\eta$ and $\frac{1}{\nu}$ are exact due to the supersymmetry Ward identity
and are within the errors given in \cite{84}. As an aside we note that the 
other $\epsilon$ expansion result of \cite{15} did not benefit from a two-sided
Pad\'{e} approach which may be the reason why that estimate for $\frac{1}{\nu}$
is low compared to \cite{84}. In terms of the overall picture there appears to 
be a consensus that the value of $\eta$ is around $0.166$ especially in the 
more recent articles that did not have the use of the supersymmetry Ward 
identity present in the $\epsilon$ expansion. The latest conformal bootstrap 
value appears to be the most accurate numerically given the precision and tight
error bars on $\eta$ and $\frac{1}{\nu}$. Indeed our exact values differ by 
around $1.3\%$ and $0.08\%$ respectively with both conformal bootstrap values 
satisfying (\ref{supscal}). For $\hat{\omega}$ the difference is roughly 
$0.5\%$.

{\begin{table}[ht]
\begin{center}
\begin{tabular}{|c|c||l|l|l|}
\hline
\rule{0pt}{12pt}
Method & Reference & $\eta$ & $\frac{1}{\nu}_{\frac{}{}}$ & $\hat{\omega}$ \\
\hline
CB & \cite{18} & $0.166667$ & $1.0902(20)$ & $0.9098(20)$ \\
FRG & \cite{35} & $0.114$ & $1.443$ & $0.796$ \\
CB & \cite{36} & $0.164$ & $(1.418)$ & -------------- \\
FRG & \cite{37} $(n=1)$ & $0.174$ & $1.385$ & $0.765$ \\
FRG & \cite{37} $(n=2)$ & $0.167$ & $1.395$ & $0.782$ \\
$\epsilon$ & \cite{15} & $0.166667$ & $1.129(1)$ & $0.871(1)$ \\
FRG & \cite{38} & -------------- & $1.1656$ & $0.8344$ \\
$\epsilon$ & \cite{84} & $0.1673(50)$ & $1.415(12)$ & -------------- \\
CB & \cite{39} & $0.168888(60)$ & $1.415556(30)$ & $0.882(9)$ \\
$\epsilon$ & This work & $0.166667$ & $1.416667$ & $0.878585$ \\
\hline
\end{tabular}
\end{center}
\begin{center}
\caption{Summary of exponent estimates by conformal bootstrap (CB), functional
renormalization group (FRG) and $\epsilon$ expansion methods.}
\label{expd3sum}
\end{center}
\end{table}}

One interesting application of considering (\ref{lagwzt}) is that the 
renormalization group functions can be deduced for Lie groups which have a
non-trivial rank $3$ fully symmetric tensor $d^{ijk}$. One such class of groups
are the $SU(\Nc)$ ones and in that case (\ref{Tdef}) reduce to
\begin{eqnarray}
T_2 &=& \frac{[ \Nc^2 - 4]}{\Nc} ~~,~~
T_5 ~=~ -~ \frac{4}{\Nc^2} [ \Nc^2 - 10 ] \nonumber \\
T_{71} &=& \frac{1}{8\Nc^3} [ \Nc^2 - 8 ] [ \Nc^4 - 8 \Nc^2 + 256 ] 
\nonumber \\
T_{94} &=& -~ [ \Nc^6 - 64 \Nc^4 + 1216 \Nc^2 - 6784 ] \frac{1}{4 \Nc^4}
\end{eqnarray}
using \cite{87}. So, for example, for $SU(3)$ we have  
\begin{eqnarray}
\left. \gamma_\Phi(a) \right|_{SU(3)} &=& \frac{5}{6} a - \frac{25}{18} a^2 
+ \frac{5}{216} [ 48 \zeta_3 + 125 ] a^3 
+ \frac{25}{648} [ 72 \zeta_4 - 530 \zeta_5 - 225 - 240 \zeta_3 ] a^4 
\nonumber \\
&& + ~\frac{25}{15552} [ 36840 \zeta_3 - 9702 \zeta_3^2 - 17640 \zeta_4
+ 137170 \zeta_5 - 59625 \zeta_6 + 78057 \zeta_7 \nonumber \\
&& ~~~~~~~~~~~~ + 19750 ] a^5 ~+~ O(a^6)
\label{gamsu3}
\end{eqnarray}
and
\begin{eqnarray}
\left. \beta(a) \right|_{SU(3)} &=& \frac{5}{2} a^2 - \frac{25}{6} a^3 
+ \frac{5}{72} [ 48 \zeta_3 + 125 ] a^4
+ \frac{25}{216} [ 72 \zeta_4 - 530 \zeta_5 - 225 - 240 \zeta_3 ] a^5
\nonumber \\
&& +~ \frac{25}{5184} [ 36840 \zeta_3 - 9702 \zeta_3^2 - 17640 \zeta_4
+ 137170 \zeta_5 - 59625 \zeta_6 + 78057 \zeta_7 \nonumber \\
&& ~~~~~~~~~~ + 19750 ] a^6 ~+~ O(a^7)
\label{betsu3}
\end{eqnarray}
which we record for later purposes. As there has also been recent interest in
Wess-Zumino models with $F_4$ symmetry, \cite{46}, we note that the 
corresponding renormalization group functions and exponents can be extracted 
from (\ref{gam5T}) and (\ref{omegaT2}) with 
\begin{eqnarray}
T_3 &=& -~ [ N - 2 ] \frac{T_2}{2[N+2]} ~~,~~
T_5 ~=~ -~ [ N^2 - 10 N - 16 ] \frac{T_2^2}{2[N+2]^2} \nonumber \\
T_{71} &=& [ N^3 - 3 N^2 + 80 N + 100 ] \frac{T_2^3}{4[N+2]^3} \nonumber \\
T_{94} &=& -~ [ N^4 - 14 N^3 - 12 N^2 - 616 N - 672 ] \frac{T_2^4}{8[N+2]^4}
\label{f4inv}
\end{eqnarray}
where $N$ is the dimension of an $F_4$ representation such as $\mathbf{2}$,
$\mathbf{5}$, $\mathbf{8}$, $\mathbf{14}$, $\mathbf{26}$, $\mathbf{27}$,
$\mathbf{90}$ or $\mathbf{324}$.

\sect{$O(N)$ Wess-Zumino model.}

As a second generalization of (\ref{lagwz}) we consider the Wess-Zumino model 
with an $O(N)$ symmetry as it will provide us with another check on our 
computation. This is because the $O(N)$ model admits a large $N$ expansion and 
the renormalization group functions have been computed to three orders in 
powers of $1/N$ in \cite{48,49}. The action in terms of bare quantities is
\begin{eqnarray}
S^{O(N)} &=& \int d^4 x \left[ \int d^2 \theta d^2 \bar{\theta} \,
\bar{\Phi}_0^i (x,\bar{\theta}) e^{-2 \theta {\partialline} \bar{\theta}} 
\Phi_0^i (x,\theta) ~+~ 
\bar{\sigma}_0 (x,\bar{\theta}) e^{-2 \theta {\partialline} \bar{\theta}} 
\sigma_0 (x,\theta) \right. \nonumber \\
&& \left. ~~~~~~~~~+~ 
\frac{{g_1}_0}{2} \int d^2 \theta \, \sigma_0 \Phi_0^i \Phi_0^i ~+~ 
\frac{{g_1}_0}{2} \int d^2 \bar{\theta} \, \bar{\sigma}_0 \bar{\Phi}_0^i 
\bar{\Phi}_0^i \right. \nonumber \\
&& \left. ~~~~~~~~~+~ 
\frac{{g_2}_0}{6} \int d^2 \theta \, \sigma_0^3 ~+~ 
\frac{{g_2}_0}{6} \int d^2 \bar{\theta} \, \bar{\sigma}_0^3 \right]
\label{lagwzon}
\end{eqnarray}
and was given in \cite{88} where $1$~$\leq$~$i$~$\leq$~$N$. We regard the 
coupling constants as real. In \cite{88} they were taken to be complex but they
will only appear as squares in the renormalization group functions. In this 
case this combination will be equivalent to the squared length of $g_1$ and 
$g_2$ respectively given in \cite{88}. The superfields $\Phi^i$ and 
$\bar{\Phi}^i$ lie in an $O(N)$ multiplet and the $\sigma$ and $\bar{\sigma}$ 
fields would equate to auxiliary fields in non-supersymmetric four
dimensional $\phi^4$ theory. In other words in that instance the quartic 
interaction can be rewritten as a cubic interaction, akin to that of 
(\ref{lagwzon}) with the $g_1$ coupling constant, and a non-kinetic quadratic 
term equivalent to that for $\sigma$ and $\bar{\sigma}$ but without the 
$\theta$ dependent exponential. For that reason one can regard the $O(N)$ 
Wess-Zumino model as a supersymmetric generalization of $O(N)$ scalar $\phi^4$ 
theory. This is apparent in the purely bosonic sector of the component 
Lagrangian (\ref{lagwzc}). Indeed it is that rewriting of the quartic
interaction that is the key to accessing the large $N$ expansion through the 
critical point formalism developed in $d$-dimensions in \cite{73,74,89} for 
scalar $\phi^4$ theory as we will show later. This was extended in \cite{48,49}
for (\ref{lagwzon}) where more background on this aspect to exploring the 
Wess-Zumino model can be found. It is also worth noting that when both 
couplings are non-zero the action is formally equivalent to that of 
non-supersymmetric $O(N)$ $\phi^3$ theory in six dimensions that was analysed
at three loops in \cite{79,90}. This is in the sense that in six dimensions 
there are two interactions that ensure the theory is renormalizable. Finally we
note that the $O(N)$ Wess-Zumino model also has only two independent 
renormalization constants which can be expressed as 
\begin{equation}
\beta_1^{O(N)}(g_i) ~=~ \frac{1}{2} g_1 
\left[ \gamma^{O(N)}_\sigma(g_i) ~+~ 
2 \gamma^{O(N)}_\Phi(g_i) \right] ~~,~~ 
\beta_2^{O(N)}(g_i) ~=~ \frac{3}{2} g_2 \gamma^{O(N)}_\sigma(g_i) 
\label{susywion}
\end{equation}
where $\gamma_\sigma(g_i)$ is the anomalous dimension of the $\sigma$ and
$\bar{\sigma}$ superfields and we use $g_i$ as shorthand for pair of couplings
$\{g_1,g_2\}$.

{\begin{table}[ht]
\begin{center}
\begin{tabular}{|c||r|r|}
\hline
$L$ & $\Phi$ & $\sigma$ \\
\hline
$1$ & $1$ & $2$ \\
$2$ & $3$ & $3$ \\
$3$ & $15$ & $20$ \\
$4$ & $109$ & $124$ \\
$5$ & $952$ & $1063$ \\
\hline
Total & $1080$ & $1212$ \\
\hline
\end{tabular}
\end{center}
\begin{center}
\caption{Number of graphs at each loop order $L$ for the $\Phi$ and $\sigma$
superfield $2$-point functions in the $O(N)$ Wess-Zumino model.}
\label{feynnumon}
\end{center}
\end{table}}

To extract the renormalization group functions for (\ref{lagwzon}) using 
{\sc Qgraf} we have generated all the supergraphs to five loops required for 
renormalizing the $\Phi^i$ and $\sigma$ $2$-point functions. The number of 
graphs that we had to compute at each loop order are listed in Table 
\ref{feynnumon}. With these graphs as input we applied the automatic 
integration routine that was outlined earlier and extracted the corresponding
renormalization group functions which are included in the attached data file. 
To five loops we found 
\begin{eqnarray}
\beta_1^{O(N)}(g_i) &=& 
\left[
\frac{1}{2} g_1 g_2^2
+ 2 g_1^3
+ \frac{1}{2} N g_1^3
\right]
+ \left[
- \frac{1}{2} g_1 g_2^4
- g_1^3 g_2^2
- 2 g_1^5
- \frac{1}{2} N g_1^3 g_2^2
- 2 N g_1^5
\right]
\nonumber \\
&& + \left[
\frac{5}{8} g_1 g_2^6
+ \frac{3}{2} g_1^3 g_2^4
+ g_1^5 g_2^2
+ 2 g_1^7
+ \frac{1}{4} N g_1^3 g_2^4
+ 4 N g_1^5 g_2^2
+ \frac{11}{2} N g_1^7
- \frac{3}{8} N^2 g_1^5 g_2^2
\right. \nonumber \\
&& \left. ~~~
+ \frac{1}{2} N^2 g_1^7
+ \frac{3}{2} \zeta_3 g_1 g_2^6
+ 12 \zeta_3 g_1^5 g_2^2
+ 12 \zeta_3 g_1^7
+ \frac{15}{2} \zeta_3 N g_1^5 g_2^2
+ 3 \zeta_3 N g_1^7
\right]
\nonumber \\
&& + \left[
- \frac{9}{8} g_1 g_2^8
- \frac{8}{3} g_1^3 g_2^6
- \frac{8}{3} g_1^5 g_2^4
- \frac{1}{3} g_1^7 g_2^2
- \frac{10}{3} g_1^9
- \frac{49}{6} N g_1^5 g_2^4
- \frac{97}{6} N g_1^7 g_2^2
- 14 N g_1^9
\right. \nonumber \\
&& \left. ~~~
+ \frac{7}{8} N^2 g_1^5 g_2^4
- \frac{1}{3} N^2 g_1^7 g_2^2
- 6 N^2 g_1^9
- \frac{1}{4} N^3 g_1^7 g_2^2
+ \frac{1}{6} N^3 g_1^9
- 10 \zeta_5 g_1 g_2^8
\right. \nonumber \\
&& \left. ~~~
- 40 \zeta_5 g_1^5 g_2^4
- 160 \zeta_5 g_1^7 g_2^2
- 80 \zeta_5 g_1^9
- 40 \zeta_5 N g_1^5 g_2^4
- 80 \zeta_5 N g_1^7 g_2^2
- 60 \zeta_5 N g_1^9
\right. \nonumber \\
&& \left. ~~~
- 10 \zeta_5 N^2 g_1^9
+ \frac{9}{4} \zeta_4 g_1 g_2^8
- \frac{3}{2} \zeta_4 g_1^3 g_2^6
+ 15 \zeta_4 g_1^5 g_2^4
+ 21 \zeta_4 g_1^7 g_2^2
+ 24 \zeta_4 g_1^9
\right. \nonumber \\
&& \left. ~~~
+ 3 \zeta_4 N g_1^3 g_2^6
+ \frac{15}{4} \zeta_4 N g_1^5 g_2^4
+ \frac{39}{2} \zeta_4 N g_1^7 g_2^2
+ 12 \zeta_4 N g_1^9
+ \frac{15}{2} \zeta_4 N^2 g_1^7 g_2^2
\right. \nonumber \\
&& \left. ~~~
+ \frac{3}{2} \zeta_4 N^2 g_1^9
- \frac{15}{2} \zeta_3 g_1 g_2^8
- \frac{7}{2} \zeta_3 g_1^3 g_2^6
- 30 \zeta_3 g_1^5 g_2^4
- 64 \zeta_3 g_1^7 g_2^2
- 72 \zeta_3 g_1^9
\right. \nonumber \\
&& \left. ~~~
- \frac{11}{2} \zeta_3 N g_1^3 g_2^6
- \frac{47}{2} \zeta_3 N g_1^5 g_2^4
- 89 \zeta_3 N g_1^7 g_2^2
- 48 \zeta_3 N g_1^9
+ \zeta_3 N^2 g_1^5 g_2^4
\right. \nonumber \\
&& \left. ~~~
- \frac{31}{2} \zeta_3 N^2 g_1^7 g_2^2
- 3 \zeta_3 N^2 g_1^9
+ \frac{1}{2} \zeta_3 N^3 g_1^7 g_2^2
\right]
\nonumber \\
&& + \left[
\frac{79}{32} g_1 g_2^{10}
+ 6 g_1^3 g_2^8
+ \frac{67}{12} g_1^5 g_2^6
+ \frac{17}{6} g_1^7 g_2^4
- \frac{1}{3} g_1^9 g_2^2
+ \frac{20}{3} g_1^{11}
- \frac{7}{8} N g_1^3 g_2^8
\right. \nonumber \\
&& \left. ~~~
+ \frac{1021}{48} N g_1^5 g_2^6
+ \frac{351}{8} N g_1^7 g_2^4
+ \frac{587}{12} N g_1^9 g_2^2
+ 38 N g_1^{11}
- \frac{37}{16} N^2 g_1^5 g_2^6
\right. \nonumber \\
&& \left. ~~~
- \frac{19}{48} N^2 g_1^7 g_2^4
+ \frac{173}{6} N^2 g_1^9 g_2^2
+ \frac{145}{4} N^2 g_1^{11}
+ \frac{7}{8} N^3 g_1^7 g_2^4
- \frac{77}{48} N^3 g_1^9 g_2^2
\right. \nonumber \\
&& \left. ~~~
+ \frac{25}{24} N^3 g_1^{11}
- \frac{5}{32} N^4 g_1^9 g_2^2
+ \frac{1}{16} N^4 g_1^{11}
+ \frac{1323}{16} \zeta_7 g_1 g_2^{10}
+ \frac{441}{2} \zeta_7 g_1^5 g_2^6
\right. \nonumber \\
&& \left. ~~~
+ \frac{3969}{4} \zeta_7 g_1^7 g_2^4
+ 2205 \zeta_7 g_1^9 g_2^2
+ 882 \zeta_7 g_1^{11}
+ \frac{4851}{16} \zeta_7 N g_1^5 g_2^6
+ \frac{11907}{16} \zeta_7 N g_1^7 g_2^4
\right. \nonumber \\
&& \left. ~~~
+ \frac{22491}{16} \zeta_7 N g_1^9 g_2^2
+ \frac{3087}{4} \zeta_7 N g_1^{11}
+ \frac{3087}{16} \zeta_7 N^2 g_1^9 g_2^2
+ \frac{2205}{16} \zeta_7 N^2 g_1^{11}
\right. \nonumber \\
&& \left. ~~~
- \frac{225}{8} \zeta_6 g_1 g_2^{10}
+ \frac{25}{2} \zeta_6 g_1^3 g_2^8
- 100 \zeta_6 g_1^5 g_2^6
- 350 \zeta_6 g_1^7 g_2^4
- 500 \zeta_6 g_1^9 g_2^2
\right. \nonumber \\
&& \left. ~~~
- 300 \zeta_6 g_1^{11}
- \frac{275}{8} \zeta_6 N g_1^3 g_2^8
- \frac{125}{2} \zeta_6 N g_1^5 g_2^6
- \frac{425}{2} \zeta_6 N g_1^7 g_2^4
- \frac{2025}{4} \zeta_6 N g_1^9 g_2^2
\right. \nonumber \\
&& \left. ~~~
- 300 \zeta_6 N g_1^{11}
- \frac{175}{2} \zeta_6 N^2 g_1^7 g_2^4
- \frac{1025}{8} \zeta_6 N^2 g_1^9 g_2^2
- \frac{375}{4} \zeta_6 N^2 g_1^{11}
\right. \nonumber \\
&& \left. ~~~
- \frac{75}{8} \zeta_6 N^3 g_1^{11}
+ \frac{457}{8} \zeta_5 g_1 g_2^{10}
+ 11 \zeta_5 g_1^3 g_2^8
+ \frac{355}{2} \zeta_5 g_1^5 g_2^6
+ 531 \zeta_5 g_1^7 g_2^4
\right. \nonumber \\
&& \left. ~~~
+ \frac{2143}{2} \zeta_5 g_1^9 g_2^2
+ 693 \zeta_5 g_1^{11}
+ \frac{193}{4} \zeta_5 N g_1^3 g_2^8
+ \frac{277}{8} \zeta_5 N g_1^5 g_2^6
+ \frac{3979}{4} \zeta_5 N g_1^7 g_2^4
\right. \nonumber \\
&& \left. ~~~
+ \frac{3379}{4} \zeta_5 N g_1^9 g_2^2
+ 451 \zeta_5 N g_1^{11}
- \frac{19}{8} \zeta_5 N^2 g_1^5 g_2^6
- \frac{105}{4} \zeta_5 N^2 g_1^7 g_2^4
\right. \nonumber \\
&& \left. ~~~
+ \frac{1601}{4} \zeta_5 N^2 g_1^9 g_2^2
+ \frac{991}{4} \zeta_5 N^2 g_1^{11}
- \frac{555}{8} \zeta_5 N^3 g_1^9 g_2^2
+ \frac{39}{2} \zeta_5 N^3 g_1^{11}
\right. \nonumber \\
&& \left. ~~~
- \frac{441}{32} \zeta_4 g_1 g_2^{10}
+ \frac{33}{16} \zeta_4 g_1^3 g_2^8
- 57 \zeta_4 g_1^5 g_2^6
- \frac{219}{2} \zeta_4 g_1^7 g_2^4
- 162 \zeta_4 g_1^9 g_2^2
- 174 \zeta_4 g_1^{11}
\right. \nonumber \\
&& \left. ~~~
- \frac{393}{16} \zeta_4 N g_1^3 g_2^8
- \frac{963}{32} \zeta_4 N g_1^5 g_2^6
- \frac{1077}{8} \zeta_4 N g_1^7 g_2^4
- \frac{1917}{8} \zeta_4 N g_1^9 g_2^2
\right. \nonumber \\
&& \left. ~~~
- \frac{327}{2} \zeta_4 N g_1^{11}
- \frac{267}{32} \zeta_4 N^2 g_1^5 g_2^6
- \frac{1293}{32} \zeta_4 N^2 g_1^7 g_2^4
- \frac{471}{4} \zeta_4 N^2 g_1^9 g_2^2
\right. \nonumber \\
&& \left. ~~~
- \frac{147}{4} \zeta_4 N^2 g_1^{11}
+ \frac{51}{32} \zeta_4 N^3 g_1^7 g_2^4
- \frac{27}{2} \zeta_4 N^3 g_1^9 g_2^2
- \frac{27}{16} \zeta_4 N^3 g_1^{11}
+ \frac{15}{32} \zeta_4 N^4 g_1^9 g_2^2
\right. \nonumber \\
&& \left. ~~~
+ \frac{429}{16} \zeta_3 g_1 g_2^{10}
+ \frac{177}{8} \zeta_3 g_1^3 g_2^8
+ 90 \zeta_3 g_1^5 g_2^6
+ 138 \zeta_3 g_1^7 g_2^4
+ 252 \zeta_3 g_1^9 g_2^2
+ 268 \zeta_3 g_1^{11}
\right. \nonumber \\
&& \left. ~~~
+ \frac{47}{2} \zeta_3 N g_1^3 g_2^8
+ \frac{1911}{16} \zeta_3 N g_1^5 g_2^6
+ \frac{1193}{4} \zeta_3 N g_1^7 g_2^4
+ \frac{2521}{4} \zeta_3 N g_1^9 g_2^2
\right. \nonumber \\
&& \left. ~~~
+ 377 \zeta_3 N g_1^{11}
- \frac{53}{16} \zeta_3 N^2 g_1^5 g_2^6
+ \frac{869}{16} \zeta_3 N^2 g_1^7 g_2^4
+ \frac{423}{2} \zeta_3 N^2 g_1^9 g_2^2
+ 68 \zeta_3 N^2 g_1^{11}
\right. \nonumber \\
&& \left. ~~~
- \frac{55}{16} \zeta_3 N^3 g_1^7 g_2^4
+ \frac{3}{8} \zeta_3 N^3 g_1^9 g_2^2
+ \frac{13}{8} \zeta_3 N^3 g_1^{11}
- \frac{5}{16} \zeta_3 N^4 g_1^9 g_2^2
- \frac{1}{8} \zeta_3 N^4 g_1^{11}
\right. \nonumber \\
&& \left. ~~~
+ \frac{63}{4} \zeta_3^2 g_1 g_2^{10}
- \zeta_3^2 g_1^3 g_2^8
+ 44 \zeta_3^2 g_1^5 g_2^6
+ 172 \zeta_3^2 g_1^7 g_2^4
+ 448 \zeta_3^2 g_1^9 g_2^2
+ 288 \zeta_3^2 g_1^{11}
\right. \nonumber \\
&& \left. ~~~
- \frac{25}{4} \zeta_3^2 N g_1^3 g_2^8
+ 86 \zeta_3^2 N g_1^5 g_2^6
+ 71 \zeta_3^2 N g_1^7 g_2^4
+ \frac{693}{2} \zeta_3^2 N g_1^9 g_2^2
+ 18 \zeta_3^2 N g_1^{11}
\right. \nonumber \\
&& \left. ~~~
- 11 \zeta_3^2 N^2 g_1^7 g_2^4
+ \frac{263}{4} \zeta_3^2 N^2 g_1^9 g_2^2
- \frac{45}{2} \zeta_3^2 N^2 g_1^{11}
- \frac{9}{4} \zeta_3^2 N^3 g_1^{11}
\right] ~+~ O \left( g_i^{13} \right)
\label{beta1on5}
\end{eqnarray}
and 
\begin{eqnarray}
\beta_2^{O(N)}(g_i) &=& 
\left[
\frac{3}{2} g_2^3
+ \frac{3}{2} N g_1^2 g_2
\right]
+ \left[
- \frac{3}{2} g_2^5
- \frac{3}{2} N g_1^2 g_2^3
- 3 N g_1^4 g_2
\right]
\nonumber \\
&& + \left[
\frac{15}{8} g_2^7
+ \frac{3}{4} N g_1^2 g_2^5
+ 9 N g_1^4 g_2^3
+ \frac{3}{2} N g_1^6 g_2
- \frac{9}{8} N^2 g_1^4 g_2^3
+ 3 N^2 g_1^6 g_2
+ \frac{9}{2} \zeta_3 g_2^7
\right. \nonumber \\
&& \left. ~~~
+ \frac{45}{2} \zeta_3 N g_1^4 g_2^3
+ 9 \zeta_3 N g_1^6 g_2
\right]
\nonumber \\
&& + \left[
- \frac{27}{8} g_2^9
- \frac{91}{4} N g_1^4 g_2^5
- \frac{13}{2} N g_1^6 g_2^3
- 4 N g_1^8 g_2
+ \frac{21}{8} N^2 g_1^4 g_2^5
- \frac{13}{2} N^2 g_1^6 g_2^3
\right. \nonumber \\
&& \left. ~~~
- 14 N^2 g_1^8 g_2
- \frac{3}{4} N^3 g_1^6 g_2^3
+ \frac{5}{4} N^3 g_1^8 g_2
- 30 \zeta_5 g_2^9
- 120 \zeta_5 N g_1^4 g_2^5
- 240 \zeta_5 N g_1^6 g_2^3
\right. \nonumber \\
&& \left. ~~~
- 60 \zeta_5 N g_1^8 g_2
- 30 \zeta_5 N^2 g_1^8 g_2
+ \frac{27}{4} \zeta_4 g_2^9
+ 9 \zeta_4 N g_1^2 g_2^7
+ \frac{45}{4} \zeta_4 N g_1^4 g_2^5
\right. \nonumber \\
&& \left. ~~~
+ 36 \zeta_4 N g_1^6 g_2^3
+ 18 \zeta_4 N g_1^8 g_2
+ \frac{45}{2} \zeta_4 N^2 g_1^6 g_2^3
+ \frac{9}{2} \zeta_4 N^2 g_1^8 g_2
- \frac{45}{2} \zeta_3 g_2^9
\right. \nonumber \\
&& \left. ~~~
- \frac{33}{2} \zeta_3 N g_1^2 g_2^7
- 69 \zeta_3 N g_1^4 g_2^5
- 132 \zeta_3 N g_1^6 g_2^3
- 66 \zeta_3 N g_1^8 g_2
+ 3 \zeta_3 N^2 g_1^4 g_2^5
\right. \nonumber \\
&& \left. ~~~
- 48 \zeta_3 N^2 g_1^6 g_2^3
- 9 \zeta_3 N^2 g_1^8 g_2
+ \frac{3}{2} \zeta_3 N^3 g_1^6 g_2^3
- \frac{3}{2} \zeta_3 N^3 g_1^8 g_2
\right]
\nonumber \\
&& + \left[
\frac{237}{32} g_2^{11}
- \frac{21}{8} N g_1^2 g_2^9
+ \frac{1039}{16} N g_1^4 g_2^7
+ \frac{123}{8} N g_1^6 g_2^5
+ \frac{93}{4} N g_1^8 g_2^3
+ 7 N g_1^{10} g_2
\right. \nonumber \\
&& \left. ~~~
- \frac{111}{16} N^2 g_1^4 g_2^7
+ \frac{215}{16} N^2 g_1^6 g_2^5
+ \frac{313}{4} N^2 g_1^8 g_2^3
+ \frac{131}{4} N^2 g_1^{10} g_2
+ \frac{21}{8} N^3 g_1^6 g_2^5
\right. \nonumber \\
&& \left. ~~~
- \frac{143}{16} N^3 g_1^8 g_2^3
+ \frac{83}{8} N^3 g_1^{10} g_2
- \frac{15}{32} N^4 g_1^8 g_2^3
+ \frac{9}{16} N^4 g_1^{10} g_2
+ \frac{3969}{16} \zeta_7 g_2^{11}
\right. \nonumber \\
&& \left. ~~~
+ \frac{14553}{16} \zeta_7 N g_1^4 g_2^7
+ \frac{35721}{16} \zeta_7 N g_1^6 g_2^5
+ \frac{46305}{16} \zeta_7 N g_1^8 g_2^3
+ \frac{1323}{2} \zeta_7 N g_1^{10} g_2
\right. \nonumber \\
&& \left. ~~~
+ \frac{9261}{16} \zeta_7 N^2 g_1^8 g_2^3
+ \frac{6615}{16} \zeta_7 N^2 g_1^{10} g_2
- \frac{675}{8} \zeta_6 g_2^{11}
- \frac{825}{8} \zeta_6 N g_1^2 g_2^9
\right. \nonumber \\
&& \left. ~~~
- \frac{375}{2} \zeta_6 N g_1^4 g_2^7
- \frac{975}{2} \zeta_6 N g_1^6 g_2^5
- \frac{3075}{4} \zeta_6 N g_1^8 g_2^3
- 225 \zeta_6 N g_1^{10} g_2
\right. \nonumber \\
&& \left. ~~~
- \frac{525}{2} \zeta_6 N^2 g_1^6 g_2^5
- \frac{3075}{8} \zeta_6 N^2 g_1^8 g_2^3
- \frac{675}{4} \zeta_6 N^2 g_1^{10} g_2
- \frac{225}{8} \zeta_6 N^3 g_1^{10} g_2
\right. \nonumber \\
&& \left. ~~~
+ \frac{1371}{8} \zeta_5 g_2^{11}
+ \frac{579}{4} \zeta_5 N g_1^2 g_2^9
+ \frac{879}{8} \zeta_5 N g_1^4 g_2^7
+ \frac{9207}{4} \zeta_5 N g_1^6 g_2^5
+ \frac{4173}{4} \zeta_5 N g_1^8 g_2^3
\right. \nonumber \\
&& \left. ~~~
+ \frac{717}{2} \zeta_5 N g_1^{10} g_2
- \frac{57}{8} \zeta_5 N^2 g_1^4 g_2^7
- \frac{315}{4} \zeta_5 N^2 g_1^6 g_2^5
+ \frac{4803}{4} \zeta_5 N^2 g_1^8 g_2^3
\right. \nonumber \\
&& \left. ~~~
+ \frac{1557}{4} \zeta_5 N^2 g_1^{10} g_2
- \frac{1665}{8} \zeta_5 N^3 g_1^8 g_2^3
+ \frac{117}{2} \zeta_5 N^3 g_1^{10} g_2
- \frac{1323}{32} \zeta_4 g_2^{11}
\right. \nonumber \\
&& \left. ~~~
- \frac{1179}{16} \zeta_4 N g_1^2 g_2^9
- \frac{2727}{32} \zeta_4 N g_1^4 g_2^7
- \frac{1341}{8} \zeta_4 N g_1^6 g_2^5
- \frac{2295}{8} \zeta_4 N g_1^8 g_2^3
\right. \nonumber \\
&& \left. ~~~
- \frac{297}{2} \zeta_4 N g_1^{10} g_2
- \frac{801}{32} \zeta_4 N^2 g_1^4 g_2^7
- \frac{3789}{32} \zeta_4 N^2 g_1^6 g_2^5
- \frac{2097}{8} \zeta_4 N^2 g_1^8 g_2^3
\right. \nonumber \\
&& \left. ~~~
- \frac{135}{2} \zeta_4 N^2 g_1^{10} g_2
+ \frac{153}{32} \zeta_4 N^3 g_1^6 g_2^5
- \frac{675}{16} \zeta_4 N^3 g_1^8 g_2^3
- \frac{153}{16} \zeta_4 N^3 g_1^{10} g_2
\right. \nonumber \\
&& \left. ~~~
+ \frac{45}{32} \zeta_4 N^4 g_1^8 g_2^3
- \frac{9}{8} \zeta_4 N^4 g_1^{10} g_2
+ \frac{1287}{16} \zeta_3 g_2^{11}
+ \frac{141}{2} \zeta_3 N g_1^2 g_2^9
+ \frac{4959}{16} \zeta_3 N g_1^4 g_2^7
\right. \nonumber \\
&& \left. ~~~
+ \frac{1707}{4} \zeta_3 N g_1^6 g_2^5
+ \frac{2247}{4} \zeta_3 N g_1^8 g_2^3
+ 243 \zeta_3 N g_1^{10} g_2
- \frac{159}{16} \zeta_3 N^2 g_1^4 g_2^7
\right. \nonumber \\
&& \left. ~~~
+ \frac{2745}{16} \zeta_3 N^2 g_1^6 g_2^5
+ \frac{2115}{4} \zeta_3 N^2 g_1^8 g_2^3
+ \frac{357}{2} \zeta_3 N^2 g_1^{10} g_2
- \frac{165}{16} \zeta_3 N^3 g_1^6 g_2^5
\right. \nonumber \\
&& \left. ~~~
+ \frac{45}{4} \zeta_3 N^3 g_1^8 g_2^3
+ \frac{99}{8} \zeta_3 N^3 g_1^{10} g_2
- \frac{15}{16} \zeta_3 N^4 g_1^8 g_2^3
+ \frac{3}{8} \zeta_3 N^4 g_1^{10} g_2
+ \frac{189}{4} \zeta_3^2 g_2^{11}
\right. \nonumber \\
&& \left. ~~~
- \frac{75}{4} \zeta_3^2 N g_1^2 g_2^9
+ 258 \zeta_3^2 N g_1^4 g_2^7
+ 309 \zeta_3^2 N g_1^6 g_2^5
+ \frac{1167}{2} \zeta_3^2 N g_1^8 g_2^3
+ 216 \zeta_3^2 N g_1^{10} g_2
\right. \nonumber \\
&& \left. ~~~
- 33 \zeta_3^2 N^2 g_1^6 g_2^5
+ \frac{789}{4} \zeta_3^2 N^2 g_1^8 g_2^3
- \frac{81}{2} \zeta_3^2 N^2 g_1^{10} g_2
- \frac{27}{4} \zeta_3^2 N^3 g_1^{10} g_2
\right] \nonumber \\
&& +~ O \left( g_i^{13} \right)
\label{beta2on5}
\end{eqnarray}
for the $\beta$-functions in the $\MSbar$ scheme where the terms have been
bracketed by loop order when there is more than one contribution. As the 
anomalous dimensions of both fields in the $O(N)$ model have not been recorded 
before we found
\begin{eqnarray}
\gamma_\Phi^{O(N)}(g_i) &=& 
2 g_1^2
+ \left[
- g_1^2 g_2^2
- 2 g_1^4
- N g_1^4
\right]
\nonumber \\
&&
+ \left[
\frac{3}{2} g_1^2 g_2^4
+ g_1^4 g_2^2
+ 2 g_1^6
+ N g_1^4 g_2^2
+ 5 N g_1^6
- \frac{1}{2} N^2 g_1^6
+ 12 \zeta_3 g_1^4 g_2^2
+ 12 \zeta_3 g_1^6
\right]
\nonumber \\
&& + \left[
- \frac{8}{3} g_1^2 g_2^6
- \frac{8}{3} g_1^4 g_2^4
- \frac{1}{3} g_1^6 g_2^2
- \frac{10}{3} g_1^8
- \frac{7}{12} N g_1^4 g_2^4
- 14 N g_1^6 g_2^2
- \frac{38}{3} N g_1^8
\right. \nonumber \\
&& \left. ~~~
+ \frac{11}{6} N^2 g_1^6 g_2^2
- \frac{4}{3} N^2 g_1^8
- \frac{1}{4} N^3 g_1^8
- 40 \zeta_5 g_1^4 g_2^4
- 160 \zeta_5 g_1^6 g_2^2
- 80 \zeta_5 g_1^8
\right. \nonumber \\
&& \left. ~~~
- 40 \zeta_5 N g_1^8
- \frac{3}{2} \zeta_4 g_1^2 g_2^6
+ 15 \zeta_4 g_1^4 g_2^4
+ 21 \zeta_4 g_1^6 g_2^2
+ 24 \zeta_4 g_1^8
+ \frac{15}{2} \zeta_4 N g_1^6 g_2^2
\right. \nonumber \\
&& \left. ~~~
+ 6 \zeta_4 N g_1^8
- \frac{7}{2} \zeta_3 g_1^2 g_2^6
- 30 \zeta_3 g_1^4 g_2^4
- 64 \zeta_3 g_1^6 g_2^2
- 72 \zeta_3 g_1^8
- \frac{1}{2} \zeta_3 N g_1^4 g_2^4
\right. \nonumber \\
&& \left. ~~~
- 45 \zeta_3 N g_1^6 g_2^2
- 26 \zeta_3 N g_1^8
+ \frac{1}{2} \zeta_3 N^2 g_1^6 g_2^2
+ \frac{1}{2} \zeta_3 N^3 g_1^8
\right]
\nonumber \\
&& + \left[
6 g_1^2 g_2^8
+ \frac{67}{12} g_1^4 g_2^6
+ \frac{17}{6} g_1^6 g_2^4
- \frac{1}{3} g_1^8 g_2^2
+ \frac{20}{3} g_1^{10}
- \frac{3}{8} N g_1^4 g_2^6
+ \frac{155}{4} N g_1^6 g_2^4
\right. \nonumber \\
&& \left. ~~~
+ \frac{247}{6} N g_1^8 g_2^2
+ \frac{107}{3} N g_1^{10}
- \frac{39}{8} N^2 g_1^6 g_2^4
+ \frac{11}{4} N^2 g_1^8 g_2^2
+ \frac{76}{3} N^2 g_1^{10}
\right. \nonumber \\
&& \left. ~~~
+ \frac{11}{8} N^3 g_1^8 g_2^2
- \frac{29}{12} N^3 g_1^{10}
- \frac{1}{8} N^4 g_1^{10}
+ \frac{441}{2} \zeta_7 g_1^4 g_2^6
+ \frac{3969}{4} \zeta_7 g_1^6 g_2^4
\right. \nonumber \\
&& \left. ~~~
+ 2205 \zeta_7 g_1^8 g_2^2
+ 882 \zeta_7 g_1^{10}
+ 441 \zeta_7 N g_1^8 g_2^2
+ \frac{2205}{4} \zeta_7 N g_1^{10}
+ \frac{25}{2} \zeta_6 g_1^2 g_2^8
\right. \nonumber \\
&& \left. ~~~
- 100 \zeta_6 g_1^4 g_2^6
- 350 \zeta_6 g_1^6 g_2^4
- 500 \zeta_6 g_1^8 g_2^2
- 300 \zeta_6 g_1^{10}
- 50 \zeta_6 N g_1^6 g_2^4
\right. \nonumber \\
&& \left. ~~~
- 250 \zeta_6 N g_1^8 g_2^2
- 225 \zeta_6 N g_1^{10}
- \frac{75}{2} \zeta_6 N^2 g_1^{10}
+ 11 \zeta_5 g_1^2 g_2^8
+ \frac{355}{2} \zeta_5 g_1^4 g_2^6
\right. \nonumber \\
&& \left. ~~~
+ 531 \zeta_5 g_1^6 g_2^4
+ \frac{2143}{2} \zeta_5 g_1^8 g_2^2
+ 693 \zeta_5 g_1^{10}
- 2 \zeta_5 N g_1^4 g_2^6
+ \frac{455}{2} \zeta_5 N g_1^6 g_2^4
\right. \nonumber \\
&& \left. ~~~
+ 497 \zeta_5 N g_1^8 g_2^2
+ \frac{663}{2} \zeta_5 N g_1^{10}
+ 118 \zeta_5 N^2 g_1^{10}
+ \frac{33}{16} \zeta_4 g_1^2 g_2^8
- 57 \zeta_4 g_1^4 g_2^6
\right. \nonumber \\
&& \left. ~~~
- \frac{219}{2} \zeta_4 g_1^6 g_2^4
- 162 \zeta_4 g_1^8 g_2^2
- 174 \zeta_4 g_1^{10}
- \frac{27}{16} \zeta_4 N g_1^4 g_2^6
- \frac{315}{4} \zeta_4 N g_1^6 g_2^4
\right. \nonumber \\
&& \left. ~~~
- 144 \zeta_4 N g_1^8 g_2^2
- 114 \zeta_4 N g_1^{10}
- \frac{15}{16} \zeta_4 N^2 g_1^6 g_2^4
- \frac{243}{8} \zeta_4 N^2 g_1^8 g_2^2
\right. \nonumber \\
&& \left. ~~~
- \frac{57}{4} \zeta_4 N^2 g_1^{10}
+ \frac{9}{16} \zeta_4 N^3 g_1^8 g_2^2
+ \frac{3}{2} \zeta_4 N^3 g_1^{10}
+ \frac{3}{8} \zeta_4 N^4 g_1^{10}
+ \frac{177}{8} \zeta_3 g_1^2 g_2^8
\right. \nonumber \\
&& \left. ~~~
+ 90 \zeta_3 g_1^4 g_2^6
+ 138 \zeta_3 g_1^6 g_2^4
+ 252 \zeta_3 g_1^8 g_2^2
+ 268 \zeta_3 g_1^{10}
+ \frac{129}{8} \zeta_3 N g_1^4 g_2^6
\right. \nonumber \\
&& \left. ~~~
+ 156 \zeta_3 N g_1^6 g_2^4
+ 443 \zeta_3 N g_1^8 g_2^2
+ 296 \zeta_3 N g_1^{10}
- \frac{23}{8} \zeta_3 N^2 g_1^6 g_2^4
\right. \nonumber \\
&& \left. ~~~
+ \frac{141}{4} \zeta_3 N^2 g_1^8 g_2^2
+ \frac{17}{2} \zeta_3 N^2 g_1^{10}
- \frac{27}{8} \zeta_3 N^3 g_1^8 g_2^2
- \frac{5}{2} \zeta_3 N^3 g_1^{10}
\right. \nonumber \\
&& \left. ~~~
- \frac{1}{4} \zeta_3 N^4 g_1^{10}
- \zeta_3^2 g_1^2 g_2^8
+ 44 \zeta_3^2 g_1^4 g_2^6
+ 172 \zeta_3^2 g_1^6 g_2^4
+ 448 \zeta_3^2 g_1^8 g_2^2
+ 288 \zeta_3^2 g_1^{10}
\right. \nonumber \\
&& \left. ~~~
- 32 \zeta_3^2 N g_1^6 g_2^4
+ 152 \zeta_3^2 N g_1^8 g_2^2
- 54 \zeta_3^2 N g_1^{10}
- 9 \zeta_3^2 N^2 g_1^{10}
\right] ~+~ O \left( g_i^{13} \right)
\label{gampon5}
\end{eqnarray}
and 
\begin{eqnarray}
\gamma_\sigma^{O(N)}(g_i) &=&
\left[
g_2^2
+ N g_1^2
\right]
+ \left[
- g_2^4
- N g_1^2 g_2^2
- 2 N g_1^4
\right]
\nonumber \\
&& + \left[
\frac{5}{4} g_2^6
+ \frac{1}{2} N g_1^2 g_2^4
+ 6 N g_1^4 g_2^2
+ N g_1^6
- \frac{3}{4} N^2 g_1^4 g_2^2
+ 2 N^2 g_1^6
+ 3 \zeta_3 g_2^6
\right. \nonumber \\
&& \left. ~~~
+ 15 \zeta_3 N g_1^4 g_2^2
+ 6 \zeta_3 N g_1^6
\right]
\nonumber \\
&& + \left[
- \frac{9}{4} g_2^8
- \frac{91}{6} N g_1^4 g_2^4
- \frac{13}{3} N g_1^6 g_2^2
- \frac{8}{3} N g_1^8
+ \frac{7}{4} N^2 g_1^4 g_2^4
- \frac{13}{3} N^2 g_1^6 g_2^2
- \frac{28}{3} N^2 g_1^8
\right. \nonumber \\
&& \left. ~~~
- \frac{1}{2} N^3 g_1^6 g_2^2
+ \frac{5}{6} N^3 g_1^8
- 20 \zeta_5 g_2^8
- 80 \zeta_5 N g_1^4 g_2^4
- 160 \zeta_5 N g_1^6 g_2^2
- 40 \zeta_5 N g_1^8
\right. \nonumber \\
&& \left. ~~~
- 20 \zeta_5 N^2 g_1^8
+ \frac{9}{2} \zeta_4 g_2^8
+ 6 \zeta_4 N g_1^2 g_2^6
+ \frac{15}{2} \zeta_4 N g_1^4 g_2^4
+ 24 \zeta_4 N g_1^6 g_2^2
+ 12 \zeta_4 N g_1^8
\right. \nonumber \\
&& \left. ~~~
+ 15 \zeta_4 N^2 g_1^6 g_2^2
+ 3 \zeta_4 N^2 g_1^8
- 15 \zeta_3 g_2^8
- 11 \zeta_3 N g_1^2 g_2^6
- 46 \zeta_3 N g_1^4 g_2^4
- 88 \zeta_3 N g_1^6 g_2^2
\right. \nonumber \\
&& \left. ~~~
- 44 \zeta_3 N g_1^8
+ 2 \zeta_3 N^2 g_1^4 g_2^4
- 32 \zeta_3 N^2 g_1^6 g_2^2
- 6 \zeta_3 N^2 g_1^8
+ \zeta_3 N^3 g_1^6 g_2^2
- \zeta_3 N^3 g_1^8
\right]
\nonumber \\
&& + \left[
\frac{79}{16} g_2^{10}
- \frac{7}{4} N g_1^2 g_2^8
+ \frac{1039}{24} N g_1^4 g_2^6
+ \frac{41}{4} N g_1^6 g_2^4
+ \frac{31}{2} N g_1^8 g_2^2
+ \frac{14}{3} N g_1^{10}
\right. \nonumber \\
&& \left. ~~~
- \frac{37}{8} N^2 g_1^4 g_2^6
+ \frac{215}{24} N^2 g_1^6 g_2^4
+ \frac{313}{6} N^2 g_1^8 g_2^2
+ \frac{131}{6} N^2 g_1^{10}
+ \frac{7}{4} N^3 g_1^6 g_2^4
\right. \nonumber \\
&& \left. ~~~
- \frac{143}{24} N^3 g_1^8 g_2^2
+ \frac{83}{12} N^3 g_1^{10}
- \frac{5}{16} N^4 g_1^8 g_2^2
+ \frac{3}{8} N^4 g_1^{10}
+ \frac{1323}{8} \zeta_7 g_2^{10}
\right. \nonumber \\
&& \left. ~~~
+ \frac{4851}{8} \zeta_7 N g_1^4 g_2^6
+ \frac{11907}{8} \zeta_7 N g_1^6 g_2^4
+ \frac{15435}{8} \zeta_7 N g_1^8 g_2^2
+ 441 \zeta_7 N g_1^{10}
\right. \nonumber \\
&& \left. ~~~
+ \frac{3087}{8} \zeta_7 N^2 g_1^8 g_2^2
+ \frac{2205}{8} \zeta_7 N^2 g_1^{10}
- \frac{225}{4} \zeta_6 g_2^{10}
- \frac{275}{4} \zeta_6 N g_1^2 g_2^8
- 125 \zeta_6 N g_1^4 g_2^6
\right. \nonumber \\
&& \left. ~~~
- 325 \zeta_6 N g_1^6 g_2^4
- \frac{1025}{2} \zeta_6 N g_1^8 g_2^2
- 150 \zeta_6 N g_1^{10}
- 175 \zeta_6 N^2 g_1^6 g_2^4
\right. \nonumber \\
&& \left. ~~~
- \frac{1025}{4} \zeta_6 N^2 g_1^8 g_2^2
- \frac{225}{2} \zeta_6 N^2 g_1^{10}
- \frac{75}{4} \zeta_6 N^3 g_1^{10}
+ \frac{457}{4} \zeta_5 g_2^{10}
+ \frac{193}{2} \zeta_5 N g_1^2 g_2^8
\right. \nonumber \\
&& \left. ~~~
+ \frac{293}{4} \zeta_5 N g_1^4 g_2^6
+ \frac{3069}{2} \zeta_5 N g_1^6 g_2^4
+ \frac{1391}{2} \zeta_5 N g_1^8 g_2^2
+ 239 \zeta_5 N g_1^{10}
- \frac{19}{4} \zeta_5 N^2 g_1^4 g_2^6
\right. \nonumber \\
&& \left. ~~~
- \frac{105}{2} \zeta_5 N^2 g_1^6 g_2^4
+ \frac{1601}{2} \zeta_5 N^2 g_1^8 g_2^2
+ \frac{519}{2} \zeta_5 N^2 g_1^{10}
- \frac{555}{4} \zeta_5 N^3 g_1^8 g_2^2
\right. \nonumber \\
&& \left. ~~~
+ 39 \zeta_5 N^3 g_1^{10}
- \frac{441}{16} \zeta_4 g_2^{10}
- \frac{393}{8} \zeta_4 N g_1^2 g_2^8
- \frac{909}{16} \zeta_4 N g_1^4 g_2^6
- \frac{447}{4} \zeta_4 N g_1^6 g_2^4
\right. \nonumber \\
&& \left. ~~~
- \frac{765}{4} \zeta_4 N g_1^8 g_2^2
- 99 \zeta_4 N g_1^{10}
- \frac{267}{16} \zeta_4 N^2 g_1^4 g_2^6
- \frac{1263}{16} \zeta_4 N^2 g_1^6 g_2^4
\right. \nonumber \\
&& \left. ~~~
- \frac{699}{4} \zeta_4 N^2 g_1^8 g_2^2
- 45 \zeta_4 N^2 g_1^{10}
+ \frac{51}{16} \zeta_4 N^3 g_1^6 g_2^4
- \frac{225}{8} \zeta_4 N^3 g_1^8 g_2^2
- \frac{51}{8} \zeta_4 N^3 g_1^{10}
\right. \nonumber \\
&& \left. ~~~
+ \frac{15}{16} \zeta_4 N^4 g_1^8 g_2^2
- \frac{3}{4} \zeta_4 N^4 g_1^{10}
+ \frac{429}{8} \zeta_3 g_2^{10}
+ 47 \zeta_3 N g_1^2 g_2^8
+ \frac{1653}{8} \zeta_3 N g_1^4 g_2^6
\right. \nonumber \\
&& \left. ~~~
+ \frac{569}{2} \zeta_3 N g_1^6 g_2^4
+ \frac{749}{2} \zeta_3 N g_1^8 g_2^2
+ 162 \zeta_3 N g_1^{10}
- \frac{53}{8} \zeta_3 N^2 g_1^4 g_2^6
+ \frac{915}{8} \zeta_3 N^2 g_1^6 g_2^4
\right. \nonumber \\
&& \left. ~~~
+ \frac{705}{2} \zeta_3 N^2 g_1^8 g_2^2
+ 119 \zeta_3 N^2 g_1^{10}
- \frac{55}{8} \zeta_3 N^3 g_1^6 g_2^4
+ \frac{15}{2} \zeta_3 N^3 g_1^8 g_2^2
+ \frac{33}{4} \zeta_3 N^3 g_1^{10}
\right. \nonumber \\
&& \left. ~~~
- \frac{5}{8} \zeta_3 N^4 g_1^8 g_2^2
+ \frac{1}{4} \zeta_3 N^4 g_1^{10}
+ \frac{63}{2} \zeta_3^2 g_2^{10}
- \frac{25}{2} \zeta_3^2 N g_1^2 g_2^8
+ 172 \zeta_3^2 N g_1^4 g_2^6
\right. \nonumber \\
&& \left. ~~~
+ 206 \zeta_3^2 N g_1^6 g_2^4
+ 389 \zeta_3^2 N g_1^8 g_2^2
+ 144 \zeta_3^2 N g_1^{10}
- 22 \zeta_3^2 N^2 g_1^6 g_2^4
+ \frac{263}{2} \zeta_3^2 N^2 g_1^8 g_2^2
\right. \nonumber \\
&& \left. ~~~
- 27 \zeta_3^2 N^2 g_1^{10}
- \frac{9}{2} \zeta_3^2 N^3 g_1^{10}
\right] ~+~ O \left( g_i^{13} \right)
\label{gamson5}
\end{eqnarray}
in the same scheme. We note that the first two loop orders of each 
$\beta$-function were recorded in \cite{88} with which we are in agreement. In 
\cite{88} the higher loop terms were deduced from the four loop results of
\cite{91}. Therefore the results (\ref{beta1on5}), (\ref{beta2on5}), 
(\ref{gampon5}) and (\ref{gamson5}) are the first direct calculation of the 
$O(N)$ theory renormalization group functions including $\gamma_\Phi(g_i)$ and 
$\gamma_\sigma(g_i)$.

We recall from \cite{88} that there are four different fixed points given by 
the solutions of $\beta_i(g_j)$~$=$~$0$ in $d$~$=$~$4$~$-$~$2\epsilon$. 
Explicit expressions to two loops are recorded in equation (2.4) of \cite{88}. 
One of these is the trivial Gaussian one while two involve one or other of the 
couplings being zero. The remaining fixed point has both $g_1$ and $g_2$ 
non-zero which only exists for $N$~$\leq$~$2$. In this instance when 
$N$~$=$~$2$ the solution for the critical couplings reduces to to the 
$g_1$~$=$~$0$ solution, \cite{88}. In the other case with $N$~$=$~$1$ both 
critical couplings are equal and this corresponds to the emergent 
supersymmetric fixed point in the Gross-Neveu-Yukawa theory. This can be seen 
by computing the eigenvalues of the matrix
\begin{equation}
\beta_{ij}(g_1,g_2) ~=~ \left( 
\frac{\partial \beta_i(g_1,g_2)}{\partial g_j} \right) 
\end{equation}
at the critical point. We find these are
\begin{eqnarray}
\hat{\omega}_1 &=& 2 \epsilon ~-~ \frac{4}{3} \epsilon^2 ~+~ 
\frac{4}{9} [12 \zeta_3 + 1 ] \epsilon^3 ~+~ 
\frac{4}{27} [ 54 \zeta_4 - 84 \zeta_3 - 240 \zeta_5 - 7 ] \epsilon^4 
\nonumber \\
&& +~ \frac{4}{81} [ 576 \zeta_3^2 + 396 \zeta_3 - 378 \zeta_4 + 1416 \zeta_5
- 1800 \zeta_6 + 5292 \zeta_7 + 19 ] \epsilon^5 ~+~ O(\epsilon^6) \nonumber \\
\hat{\omega}_2 &=& \frac{2}{3} \epsilon ~+~ O(\epsilon^6) 
\label{omega12on}
\end{eqnarray}
where the first is equivalent to (\ref{omeps}) and the second would appear to
be exact. 

While we have already noted several internal consistency checks on the earlier
five loop renormalization it is also possible to check the computation via the 
$O(N)$ fixed point given by $g_2$~$=$~$0$. To assist with this we record the 
renormalization group functions for that and note
\begin{eqnarray}
\gamma^{O(N)}_\Phi(g_1,0) &=& 2 g_1^2 ~-~ [N+2] g_1^4 ~-~
\left[ N^2 - 10 N - 4 - 24 \zeta_3 \right] \frac{g_1^6}{2} \nonumber \\
&& +~ \left[ [ 6 \zeta_3 - 3 ] N^3 - 16 N^2 
+ [ 72 \zeta_4 - 152 - 312 \zeta_3 - 480 \zeta_5 ] N - 40 - 864 \zeta_3 
\right. \nonumber \\
&& \left. ~~~~
+ 288 \zeta_4 - 960 \zeta_5 \right] \frac{g_1^8}{12}
\nonumber \\
&& +~ \left[ 9 \zeta_4 N^4 - 6 \zeta_3 N^4 - 3 N^4 - 60 \zeta_3 N^3 
+ 36 \zeta_4 N^3 - 58 N^3 - 216 \zeta_3^2 N^2 
\right. \nonumber \\
&& \left. ~~~~
+ 204 \zeta_3 N^2 - 342 \zeta_4 N^2 + 2832 \zeta_5 N^2 - 900 \zeta_6 N^2 
+ 608 N^2 - 1296 \zeta_3^2 N 
\right. \nonumber \\
&& \left. ~~~~
+ 7104 \zeta_3 N - 2736 \zeta_4 N + 7956 \zeta_5 N - 5400 \zeta_6 N 
+ 13230 \zeta_7 N + 856 N 
\right. \nonumber \\
&& \left. ~~~~
+ 6912 \zeta_3^2 
+ 6432 \zeta_3 - 4176 \zeta_4 + 16632 \zeta_5 - 7200 \zeta_6 + 21168 \zeta_7 
+ 160 \right] \frac{g_1^{10}}{24} \nonumber \\
&& +~ O \left( g_1^{12} \right)
\label{gamp5on}
\end{eqnarray}
and
\begin{eqnarray}
\gamma^{O(N)}_\sigma(g_1,0) &=& N g_1^2 ~-~ 2 N g_1^4 ~+~
N \left[ 2 N + 1 + 6 \zeta_3 \right] g_1^6 \nonumber \\
&& +~ N \left[ [ 5 - 6 \zeta_3 ] N^2 + [ 18 \zeta_4 - 56 - 36 \zeta_3 
- 120 \zeta_5 ] N 
\right. \nonumber \\
&& \left. ~~~~~~
+ [ 72 \zeta_4 - 16 - 264 \zeta_3 - 240 \zeta_5 ] \right] \frac{g_1^8}{6}
\nonumber \\
&& +~ N \left[ 6 \zeta_3 N^3 - 18 \zeta_4 N^3 + 9 N^3 - 108 \zeta_3^2 N^2 
+ 198 \zeta_3 N^2 - 153 \zeta_4 N^2 + 936 \zeta_5 N^2
\right. \nonumber \\
&& \left. ~~~~~~~~
- 450 \zeta_6 N^2 + 166 N^2 - 648 \zeta_3^2 N + 2856 \zeta_3 N - 1080 \zeta_4 N
+ 6228 \zeta_5 N 
\right. \nonumber \\
&& \left. ~~~~~~
- 2700 \zeta_6 N + 6615 \zeta_7 N + 524 N + 3456 \zeta_3^2 + 3888 \zeta_3 
- 2376 \zeta_4 + 5736 \zeta_5 
\right. \nonumber \\
&& \left. ~~~~~~
- 3600 \zeta_6 + 10584 \zeta_7 + 112 \right] \frac{g_1^{10}}{24} ~+~ 
O \left( g_1^{12} \right)
\label{gams5on}
\end{eqnarray}
for the two field anomalous dimensions. The non-trivial $\beta$-function is
\begin{eqnarray}
\beta_1^{O(N)}(g_1,0) &=& [ N + 4 ] \frac{g_1^3}{2} ~-~ 2 [ N + 1 ] g_1^5 ~+~
\left[ N^2 + [ 11 + 6 \zeta_3 ] N + 4 + 24 \zeta_3 \right] \frac{g_1^7}{2}
\nonumber \\
&& +~ \left[ N^3 + [ 9 \zeta_4 - 36 - 18 \zeta_3 - 60 \zeta_5 ] N^2 
+ [ 72 \zeta_4 - 84 - 288 \zeta_3 - 360 \zeta_5 ] N 
\right. \nonumber \\
&& \left. ~~~~
- 20 - 432 \zeta_3 + 144 \zeta_4 - 480 \zeta_5
\right] \frac{g_1^9}{6} \nonumber \\
&& +~ \left[ 3 N^4 - 6 \zeta_3 N^4 - 108 \zeta_3^2 N^3 + 78 \zeta_3 N^3 
- 81 \zeta_4 N^3 + 936 \zeta_5 N^3 - 450 \zeta_6 N^3 
\right. \nonumber \\
&& \left. ~~~~
+ 50 N^3 - 1080 \zeta_3^2 N^2 + 3264 \zeta_3 N^2 - 1764 \zeta_4 N^2 
+ 11892 \zeta_5 N^2 - 4500 \zeta_6 N^2 
\right. \nonumber \\
&& \left. ~~~~
+ 6615 \zeta_7 N^2 + 1740 N^2 + 864 \zeta_3^2 N + 18096 \zeta_3 N 
- 7848 \zeta_4 N + 21648 \zeta_5 N 
\right. \nonumber \\
&& \left. ~~~~
- 14400 \zeta_6 N + 37044 \zeta_7 N + 1824 N + 13824 \zeta_3^2 + 12864 \zeta_3 
- 8352 \zeta_4 
\right. \nonumber \\
&& \left. ~~~~
+ 33264 \zeta_5 - 14400 \zeta_6 + 42336 \zeta_7 
+ 320 \right] \frac{g_1^{11}}{48} ~+~ O \left( g_1^{13} \right) ~.
\label{bet5on}
\end{eqnarray}

We recall that the $O(N)$ Wess-Zumino model renormalization group functions are
known to several orders in the $1/N$ expansion, \cite{47,48,49}. The $O(1/N^2)$
correction to the $\beta$-function and the $O(1/N^3)$ ones for $\gamma_\Phi(a)$
were computed by exploiting the scaling properties of the propagators at the 
Wilson-Fisher fixed point in $d$-dimensions using the large $N$ formalism 
developed in \cite{73,73,89} for the non-supersymmetric version of 
(\ref{lagwz}) which is the $O(N)$ non-linear sigma model. That model is in the 
same universality class of $O(N)$ $\phi^4$ theory in four dimensions. In order 
to check (\ref{gamp5on}) and (\ref{bet5on}) in large $N$ we compute the 
critical exponents 
$\eta_\Phi^{O(N)}$~$=$~$\half \gamma^{O(N)}_\Phi(g_1^\ast,0)$ and 
$\hat{\omega}^{O(N)}$~$=$~$\half \left(\beta_1^{O(N)}\right)^\prime 
(g_1^\ast,0)$ where $g_1^\ast$ is the value of the coupling constant at the 
Wilson-Fisher critical point in $d$-dimensions and the factor of $2$ has been
omitted here to be consistent with the definition used in \cite{48}. From 
(\ref{bet5on}) we have
\begin{eqnarray}
{g_1^\ast}^2 &=& \frac{2\epsilon}{N}
+ \left[
- 8 \epsilon
+ 16 \epsilon^2
- 8 \epsilon^3
- \frac{16}{3} \epsilon^4
+ [ 8 \zeta_3 - 4 ] \epsilon^5
\right] \frac{1}{N}
\nonumber \\
&&
+ \left[
32 \epsilon
- 176 \epsilon^2
+ [ 296
- 48 \zeta_3 ] \epsilon^3
+ [ 320 \zeta_5
- \frac{64}{3}
- 48 \zeta_4
+ 96 \zeta_3 ] \epsilon^4
\right. \nonumber \\
&& \left. ~~~
+ [ 600 \zeta_6
- \frac{392}{3}
- 1248 \zeta_5
+ 108 \zeta_4
- 296 \zeta_3
+ 144 \zeta_3^2 ] \epsilon^5
\right] \frac{1}{N^2} ~+~ O \left( \epsilon^6; \frac{1}{N^3} \right) ~~~~~
\label{critccon}
\end{eqnarray}
to the necessary orders in powers of $1/N$ that are needed to compare with
\cite{47,48,49}. Thus we have 
\begin{eqnarray}
\eta_\Phi^{O(N)} &=&
\left[
2 \epsilon
- 2 \epsilon^2
- 2 \epsilon^3
+ [ 4 \zeta_3 - 2 ] \epsilon^4
+ [ 6 \zeta_4 - 2 - 4 \zeta_3 ] \epsilon^5 \right] \frac{1}{N}
\nonumber \\
&&
+ \left[
- 8 \epsilon
+ 28 \epsilon^2
+ 4 \epsilon^3 - [ 16
+ 64 \zeta_3 ] \epsilon^4
+ [ 176 \zeta_3 - 32 - 95 \zeta_4 ] \epsilon^5 \right] \frac{1}{N^2}
\nonumber \\
&&
+ \left[
32 \epsilon
- 240 \epsilon^2
+ 288 \epsilon^3
+ [ 368 + 624 \zeta_3 ] \epsilon^4
+ [ 144 + 936 \zeta_4 - 3312 \zeta_3 ] \epsilon^5 \right] \frac{1}{N^3}
\nonumber \\
&& +~ O \left( \epsilon^6; \frac{1}{N^4} \right)
\end{eqnarray}
and
\begin{eqnarray}
\hat{\omega}^{O(N)} &=&
\epsilon
+ \left[
- 8 \epsilon^2
+ 8 \epsilon^3
+ 8 \epsilon^4
+ [ 8 - 16 \zeta_3 ] \epsilon^5
\right] \frac{1}{N}
\nonumber \\
&&
+ \left[
56 \epsilon^2
+ [ 48 \zeta_3 - 136 ] \epsilon^3
+ [ 72 \zeta_4 - 160 - 480 \zeta_5 - 144 \zeta_3 ] \epsilon^4
\right. \nonumber \\
&& \left. ~~~
+ [ 176 - 1200 \zeta_6 + 1696 \zeta_5
- 216 \zeta_4 + 528 \zeta_3
- 288 \zeta_3^2 ] \epsilon^5 \right] \frac{1}{N^2} \nonumber \\
&& +~ O \left( \epsilon^6; \frac{1}{N^3} \right) ~.
\end{eqnarray}
If one expands the $d$-dimensional expressions for $\eta$ and $\hat{\omega}$ of
\cite{48,49} in powers of $\epsilon$ we find precise agreement. This is the
other non-trivial check on our perturbative computation, that we referred to
earlier, since the higher order large $N$ calculations involve the three and 
four loop primitive topologies. Hence several of the dressed propagator graphs 
of Figures \ref{figtw53} and \ref{figtw54} arise in the higher order large $N$ 
exponent calculations. The critical exponent associated with 
$\eta^{O(N)}_\sigma$~$=$~$\gamma^{O(N)}_\sigma(g_1^\ast,0)$ is also in 
agreement. However this is a trivial check since the vertex of (\ref{lagwzon})
is not renormalized due to the supersymmetry Ward identity. Thus at the 
critical point this implies that the vertex anomalous dimension exponent is 
zero to all orders and so $\eta^{O(N)}_\sigma$ is not independent of 
$\eta^{O(N)}_\Phi$. We have checked that this is indeed the case to five loops 
and $O(1/N^3)$. In fact given this identity the Wess-Zumino model is perhaps 
the first case where the anomalous dimension of the linear field in the cubic 
interaction of the class of large $N$ expandable theories using the technology 
of \cite{73,74,89} is available at $O(1/N^3)$ rather than $O(1/N^2)$.

One observation in respect of the connection between the Wess-Zumino model and 
the emergent supersymmetry of the Gross-Neveu-Yukawa Lagrangian needs to be 
made in the context of the large $N$ expansion. First we set some notation and 
denote the $O(1/N^r)$ term of the matter field anomalous dimension by $\eta_r$ 
for both theories. By matter field we mean $\Phi^i$ of (\ref{lagwzon}) and 
$\psi^i$ of the $O(N)$ extension of (\ref{lagwzc}) when an $O(N)$ symmetry is 
included. For background to this point we recall that in the scalar $O(N)$ 
universality class containing four dimensional $\phi^4$ theory the 
$d$-dimensional expression for $\eta_3$, \cite{89}, involved a function 
$I(\mu)$ which was related to an ${}_4 F_3$ hypergeometric function in
\cite{92,93}. Its $\epsilon$ expansion near four dimensions involves multiple 
zeta values, \cite{89,92,94}, and implies that such irrationals will appear at 
high loop order in the renormalization group functions. The same function 
appears in $\eta_3$ in various other models including the $O(N)$ Gross-Neveu 
model, \cite{95,96}, and its ${\cal N}$~$=$~$1$ supersymmetric 
extension \cite{97}. What was unusual about $\eta_3$ computed for 
(\ref{lagwzon}) in \cite{49} was that the integral $I(\mu)$ did not appear. 
This was attributed to either the presence of supersymmetry, since 
simplifications in the renormalization group functions are known to occur when 
this symmetry is present, or chiral symmetry. Alternatively both symmetries 
could have equally conspired to exclude the underlying topologies that would 
have led to $I(\mu)$. The key point is that to $O(1/N^3)$ no multiple zeta 
irrationals will appear in $\gamma^{O(N)}_\Phi(a)$. Since the simple $O(N)$ 
Gross-Neveu model $\eta_3$ contains $I(\mu)$, \cite{95,96}, one question that 
was recently addressed, \cite{98}, was whether $I(\mu)$ would be present in 
$\eta_3$ of the non-supersymmetric chiral XY or chiral Gross-Neveu model 
universality class where the theory has a $U(1)$ symmetry. This was 
particularly relevant since the four dimensional theory has an emergent 
supersymmetry. It transpires that the $d$-dimensional expression for $\eta_3$ 
in the chiral Gross-Neveu theory does {\em not} contain $I(\mu)$, \cite{98}. 
Although the emergent supersymmetry occurs for a specific value of $N$ that is 
low, the large $N$ critical exponent $\eta_3$ contains information on the 
renormalization group functions. While the absence of $I(\mu)$ in the chiral 
Gross-Neveu model at $O(1/N^3)$ is an indirect indication of the structural 
similarities of both models at criticality it also suggests that the absence of
$I(\mu)$ is perhaps due to the chiral symmetry. One final comment needs to be 
made concerning the multiple zeta irrationals. The absence of such numbers at 
$O(1/N^3)$ does not necessarily imply that they are absent for all orders in 
large $N$ or perturbation theory. They could arise at much higher order. In 
perturbation theory for example the first multiple zeta, $\zeta_{3,5}$, appears
at six loops in $\phi^4$ theory $\beta$-function. That term would be present in
the critical $\beta$-function exponent at $O(1/N^3)$ in the large $N$ expansion
of the $O(N)$ extension of that model, \cite{89,93}.

At the end of this section we pause to discuss a potential connection with the
large $N$ expansion technique mentioned here in relation to the renormalization
group functions and the Hopf algebra solution of the Dyson-Schwinger equations
of \cite{81}. Indeed the large $N$ methods of \cite{73,74} also relies upon the
solution of the Dyson-Schwinger equation in the critical region close to the 
Wilson-Fisher fixed point. In the latter approach the use of the group 
invariants has allowed us to identify that solution with a seemingly parallel 
bubble expansion. This is effected through the group factor $T_2$. For instance
the $\epsilon$ expansion of the correction to scaling exponent was given in 
(\ref{omegaT2}) through the critical coupling (\ref{critccT2}) and both have a
similar structure to each other. Both actions (\ref{lagwzt}) and 
(\ref{lagwzon}), however, are different in that the former involves one field 
whereas the latter has an $O(N)$ multiplet of fields in addition to a scalar 
field. Indeed the interaction connecting both fields is akin to the force 
matter one of QCD which is a theory of $\Nf$ quarks with gluons that are 
elements of the adjoint representation of the $SU(\Nc)$ Lie group with 
$\Nc$~$=$~$3$. In addition to canonical perturbation theory it admits both a 
large $\Nf$ and large $\Nc$ expansion with the former being achieved using the
same techniques as \cite{73,74}. The large $\Nc$ properties have also been 
widely investigated where background to the issues are given in \cite{99,100}. 
There could not be a greater difference though in how the Feynman graphs of 
each expansion are ordered. For instance in the solution of the large $\Nf$ 
Dyson-Schwinger equations at criticality there is a finite and small number of 
graphs at leading order. By contrast in the large $\Nc$ case it is known that 
there are an infinite number of graphs at leading order, \cite{20,21}. This is 
evident in the structure of the QCD $\beta$-function. To two loops it is linear
in $\Nf$ which means the leading large $\Nf$ term of the critical coupling at 
the Wilson-Fisher fixed point has a finite number of terms in $\epsilon$. In 
fact there is only one. The $\Nc$ dependence for the $SU(\Nc)$ colour group by 
contrast is different in that the coefficient of the leading order $1/\Nc$ term
of the critical coupling is an infinite series in $\epsilon$. In the absence of
 the all orders $\beta$-function it therefore remains unavailable. These two 
situations have parallels in the two actions (\ref{lagwzt}) and 
(\ref{lagwzon}). Clearly the large $N$ expansion discussed in this section is 
completely the same as the large $\Nf$ one of QCD given the common use of
\cite{73,74} in finding the $d$-dimensional critical exponents. Indeed the 
critical coupling (\ref{critccon}) has only one term at leading order as the 
$\beta$-function (\ref{bet5on}) is linear in $N$. By contrast the 
$\beta$-function of the other action, (\ref{gam5T}) is not linear in $T_2$ 
which leads to an infinite number of terms in $\epsilon$ at leading order in 
the $1/T_2$ expansion of the critical coupling (\ref{critccT2}). Equally the 
correction to scaling exponent has the same property in complete parallel with 
the large $\Nc$ expansion.

This suggests that the $1/T_2$ expansion of the renormalization group functions
of (\ref{lagwzt}) using the Hopf algebra solution of the Dyson-Schwinger 
equation is a potential way of carrying out a large $\Nc$ expansion of the 
$\beta$-function of QCD. It is worth outlining the ingredients needed for such 
an exercise. Indeed there are many challenges that would need to be resolved.
First, the Wess-Zumino model has a supersymmetry Ward identity that allows the 
$\beta$-function to be deduced from the field anomalous dimension. So the
Dyson-Schwinger equation for the vertex function would need to be analysed in
the Hopf algebra formalism. This could be played out in the same laboratory of
$\phi^3$ and scalar-Yukawa theory \cite{82,83} where the field anomalous 
dimension was examined in the first instance. Next in the QCD case there is the
complication of gauge symmetry. Even for Yang-Mills theory one would have more 
Dyson-Schwinger equations to consider. Aside from treating the transverse and 
longitudinal contributions to the gluon equations separately, unless the focus 
was on the Landau gauge, the Faddeev-Popov ghost Dyson-Schwinger equation would
play a non-trivial role. The use of the Landau gauge may have the advantage 
that the $\beta$-function could be accessible in the Hopf approach since the 
ghost-gluon vertex is finite in this gauge due to Taylor's theorem, \cite{101}.
This would be a parallel to the non-renormalization of the Wess-Zumino vertex 
here due to the supersymmetry Ward identity. While these observations have in
the main concentrated on the close similarities there are inevitably several 
technical differences. The obvious one is that the set of basic Feynman graphs 
of the Wess-Zumino model is smaller than the QCD one. By set we mean the 
underlying graph topology and the difference lies in the absence of one loop 
subgraphs with an odd number of propagators as well as no quartic interaction. 
In turn this means that the group invariant designation $T_i$ does not have the
same parallels as the group Casimirs in QCD. This is understandable since the 
core tensor of (\ref{lagwzt}) is symmetric in contrast to the antisymmetric 
structure constants of the $SU(\Nc)$ Lie colour group. In this case while $T_5$
does have a partner group theory combination in Yang-Mills, since the two loop 
non-planar vertex function has subgraphs with an even number of propagators, it
is actually zero in the adjoint representation in Yang-Mills theory. Instead 
$T_{71}$ would be the first topology that non-trivially connects with graphs in
QCD where they would equate with the so-called four loop light-by-light graphs.
Despite these issues that we have outlined it would seem that the Hopf algebra 
approach offers a viable way of probing ideas concerning the renormalization 
group functions of QCD in the $1/\Nc$ expansion in parallel with potentially 
the same benefit as the large $\Nf$ $d$-dimensional critical exponents. Finally
we remark that there is also the potential for the Hopf algebra constuction 
given in \cite{81} to be extended to the next order for the Wess-Zumino model. 
From the location of $T_5$ in (\ref{critccT2}) and (\ref{omegaT2}) it is clear 
that the next topology to consider beyond the iteration of the one loop bubble 
used in \cite{81} is the bubble decoration of the non-planar primitive of 
Figure \ref{figtw3}. The Chebyshev polynomial approach to evaluate this graph 
given in the appendix of \cite{4} should be useful in this respect. 

\sect{Tensor $O(N)$ Wess-Zumino model.}

We now turn to an alternative version of the $O(N)$ theory which we will term
the tensor $O(N)$ Wess-Zumino model as it also has an origin in 
non-supersymmetric $O(N)$ $\phi^4$ theory. In that case the interaction 
$(\phi^2)^2$ can be rewritten in terms of an auxiliary field $\sigma$ which 
leads to the cubic interaction akin to that of (\ref{lagwzon}). As pointed out 
in \cite{50,102} this is not the only way of decomposing the quartic 
interaction since one can introduce a tensor channel rather than a scalar one. 
In this case the auxiliary field is a vector in the $O(N)$ group and denoted by
$\sigma^a$ where $1$~$\leq$~$a$~$\leq$~$\NA$ with $\NA$~$=$~$\half (N-1)(N+2)$.
Since this decomposition has parallels with the canonical one of 
(\ref{lagwzon}) it can also be incorporated in the Wess-Zumino case as well. 
This is the focus of this section and we note the bare action is 
\begin{eqnarray}
S^{O_\Lambda(N)} &=& \int d^4 x \left[ \int d^2 \theta d^2 \bar{\theta} \,
\bar{\Phi}_0 (x,\bar{\theta}) e^{-2 \theta {\partialline} \bar{\theta}} 
\Phi_0 (x,\theta) ~+~ 
\bar{\sigma}^a_0 (x,\bar{\theta}) e^{-2 \theta {\partialline} \bar{\theta}} 
\sigma^a_0 (x,\theta) \right. \nonumber \\
&& \left. ~~~~~~~~~+~ 
\frac{{g_1}_0}{2} \int d^2 \theta \, 
\sigma^a_0 \Phi_0^i \Lambda^a_{ij} \Phi_0^j ~+~ 
\frac{{g_1}_0}{2} \int d^2 \bar{\theta} \, \bar{\sigma}^a_0 \bar{\Phi}_0^i 
\Lambda^a_{ij} \bar{\Phi}_0^j \right. \nonumber \\
&& \left. ~~~~~~~~~+~ 
\frac{{g_2}_0}{6} d_3^{abc} \int d^2 \theta \, 
\sigma^a_0 \sigma^b_0 \sigma^c_0 ~+~ 
\frac{{g_2}_0}{6} d_3^{abc} \int d^2 \bar{\theta} \, \bar{\sigma}^a_0
\bar{\sigma}^b_0 \bar{\sigma}^c_0 \right]
\label{lagwzontens}
\end{eqnarray}
where the fully symmetric rank $3$ tensor depends on the $\NA$ real, symmetric,
traceless matrices $\Lambda^a_{ij}$ via
\begin{equation}
d_3^{abc} ~=~ \mbox{Tr} \left( \Lambda^a \Lambda^b \Lambda^c \right) 
\end{equation}
which formally has similar interactions to the non-supersymmetric scalar
tensor $O(N)$ cubic theory that is renormalizable in six dimensions
\cite{50,102}.

With this action we have constructed the five loop renormalization group
functions using an extension of the algorithm for the scalar decomposition of
the previous section. The supersymmetry Ward identities (\ref{susywion}) remain
the same. So all that is entailed is to append a {\sc Form} group theory module
to handle the presence of the matrix. Useful in implementing this is the 
relation \cite{50}
\begin{equation}
\Lambda^a_{ij} \Lambda^a_{kl} ~=~ \delta_{ik} \delta_{jl} ~+~
\delta_{il} \delta_{jk} ~-~ \frac{2}{N} \delta_{ij} \delta_{kl} ~.
\end{equation}
Like \cite{52} the expressions for the renormalization group functions for 
arbitrary $N$ are sizeable and included in the attached data file. However it 
is valuable to record them for one particular value of $N$. For instance when 
$N$~$=$~$3$ we have 
\begin{eqnarray}
\left. \gamma_\Phi(g_i) \right|_{N=3} &=&
\frac{20}{3} g_1^2
+ \frac{20}{9} \left[ 
- 13 g_1^2 
- 7 g_2^2 \right] g_1^2 
\nonumber \\
&&
+ \frac{20}{27} \left[ 276 \zeta_3 g_1^4 
+ 241 g_1^4 
- 84 \zeta_3 g_1^2 g_2^2 
+ 112 g_1^2 g_2^2 
+ 147 g_2^4 \right] g_1^2 
\nonumber \\
&&
+ \frac{20}{243} \left[ 
- 61254 \zeta_3 g_1^6 
+ 19044 \zeta_4 g_1^6 
- 61680 \zeta_5 g_1^6 
- 17201 g_1^6 
+ 2940 \zeta_3 g_1^4 g_2^2 
\right. \nonumber \\
&& \left. ~~~~~~~~
+ 5229 \zeta_4 g_1^4 g_2^2 
- 26880 \zeta_5 g_1^4 g_2^2 
- 16954 g_1^4 g_2^2 
+ 7938 \zeta_3 g_1^2 g_2^4 
- 4410 \zeta_4 g_1^2 g_2^4 
\right. \nonumber \\
&& \left. ~~~~~~~~
- 38640 \zeta_5 g_1^2 g_2^4 
- 8869 g_1^2 g_2^4 
- 7224 \zeta_3 g_2^6 
- 2583 \zeta_4 g_2^6 
- 10976 g_2^6 \right] g_1^2 
\nonumber \\
&&
+ \frac{5}{729} \left[ 2017008 \zeta_3^2 g_1^8 
+ 12797088 \zeta_3 g_1^8 
- 6943608 \zeta_4 g_1^8 
+ 21262968 \zeta_5 g_1^8 
\right. \nonumber \\
&& \left. ~~~~~~~~
- 10639800 \zeta_6 g_1^8 
+ 20806821 \zeta_7 g_1^8 
+ 2198908 g_1^8 
- 3786048 \zeta_3^2 g_1^6 g_2^2 
\right. \nonumber \\
&& \left. ~~~~~~~~
+ 2103360 \zeta_3 g_1^6 g_2^2 
- 2771496 \zeta_4 g_1^6 g_2^2 
+ 17172792 \zeta_5 g_1^6 g_2^2 
- 7812000 \zeta_6 g_1^6 g_2^2 
\right. \nonumber \\
&& \left. ~~~~~~~~
+ 5260248 \zeta_7 g_1^6 g_2^2 
+ 2154908 g_1^6 g_2^2 
- 784896 \zeta_3^2 g_1^4 g_2^4 
+ 233436 \zeta_3 g_1^4 g_2^4 
\right. \nonumber \\
&& \left. ~~~~~~~~
+ 168462 \zeta_4 g_1^4 g_2^4 
+ 15298584 \zeta_5 g_1^4 g_2^4 
- 5913600 \zeta_6 g_1^4 g_2^4 
+ 6306741 \zeta_7 g_1^4 g_2^4 
\right. \nonumber \\
&& \left. ~~~~~~~~
+ 2861012 g_1^4 g_2^4 
- 2010624 \zeta_3^2 g_1^2 g_2^6 
- 3192 \zeta_3 g_1^2 g_2^6 
+ 1006236 \zeta_4 g_1^2 g_2^6 
\right. \nonumber \\
&& \left. ~~~~~~~~
+ 11106984 \zeta_5 g_1^2 g_2^6 
- 5409600 \zeta_6 g_1^2 g_2^6 
+ 5093550 \zeta_7 g_1^2 g_2^6 
+ 882196 g_1^2 g_2^6 
\right. \nonumber \\
&& \left. ~~~~~~~~
- 92400 \zeta_3^2 g_2^8 
+ 2552508 \zeta_3 g_2^8 
+ 73206 \zeta_4 g_2^8 
+ 1000272 \zeta_5 g_2^8 
\right. \nonumber \\
&& \left. ~~~~~~~~
+ 1155000 \zeta_6 g_2^8 
+ 1382976 g_2^8 \right] g_1^2 ~+~ O \left( g_i^{12} \right)
\end{eqnarray}
and
\begin{eqnarray}
\left. \gamma_\sigma(g_i) \right|_{N=3} &=&
\frac{2}{3} \left[ 3 g_1^2 + 7 g_2^2 \right] 
+ \frac{4}{9} \left[ 
- 30 g_1^4 
- 21 g_1^2 g_2^2 
- 49 g_2^4 \right] 
\nonumber \\
&&
+ \frac{2}{27} \left[ 828 \zeta_3 g_1^6 
+ 660 g_1^6 
- 630 \zeta_3 g_1^4 g_2^2 
+ 2331 g_1^4 g_2^2 
+ 294 g_1^2 g_2^4 
\right. \nonumber \\
&& \left. ~~~~~~~
+ 1722 \zeta_3 g_2^6 
+ 1715 g_2^6 \right] 
\nonumber \\
&&
+ \frac{4}{81} \left[ 
- 30144 \zeta_3 g_1^8 
+ 9522 \zeta_4 g_1^8 
- 30840 \zeta_5 g_1^8 
- 11950 g_1^8 
- 2856 \zeta_3 g_1^6 g_2^2 
\right. \nonumber \\
&& \left. ~~~~~~~
- 6930 \zeta_4 g_1^6 g_2^2 
- 26880 \zeta_5 g_1^6 g_2^2 
- 14938 g_1^6 g_2^2 
+ 13818 \zeta_3 g_1^4 g_2^4 
- 2205 \zeta_4 g_1^4 g_2^4 
\right. \nonumber \\
&& \left. ~~~~~~~
- 77280 \zeta_5 g_1^4 g_2^4 
- 41503 g_1^4 g_2^4 
- 16548 \zeta_3 g_1^2 g_2^6 
+ 10332 \zeta_4 g_1^2 g_2^6 
- 60270 \zeta_3 g_2^8 
\right. \nonumber \\
&& \left. ~~~~~~~
+ 18081 \zeta_4 g_2^8 
- 77000 \zeta_5 g_2^8 
- 21609 g_2^8 \right]
\nonumber \\
&&
+ \frac{1}{486} \left[ 
2017008 \zeta_3^2 g_1^{10} 
+ 12023784 \zeta_3 g_1^{10} 
- 6277068 \zeta_4 g_1^{10} 
+ 17472048 \zeta_5 g_1^{10} 
\right. \nonumber \\
&& \left. ~~~~~~~~
- 10639800 \zeta_6 g_1^{10} 
+ 20806821 \zeta_7 g_1^{10} 
+ 2440520 g_1^{10} 
- 6471360 \zeta_3^2 g_1^8 g_2^2 
\right. \nonumber \\
&& \left. ~~~~~~~~
+ 8625792 \zeta_3 g_1^8 g_2^2 
- 1047312 \zeta_4 g_1^8 g_2^2 
+ 30972984 \zeta_5 g_1^8 g_2^2 
- 9815400 \zeta_6 g_1^8 g_2^2 
\right. \nonumber \\
&& \left. ~~~~~~~~
+ 9205434 \zeta_7 g_1^8 g_2^2 
+ 7490700 g_1^8 g_2^2 
- 5869248 \zeta_3^2 g_1^6 g_2^4 
- 92316 \zeta_3 g_1^6 g_2^4 
\right. \nonumber \\
&& \left. ~~~~~~~~
+ 2862090 \zeta_4 g_1^6 g_2^4 
+ 60678240 \zeta_5 g_1^6 g_2^4 
- 24460800 \zeta_6 g_1^6 g_2^4 
\right. \nonumber \\
&& \left. ~~~~~~~~
+ 18920223 \zeta_7 g_1^6 g_2^4 
+ 3823176 g_1^6 g_2^4 
- 5540640 \zeta_3^2 g_1^4 g_2^6 
+ 5826828 \zeta_3 g_1^4 g_2^6 
\right. \nonumber \\
&& \left. ~~~~~~~~
- 2235618 \zeta_4 g_1^4 g_2^6 
+ 44301264 \zeta_5 g_1^4 g_2^6 
- 13524000 \zeta_6 g_1^4 g_2^6 
\right. \nonumber \\
&& \left. ~~~~~~~~
+ 28014525 \zeta_7 g_1^4 g_2^6 
+ 13341328 g_1^4 g_2^6 
- 2310000 \zeta_3^2 g_1^2 g_2^8 
+ 7958580 \zeta_3 g_1^2 g_2^8 
\right. \nonumber \\
&& \left. ~~~~~~~~
- 9122526 \zeta_4 g_1^2 g_2^8 
+ 17502576 \zeta_5 g_1^2 g_2^8 
- 12705000 \zeta_6 g_1^2 g_2^8 
- 806736 g_1^2 g_2^8 
\right. \nonumber \\
&& \left. ~~~~~~~~
+ 2651040 \zeta_3^2 g_2^{10} 
+ 24601332 \zeta_3 g_2^{10} 
- 12403566 \zeta_4 g_2^{10} 
+ 48544888 \zeta_5 g_2^{10} 
\right. \nonumber \\
&& \left. ~~~~~~~~
- 24255000 \zeta_6 g_2^{10} 
+ 47944197 \zeta_7 g_2^{10} 
+ 5311012 g_2^{10} \right] ~+~ O \left( g_i^{12} \right)
\end{eqnarray}
for the $\MSbar$ field anomalous dimensions and
\begin{eqnarray}
\left. \beta_1(g_i) \right|_{N=3} &=&
\frac{1}{3} \left[ 23 g_1^2 + 7 g_2^2 \right] g_1 
+ \frac{2}{9} \left[ 
- 160 g_1^4 
- 91 g_1^2 g_2^2 
- 49 g_2^4 \right] g_1 
\nonumber \\
&& 
+ \frac{1}{27} \left[ 6348 \zeta_3 g_1^6 
+ 5480 g_1^6 
- 2310 \zeta_3 g_1^4 g_2^2
+ 4571 g_1^4 g_2^2 
+ 3234 g_1^2 g_2^4 
\right. \nonumber \\
&& \left. ~~~~~~
+ 1722 \zeta_3 g_2^6 
+ 1715 g_2^6 \right] g_1 
\nonumber \\
&&
+ \frac{2}{243} \left[ 
- 702972 \zeta_3 g_1^8 
+ 219006 \zeta_4 g_1^8 
- 709320 \zeta_5 g_1^8 
- 207860 g_1^8 
+ 20832 \zeta_3 g_1^6 g_2^2 
\right. \nonumber \\
&& \left. ~~~~~~~~
+ 31500 \zeta_4 g_1^6 g_2^2 
- 349440 \zeta_5 g_1^6 g_2^2 
- 214354 g_1^6 g_2^2 
+ 120834 \zeta_3 g_1^4 g_2^4 
\right. \nonumber \\
&& \left. ~~~~~~~~
- 50715 \zeta_4 g_1^4 g_2^4 
- 618240 \zeta_5 g_1^4 g_2^4 
- 213199 g_1^4 g_2^4 
- 121884 \zeta_3 g_1^2 g_2^6 
\right. \nonumber \\
&& \left. ~~~~~~~~
+ 5166 \zeta_4 g_1^2 g_2^6 
- 109760 g_1^2 g_2^6 
- 180810 \zeta_3 g_2^8 
+ 54243 \zeta_4 g_2^8 
\right. \nonumber \\
&& \left. ~~~~~~~~
- 231000 \zeta_5 g_2^8 
- 64827 g_2^8 \right] g_1 
\nonumber \\
&&
+ \frac{1}{2916} \left[ 
46391184 \zeta_3^2 g_1^{10} 
+ 292013112 \zeta_3 g_1^{10} 
- 157703364 \zeta_4 g_1^{10} 
+ 477675504 \zeta_5 g_1^{10} 
\right. \nonumber \\
&& \left. ~~~~~~~~
- 244715400 \zeta_6 g_1^{10} 
+ 478556883 \zeta_7 g_1^{10} 
+ 51299720 g_1^{10} 
- 95135040 \zeta_3^2 g_1^8 g_2^2 
\right. \nonumber \\
&& \left. ~~~~~~~~
+ 67944576 \zeta_3 g_1^8 g_2^2 
- 58571856 \zeta_4 g_1^8 g_2^2 
+ 436374792 \zeta_5 g_1^8 g_2^2 
\right. \nonumber \\
&& \left. ~~~~~~~~
- 185686200 \zeta_6 g_1^8 g_2^2 
+ 132821262 \zeta_7 g_1^8 g_2^2 
+ 65570260 g_1^8 g_2^2 
\right. \nonumber \\
&& \left. ~~~~~~~~
- 33305664 \zeta_3^2 g_1^6 g_2^4 
+ 4391772 \zeta_3 g_1^6 g_2^4 
+ 11955510 \zeta_4 g_1^6 g_2^4 
\right. \nonumber \\
&& \left. ~~~~~~~~
+ 488006400 \zeta_5 g_1^6 g_2^4 
- 191654400 \zeta_6 g_1^6 g_2^4 
+ 182895489 \zeta_7 g_1^6 g_2^4 
\right. \nonumber \\
&& \left. ~~~~~~~~
+ 68689768 g_1^6 g_2^4 
- 56834400 \zeta_3^2 g_1^4 g_2^6 
+ 17416644 \zeta_3 g_1^4 g_2^6 
+ 13417866 \zeta_4 g_1^4 g_2^6 
\right. \nonumber \\
&& \left. ~~~~~~~~
+ 355043472 \zeta_5 g_1^4 g_2^6 
- 148764000 \zeta_6 g_1^4 g_2^6 
+ 185914575 \zeta_7 g_1^4 g_2^6 
\right. \nonumber \\
&& \left. ~~~~~~~~
+ 57667904 g_1^4 g_2^6 
- 8778000 \zeta_3^2 g_1^2 g_2^8 
+ 74925900 \zeta_3 g_1^2 g_2^8 
\right. \nonumber \\
&& \left. ~~~~~~~~
- 25903458 \zeta_4 g_1^2 g_2^8 
+ 72513168 \zeta_5 g_1^2 g_2^8 
- 15015000 \zeta_6 g_1^2 g_2^8 
\right. \nonumber \\
&& \left. ~~~~~~~~
+ 25239312 g_1^2 g_2^8 
+ 7953120 \zeta_3^2 g_2^{10} 
+ 73803996 \zeta_3 g_2^{10} 
- 37210698 \zeta_4 g_2^{10} 
\right. \nonumber \\
&& \left. ~~~~~~~~
+ 145634664 \zeta_5 g_2^{10} 
- 72765000 \zeta_6 g_2^{10} 
+ 143832591 \zeta_7 g_2^{10} 
+ 15933036 g_2^{10} \right] g_1 
\nonumber \\
&& +~ O \left( g_i^{13} \right)
\label{tenbet1n3}
\end{eqnarray}
together with
\begin{eqnarray}
\left. \beta_2(g_i) \right|_{N=3} &=&
\left[ 3 g_1^2 + 7 g_2^2 \right] g_2 
+ \frac{2}{3} \left[ 
- 30 g_1^4 
- 21 g_1^2 g_2^2 
- 49 g_2^4 \right] g_2 
\nonumber \\
&&
+ \frac{1}{9} \left[ 828 \zeta_3 g_1^6 
+ 660 g_1^6 
- 630 \zeta_3 g_1^4 g_2^2 
+ 2331 g_1^4 g_2^2 
+ 294 g_1^2 g_2^4 
\right. \nonumber \\
&& \left. ~~~~~
+ 1722 \zeta_3 g_2^6 
+ 1715 g_2^6 \right] g_2  
\nonumber \\
&&
+ \frac{2}{27} \left[ 
- 30144 \zeta_3 g_1^8 
+ 9522 \zeta_4 g_1^8 
- 30840 \zeta_5 g_1^8 
- 11950 g_1^8 
- 2856 \zeta_3 g_1^6 g_2^2 
\right. \nonumber \\
&& \left. ~~~~~~~
- 6930 \zeta_4 g_1^6 g_2^2 
- 26880 \zeta_5 g_1^6 g_2^2 
- 14938 g_1^6 g_2^2 
+ 13818 \zeta_3 g_1^4 g_2^4 
- 2205 \zeta_4 g_1^4 g_2^4 
\right. \nonumber \\
&& \left. ~~~~~~~
- 77280 \zeta_5 g_1^4 g_2^4 
- 41503 g_1^4 g_2^4 
- 16548 \zeta_3 g_1^2 g_2^6 
+ 10332 \zeta_4 g_1^2 g_2^6 
\right. \nonumber \\
&& \left. ~~~~~~~
- 60270 \zeta_3 g_2^8 
+ 18081 \zeta_4 g_2^8 
- 77000 \zeta_5 g_2^8 
- 21609 g_2^8 \right] g_2 
\nonumber \\
&&
+ \frac{1}{324} \left[
2017008 \zeta_3^2 g_1^{10} 
+ 12023784 \zeta_3 g_1^{10} 
- 6277068 \zeta_4 g_1^{10} 
+ 17472048 \zeta_5 g_1^{10} 
\right. \nonumber \\
&& \left. ~~~~~~~~
- 10639800 \zeta_6 g_1^{10} 
+ 20806821 \zeta_7 g_1^{10} 
+ 2440520 g_1^{10} 
- 6471360 \zeta_3^2 g_1^8 g_2^2 
\right. \nonumber \\
&& \left. ~~~~~~~~
+ 8625792 \zeta_3 g_1^8 g_2^2 
- 1047312 \zeta_4 g_1^8 g_2^2 
+ 30972984 \zeta_5 g_1^8 g_2^2 
- 9815400 \zeta_6 g_1^8 g_2^2 
\right. \nonumber \\
&& \left. ~~~~~~~~
+ 9205434 \zeta_7 g_1^8 g_2^2 
+ 7490700 g_1^8 g_2^2 
- 5869248 \zeta_3^2 g_1^6 g_2^4 
- 92316 \zeta_3 g_1^6 g_2^4 
\right. \nonumber \\
&& \left. ~~~~~~~~
+ 2862090 \zeta_4 g_1^6 g_2^4 
+ 60678240 \zeta_5 g_1^6 g_2^4 
- 24460800 \zeta_6 g_1^6 g_2^4 
\right. \nonumber \\
&& \left. ~~~~~~~~
+ 18920223 \zeta_7 g_1^6 g_2^4 
+ 3823176 g_1^6 g_2^4 
- 5540640 \zeta_3^2 g_1^4 g_2^6 
+ 5826828 \zeta_3 g_1^4 g_2^6 
\right. \nonumber \\
&& \left. ~~~~~~~~
- 2235618 \zeta_4 g_1^4 g_2^6 
+ 44301264 \zeta_5 g_1^4 g_2^6 
- 13524000 \zeta_6 g_1^4 g_2^6 
\right. \nonumber \\
&& \left. ~~~~~~~~
+ 28014525 \zeta_7 g_1^4 g_2^6 
+ 13341328 g_1^4 g_2^6 
- 2310000 \zeta_3^2 g_1^2 g_2^8 
+ 7958580 \zeta_3 g_1^2 g_2^8 
\right. \nonumber \\
&& \left. ~~~~~~~~
- 9122526 \zeta_4 g_1^2 g_2^8 
+ 17502576 \zeta_5 g_1^2 g_2^8 
- 12705000 \zeta_6 g_1^2 g_2^8 
- 806736 g_1^2 g_2^8 
\right. \nonumber \\
&& \left. ~~~~~~~~
+ 2651040 \zeta_3^2 g_2^{10} 
+ 24601332 \zeta_3 g_2^{10} 
- 12403566 \zeta_4 g_2^{10} 
+ 48544888 \zeta_5 g_2^{10} 
\right. \nonumber \\
&& \left. ~~~~~~~~
- 24255000 \zeta_6 g_2^{10} 
+ 47944197 \zeta_7 g_2^{10} 
+ 5311012 g_2^{10} 
\right] g_2 ~+~ O \left( g_i^{13} \right)
\label{tenbet2n3}
\end{eqnarray}
for the $\MSbar$ $\beta$-functions. 

One property of the tensor $O(N)$ model that was present in the six dimensional
non-supersymmetric cubic theory \cite{50} and was illuminated in more detail in
\cite{52} was an emergent symmetry. When $N$~$=$~$3$ then $\NA$~$=$~$5$ giving 
a total of $8$ fields. This is the same dimension as the adjoint representation
of $SU(3)$ and it was shown in \cite{52} that there is an emergent $SU(3)$ 
symmetric in the tensor $O(3)$ cubic theory in six dimensions. Given that this 
is an observation at the level of group theory it is no surprise that there is 
a similar emergent $SU(3)$ symmetry in (\ref{lagwzontens}). This occurs when 
the couplings are equal as then the action can be reorganized into one that is 
formally equivalent to (\ref{lagwzt}). In particular the field anomalous 
dimensions become equal since
\begin{eqnarray}
\left. \gamma_\Phi(g_i) \right|_{N=3,g_1=g_2} &=&
\left. \gamma_\sigma(g_i) \right|_{N=3,g_1=g_2} \nonumber \\
&=& \frac{20}{3} g_1^2 - \frac{400}{9} g_1^4
+ \frac{80}{27} [48 \zeta_3 + 125 ] g_1^6  
+ \frac{1600}{81} [ 72 \zeta_4 - 240 \zeta_3 - 530 \zeta_5 - 225 ] g_1^8  
\nonumber \\
&& 
+~ \frac{800}{243} [ 36840 \zeta_3 - 9702 \zeta_3^2 - 17640 \zeta_4
+ 137170 \zeta_5 - 59625 \zeta_6 + 78057 \zeta_7 \nonumber \\
&& ~~~~~~~~~
+ 19750 ] g_1^{10} ~+~ O \left( g_1^{12} \right) 
\end{eqnarray}
as well as the $\beta$-functions which is apparent from (\ref{tenbet1n3}) and
(\ref{tenbet2n3}) since
\begin{eqnarray}
\left. \beta(g_i) \right|_{N=3,g_1=g_2} &=&
\left. \beta(g_i) \right|_{N=3,g_1=g_2} \nonumber \\
&=& 10 g_1^3 - \frac{200}{3} g_1^5 + \frac{40}{9} [48 \zeta_3 + 125 ] g_1^7 
+ \frac{800}{27} [ 72 \zeta_4 - 240 \zeta_3 - 530 \zeta_5 - 225 ] g_1^9 
\nonumber \\
&& +~ \frac{400}{81} [ 36840 \zeta_3 - 9702 \zeta_3^2 - 17640 \zeta_4 
+ 137170 \zeta_5 - 59625 \zeta_6 + 78057 \zeta_7 \nonumber \\
&& ~~~~~~~~~ + 19750 ] g_1^{11} ~+~ O \left( g_1^{13} \right)
\end{eqnarray}
to five loops. These are clearly consistent with the direct evaluation of the
same quantities given in (\ref{gamsu3}) and (\ref{betsu3}) which affirms the
emergent $SU(3)$ symmetry.

While the emergent $SU(3)$ theory from the $O(3)$ theory is not a surprise
given that it runs parallel to the same observation in six dimensional $\phi^3$
theory, the $SU(3)$ Wess-Zumino model itself already had connections to other 
supersymmetric models in three dimensions \cite{41,42,43,44,45,46}. For 
instance in \cite{44} a duality was observed in three dimensions between an 
${\cal N}$~$=$~$2$ supersymmetric $U(1)$ gauge theory or supersymmetric Quantum 
Electrodynamics which had an infrared enhancement of flavour symmetry to 
$SU(3)$ and an ${\cal N}$~$=$~$1$ supersymmetric Wess-Zumino model with an 
adjoint $SU(3)$ symmetry corresponding to the action (\ref{lagwzt}). It was 
proposed that the latter theory has an ${\cal N}$~$=$~$2$ supersymmetry in the 
infrared in three dimensions. This symmetry enhancement had been observed 
earlier in \cite{41,43} and explored further in \cite{44,45,46}. That the 
$O(3)$ tensor model has also this connection with the $SU(3)$ Wess-Zumino model
is perhaps not surprising as \cite{46} studied various breakings and 
enhancement of this group to $SU(2)$~$\times$~$U(1)$.

We close by noting that one can in principle construct a non-supersymmetric
Lagrangian with $O(3)$ symmetry that has both $SU(3)$ and supersymmetry 
emerging simultaneously at the same fixed point. Such a Lagrangian would need 
the field content of both the $\Phi^i$ and $\sigma^a$ superfields and their 
conjugates. Consequently, the interaction Lagrangian would have a large number
of terms. A non-exhaustive representative set of the formal $3$-point vertices 
is, for example,
\begin{equation}
\left\{ \varsigma^a \psi^i \Lambda^a_{ij} \psi^j, ~ 
\pi^a \psi^i \Lambda^a_{ij} \gamma^5 \psi^j, ~ 
\phi^i \psi^j \Lambda^a_{ij} \chi^a, ~ 
\phi^i \psi^j \Lambda^a_{ij} \gamma^5 \xi^a \right\}
\end{equation}
where we have temporarily dropped the Dirac conjugate on the fermions briefly
to avoid confusion with the chiral aspect of the underlying supermultiplets.
Here $\phi^i$ and $\psi^i$ are the fields that would be in the $\Phi^i$
supermultiplet while $\varsigma^a$, $\chi^a$ and $\xi^a$ are the analogous ones
for the $\sigma^a$ multiplet with the latter two being fermions. Similarly
\begin{equation}
\left\{ \left( \phi^i \phi^i \right)^2, ~
d_3^{abc} \varsigma^b \varsigma^c \Lambda^a_{ij} \phi^i \phi^j, ~
\Lambda^a_{ik} \Lambda^b_{jk} \varsigma^a \varsigma^b \phi^i \phi^j, ~
\varsigma^b \varsigma^c \Lambda^a_{ij} \phi^i \phi^j, ~
\left( d_3^{abc} \varsigma^b \varsigma^c \right)^2 \right\}
\end{equation}
are several formal quartic vertex structures. Such a Lagrangian with distinct
couplings would be non-trivial and would therefore require a large computation 
to determine its renormalization group functions even at low loop order in 
order to explore this double emergence conjecture further.

\sect{General action.}

While we considered a generalization of the Wess-Zumino model to include
interactions with group valued tensor couplings which were real in
(\ref{lagwzt}) that was not the most general cubic supersymmetric chiral 
theory. Instead the most general action involves tensors that themselves
undergo renormalization which we will determine to five loops in this section 
extending thereby the four loop work of \cite{91}. In other words the bare 
action has the form 
\begin{equation}
S ~=~ \int d^4 x \! \left[ \int d^2 \theta d^2 \bar{\theta} \,
\bar{\Phi}_0^i (x,\bar{\theta}) e^{-2 \theta {\partialline} \bar{\theta}} 
\Phi_0^i (x,\theta) \,+\,
\frac{d_0^{ijk}}{3!} \int d^2 \theta \, \Phi_0^i \Phi_0^j \Phi_0^k \,+\,
\frac{\bar{d}_0^{ijk}}{3!} \int d^2 \bar{\theta} \, 
\bar{\Phi}_0^i \bar{\Phi}_0^j \bar{\Phi}_0^k
\right] 
\label{lagwzgend}
\end{equation}
where the tensor couplings are bare in contrast to (\ref{lagwzt}). The 
corresponding renormalized quantities are defined by
\begin{equation}
\Phi_0^i ~=~ Z^{ij} \Phi^j ~~~,~~~
\bar{\Phi}_0^i ~=~ Z_\Phi^{ij} \bar{\Phi}^j
\end{equation}
for the superfields and 
\begin{equation}
d_0^{ijk} ~=~ Z_d^{ijk|pqr} d^{pqr} ~~~,~~~
\bar{d}_0^{ijk} ~=~ Z_{\bar d}^{ijk|pqr} \bar{d}^{pqr}
\end{equation}
for the tensor couplings. However, the tensor renormalization constants are not
independent due to the supersymmetry Ward identity which implies that
$Z_d^{ijk|pqr}$ and its conjugate are constrained to satisfy 
\begin{equation}
Z_\Phi^{il} Z_\Phi^{jm} Z_\Phi^{kn} Z_d^{lmn|pqr} d^{pqr} ~=~ d^{ijk} ~.
\label{susywigen}
\end{equation}
We have determined the conditions these place on the vertex counterterms to
five loops and implemented them within our automatic {\sc Form} programme to
renormalize (\ref{lagwzgend}). Once $Z_\Phi^{ij}$ has been calculated to this
order in either the $\MSbar$ or MOM schemes then the renormalization group
functions are deduced from 
\begin{equation}
\gamma_\Phi^{ik} Z_\Phi^{kj} ~=~
\beta^{pqr} \frac{\partial ~}{\partial d^{pqr}} Z_\Phi^{ij} ~+~
\bar{\beta}^{pqr} \frac{\partial ~}{\partial \bar{d}^{pqr}} Z_\Phi^{ij} 
\end{equation}
where the $\beta$-functions are defined by
\begin{equation}
\beta^{ijk} ~=~ \mu \frac{d~}{d\mu} d^{pqr} ~~~,~~~
\bar{\beta}^{ijk} ~=~ \mu \frac{d~}{d\mu} \bar{d}^{pqr} ~.
\end{equation}
The explicit form of the tensor $\beta$-function is found via the supersymmetry
Ward identity (\ref{susywigen}) which implies, \cite{91},
\begin{equation}
\beta^{ijk} ~=~ -~ \epsilon d^{ijk} ~+~ d^{ijp} \gamma_\Phi^{kp} ~+~
d^{ipk} \gamma_\Phi^{jp} ~+~ d^{pjk} \gamma_\Phi^{ip} ~.
\end{equation}
We have followed this prescription and as a check have reproduced the four loop
$\MSbar$ result of \cite{91} for $\gamma_\Phi^{ij}$. That result was expressed 
as a sum of tensors which have a close correspondence with the individual four 
loop graphs of the superfield $2$-point function. In other words it contained 
$19$ tensors which were presented in a relatively compact way. At five loops 
there are $63$ five loop graphs as indicated in Table \ref{feynnum} and we take
a similar approach here. First if we formally define the field anomalous 
dimension tensor by 
\begin{equation}
\gamma^{ij}_\Phi ~=~ \sum_{L=1}^5 \sum_{r=1}^{k_L} c^{\cal S}_{Lr} T^{ij}_{Lr} 
\end{equation}
where ${\cal S}$ denotes the renormalization scheme, $c^{\cal S}_{Lr}$ are the 
numerical coefficients of the tensors $T^{ij}_{Lr}$, $L$ labels the loop order 
and $r$ identifies the specific tensor. The explicit expression for each tensor
is provided in Appendix A which also records the connection to the underlying 
five loop graphs of the $2$-point function. 

Having set this notation we have determined the values for each of the 
coefficients. For the $\MSbar$ scheme to four loops we have
\begin{eqnarray}
c^{\MSbars}_{11} &=& \frac{1}{2} ~~,~~
c^{\MSbars}_{21} ~=~ - \frac{1}{2} ~~,~~
c^{\MSbars}_{31} ~=~ \frac{3}{2} \zeta_3 ~~,~~
c^{\MSbars}_{32} ~=~ - \frac{1}{8} ~~,~~
c^{\MSbars}_{33} ~=~ - \frac{1}{4} ~~,~~
c^{\MSbars}_{34} ~=~ 1 \nonumber \\ 
c^{\MSbars}_{41} &=& - 10 \zeta_5 ~~,~~
c^{\MSbars}_{42} ~=~ \frac{3}{4} \zeta_4 - \frac{3}{2} \zeta_3 ~~,~~
c^{\MSbars}_{43} ~=~ \frac{3}{4} \zeta_4 - \frac{3}{2} \zeta_3 ~~,~~
c^{\MSbars}_{44} ~=~ \frac{3}{2} \zeta_4 - 3 \zeta_3 \nonumber \\
c^{\MSbars}_{45} &=& - \frac{1}{8} + \frac{1}{4} \zeta_3 ~~,~~
c^{\MSbars}_{46} ~=~ \frac{1}{3} ~~,~~
c^{\MSbars}_{47} ~=~ - \frac{3}{4} \zeta_4 - \frac{3}{2} \zeta_3 ~~,~~
c^{\MSbars}_{48} ~=~ \frac{5}{24} ~~,~~
c^{\MSbars}_{49} ~=~ \frac{1}{3} \nonumber \\
c^{\MSbars}_{410} &=& \frac{1}{3} ~~,~~
c^{\MSbars}_{411} ~=~ - \frac{1}{8} + \frac{1}{4} \zeta_3 ~~,~~
c^{\MSbars}_{412} ~=~ \frac{5}{12} - \frac{1}{2} \zeta_3 ~~,~~
c^{\MSbars}_{413} ~=~ - \frac{5}{2}
\end{eqnarray}
which are in agreement with \cite{47,91}. At five loops we find
\begin{eqnarray}
c^{\MSbars}_{51} &=& \frac{9}{2} \zeta_3^2 ~~,~~
c^{\MSbars}_{52} ~=~ - \frac{143}{16} \zeta_5 - \frac{9}{32} \zeta_4 + \frac{1}{16} \zeta_3 ~~,~~
c^{\MSbars}_{53} ~=~ - \frac{143}{16} \zeta_5 - \frac{9}{32} \zeta_4 + \frac{1}{16} \zeta_3 \nonumber \\
c^{\MSbars}_{54} &=& - \frac{1}{32} + \frac{3}{32} \zeta_4 - \frac{1}{16} \zeta_3 ~~,~~
c^{\MSbars}_{55} ~=~ \frac{67}{4} \zeta_5 - \frac{9}{8} \zeta_4 + \frac{1}{4} \zeta_3 ~~,~~
c^{\MSbars}_{56} ~=~ \frac{3}{16} + \frac{3}{32} \zeta_4 - \frac{5}{16} \zeta_3 
\nonumber \\
c^{\MSbars}_{57} &=& 18 \zeta_5 - \frac{9}{8} \zeta_4 + \frac{1}{4} \zeta_3 ~~,~~
c^{\MSbars}_{58} ~=~ 18 \zeta_5 - \frac{9}{8} \zeta_4 + \frac{1}{4} \zeta_3 ~~,~~
c^{\MSbars}_{59} ~=~ - \frac{25}{4} \zeta_6 + \frac{25}{2} \zeta_5 + \frac{1}{2} \zeta_3^2 \nonumber \\
c^{\MSbars}_{510} &=& - \frac{25}{4} \zeta_6 + \frac{25}{2} \zeta_5 + \frac{1}{2} \zeta_3^2 ~~,~~
c^{\MSbars}_{511} ~=~ \frac{441}{8} \zeta_7 ~~,~~
c^{\MSbars}_{512} ~=~ - \frac{25}{4} \zeta_6 + \frac{25}{2} \zeta_5 + \frac{1}{2} \zeta_3^2 \nonumber \\
c^{\MSbars}_{513} &=& - \frac{79}{4} \zeta_5 - \frac{9}{16} \zeta_4 + \frac{1}{8} \zeta_3 ~~,~~
c^{\MSbars}_{514} ~=~ \frac{441}{16} \zeta_7 ~~,~~
c^{\MSbars}_{515} ~=~ \frac{1}{6} \nonumber \\
c^{\MSbars}_{516} &=& - \frac{25}{4} \zeta_6 + \frac{25}{2} \zeta_5 - \frac{5}{2} \zeta_3^2 ~~,~~
c^{\MSbars}_{517} ~=~ - \frac{25}{4} \zeta_6 + \frac{25}{2} \zeta_5 - \frac{5}{2} \zeta_3^2 \nonumber \\
c^{\MSbars}_{518} &=& - \frac{25}{8} \zeta_6 + \frac{25}{4} \zeta_5 - \frac{11}{4} \zeta_3^2 ~~,~~
c^{\MSbars}_{519} ~=~ 9 \zeta_3^2 ~~,~~
c^{\MSbars}_{520} ~=~ 9 \zeta_3^2 \nonumber \\
c^{\MSbars}_{521} &=& - \frac{153}{8} \zeta_5 - \frac{9}{16} \zeta_4 + \frac{1}{8} \zeta_3 ~~,~~
c^{\MSbars}_{522} ~=~ - \frac{153}{8} \zeta_5 - \frac{9}{16} \zeta_4 + \frac{1}{8} \zeta_3 \nonumber \\
c^{\MSbars}_{523} &=& - \frac{143}{8} \zeta_5 - \frac{21}{16} \zeta_4 + \frac{31}{8} \zeta_3 ~~,~~
c^{\MSbars}_{524} ~=~ - \frac{143}{8} \zeta_5 - \frac{21}{16} \zeta_4 + \frac{31}{8} \zeta_3 \nonumber \\
c^{\MSbars}_{525} &=& 18 \zeta_5 - \frac{9}{8} \zeta_4 + \frac{1}{4} \zeta_3 ~~,~~
c^{\MSbars}_{526} ~=~ 18 \zeta_5 - \frac{9}{8} \zeta_4 + \frac{1}{4} \zeta_3 \nonumber \\
c^{\MSbars}_{527} &=& - \frac{143}{8} \zeta_5 - \frac{9}{16} \zeta_4 + \frac{1}{8} \zeta_3 ~~,~~
c^{\MSbars}_{528} ~=~ - \frac{143}{8} \zeta_5 - \frac{9}{16} \zeta_4 + \frac{1}{8} \zeta_3 \nonumber \\
c^{\MSbars}_{529} &=& \frac{41}{2} \zeta_5 - \frac{9}{8} \zeta_4 + \frac{1}{4} \zeta_3 ~~,~~
c^{\MSbars}_{530} ~=~ \frac{41}{2} \zeta_5 - \frac{21}{8} \zeta_4 + \frac{31}{4} \zeta_3 \nonumber \\
c^{\MSbars}_{531} &=& \frac{1}{2} \zeta_5 + \frac{3}{32} \zeta_4 - \frac{13}{16} \zeta_3 ~~,~~
c^{\MSbars}_{532} ~=~ 0 ~~,~~
c^{\MSbars}_{533} ~=~ \frac{3}{16} + \frac{3}{32} \zeta_4 - \frac{5}{16} \zeta_3 \nonumber \\
c^{\MSbars}_{534} &=& \frac{3}{16} + \frac{3}{32} \zeta_4 - \frac{5}{16} \zeta_3 ~~,~~
c^{\MSbars}_{535} ~=~ - \frac{1}{16} + \frac{3}{16} \zeta_4 - \frac{1}{8} \zeta_3 ~~,~~ 
c^{\MSbars}_{536} ~=~ - \frac{3}{32} \zeta_4 - \frac{1}{16} \zeta_3 \nonumber \\
c^{\MSbars}_{537} &=& - \frac{7}{6} ~~,~~
c^{\MSbars}_{538} ~=~ \frac{1}{3} ~~,~~
c^{\MSbars}_{539} ~=~ \frac{25}{4} \zeta_6 + \frac{15}{2} \zeta_5 - \frac{1}{2} \zeta_3^2 ~~,~~
c^{\MSbars}_{540} ~=~ \frac{1}{8} \zeta_5 + 2 \zeta_3 \nonumber \\
c^{\MSbars}_{541} &=& \frac{1}{8} \zeta_5 + 2 \zeta_3 ~~,~~
c^{\MSbars}_{542} ~=~ - \frac{9}{4} \zeta_5 + 4 \zeta_3 
c^{\MSbars}_{543} ~=~ \frac{3}{16} - \frac{3}{16} \zeta_4 - \frac{1}{4} \zeta_3 ~~,~~
c^{\MSbars}_{544} ~=~ - \frac{5}{6} \nonumber \\
c^{\MSbars}_{545} &=& \frac{1}{2} \zeta_5 + \frac{3}{32} \zeta_4 - \frac{13}{16} \zeta_3 ~~,~~
c^{\MSbars}_{546} ~=~ \frac{1}{2} \zeta_5 + \frac{3}{32} \zeta_4 - \frac{13}{16} \zeta_3 ~~,~~
c^{\MSbars}_{547} ~=~ 0 ~~,~~
c^{\MSbars}_{548} ~=~ 0 \nonumber \\
c^{\MSbars}_{549} &=& \frac{3}{16} + \frac{3}{32} \zeta_4 - \frac{5}{16} \zeta_3 ~~,~~
c^{\MSbars}_{550} ~=~ \frac{3}{16} + \frac{3}{32} \zeta_4 - \frac{5}{16} \zeta_3 ~~,~~
c^{\MSbars}_{551} ~=~ - \frac{3}{32} \zeta_4 - \frac{1}{16} \zeta_3 \nonumber \\
c^{\MSbars}_{552} &=& - \frac{3}{32} \zeta_4 - \frac{1}{16} \zeta_3 ~~,~~
c^{\MSbars}_{553} ~=~ - \frac{7}{6} ~~,~~
c^{\MSbars}_{554} ~=~ - \frac{7}{6} ~~,~~ 
c^{\MSbars}_{555} ~=~ - \frac{1}{16} + \frac{3}{16} \zeta_4 - \frac{1}{8} \zeta_3 \nonumber \\
c^{\MSbars}_{556} &=& \frac{3}{16} + \frac{3}{32} \zeta_4 - \frac{5}{16} \zeta_3 ~~,~~
c^{\MSbars}_{557} ~=~ - \zeta_5 + \frac{21}{16} \zeta_4 + \frac{25}{8} \zeta_3 ~~,~~
c^{\MSbars}_{558} ~=~ - \frac{1}{8} \nonumber \\
c^{\MSbars}_{559} &=& - \frac{5}{6} - \frac{3}{16} \zeta_4 + \frac{7}{8} \zeta_3 ~~,~~
c^{\MSbars}_{560} ~=~ - \frac{5}{6} - \frac{3}{16} \zeta_4 + \frac{7}{8} \zeta_3 ~~,~~
c^{\MSbars}_{561} ~=~ \frac{3}{16} - \frac{3}{8} \zeta_4 + \frac{1}{8} \zeta_3 \nonumber \\
c^{\MSbars}_{562} &=& - \frac{1}{4} + \frac{3}{16} \zeta_4 + \frac{9}{8} \zeta_3 ~~,~~
c^{\MSbars}_{563} ~=~ 7 ~.
\end{eqnarray}

We have repeated this exercise for the MOM scheme and found to four loops
\begin{eqnarray}
c^{\MOMs}_{11} &=& \frac{1}{2} ~~,~~
c^{\MOMs}_{12} ~=~ - \frac{1}{2} ~~,~~
c^{\MOMs}_{31} ~=~ \frac{3}{2} \zeta_3 ~~,~~
c^{\MOMs}_{32} ~=~ \frac{1}{4} ~~,~~
c^{\MOMs}_{33} ~=~ \frac{1}{2} \nonumber \\
c^{\MOMs}_{34} &=& 1 \nonumber \\
c^{\MOMs}_{41} &=& - 10 \zeta_5 ~~,~~
c^{\MOMs}_{42} ~=~ - \frac{3}{2} \zeta_3 ~~,~~
c^{\MOMs}_{43} ~=~ - \frac{3}{2} \zeta_3 ~~,~~
c^{\MOMs}_{44} ~=~ - 3 \zeta_3 \nonumber \\
c^{\MOMs}_{45} &=& - \frac{3}{4} + \frac{1}{2} \zeta_3 ~~,~~
c^{\MOMs}_{46} ~=~ - \frac{5}{4} + \frac{1}{2} \zeta_3 ~~,~~
c^{\MOMs}_{47} ~=~ - \frac{3}{2} \zeta_3 ~~,~~
c^{\MOMs}_{48} ~=~ - \frac{3}{4} \nonumber \\
c^{\MOMs}_{49} &=& - \frac{5}{4} ~~,~~
c^{\MOMs}_{410} ~=~ - \frac{5}{4} ~~,~~
c^{\MOMs}_{411} ~=~ - \frac{3}{4} ~~,~~
c^{\MOMs}_{412} ~=~ - \frac{3}{2} \nonumber \\
c^{\MOMs}_{413} &=& - \frac{5}{2} ~.
\end{eqnarray}
To two loops the respective coefficients are the same as those of the $\MSbar$ 
scheme consistent with earlier expectations. At three and four loops a few of
the coefficients also match between schemes aside from the primitive graphs. At
next order the coefficients are
\begin{eqnarray}
c^{\MOMs}_{51} &=& \frac{9}{2} \zeta_3^2 ~~,~~;
c^{\MOMs}_{52} ~=~ - \frac{75}{8} \zeta_5 + \frac{3}{4} \zeta_3 ~~,~~
c^{\MOMs}_{53} ~=~ - \frac{75}{8} \zeta_5 + \frac{3}{4} \zeta_3 \nonumber \\
c^{\MOMs}_{54} &=& \frac{3}{4} - \frac{1}{2} \zeta_3 ~~,~~
c^{\MOMs}_{55} ~=~ 15 \zeta_5 + 3 \zeta_3 ~~,~~
c^{\MOMs}_{56} ~=~ \frac{9}{4} - \frac{3}{2} \zeta_3 ~~,~~
c^{\MOMs}_{57} ~=~ 15 \zeta_5 + 3 \zeta_3 \nonumber \\
c^{\MOMs}_{58} &=& 15 \zeta_5 + 3 \zeta_3 ~~,~~
c^{\MOMs}_{59} ~=~ 10 \zeta_5 ~~,~~
c^{\MOMs}_{510} ~=~ 10 \zeta_5 ~~,~~
c^{\MOMs}_{511} ~=~ \frac{441}{8} \zeta_7 \nonumber \\
c^{\MOMs}_{512} &=& 10 \zeta_5 ~~,~~
c^{\MOMs}_{513} ~=~ - \frac{175}{8} \zeta_5 + \frac{3}{2} \zeta_3 ~~,~~
c^{\MOMs}_{514} ~=~ \frac{441}{16} \zeta_7 ~~,~~
c^{\MOMs}_{515} ~=~ \frac{7}{4} - \zeta_3 \nonumber \\
c^{\MOMs}_{516} &=& 10 \zeta_5 - 3 \zeta_3^2 ~~,~~
c^{\MOMs}_{517} ~=~ 10 \zeta_5 - 3 \zeta_3^2 ~~,~~
c^{\MOMs}_{518} ~=~ 5 \zeta_5 - 3 \zeta_3^2 ~~,~~
c^{\MOMs}_{519} ~=~ 9 \zeta_3^2 \nonumber \\
c^{\MOMs}_{520} &=& 9 \zeta_3^2 ~~,~~
c^{\MOMs}_{521} ~=~ - \frac{85}{4} \zeta_5 + \frac{3}{2} \zeta_3 ~~,~~
c^{\MOMs}_{522} ~=~ - \frac{85}{4} \zeta_5 + \frac{3}{2} \zeta_3 \nonumber \\
c^{\MOMs}_{523} &=& - \frac{75}{4} \zeta_5 + 3 \zeta_3 ~~,~~
c^{\MOMs}_{524} ~=~ - \frac{75}{4} \zeta_5 + 3 \zeta_3 ~~,~~
c^{\MOMs}_{525} ~=~ 15 \zeta_5 + 3 \zeta_3 \nonumber \\
c^{\MOMs}_{527} &=& - \frac{75}{4} \zeta_5 + \frac{3}{2} \zeta_3 ~~,~~
c^{\MOMs}_{528} ~=~ - \frac{75}{4} \zeta_5 + \frac{3}{2} \zeta_3 ~~,~~
c^{\MOMs}_{529} ~=~ \frac{65}{4} \zeta_5 + 3 \zeta_3 \nonumber \\
c^{\MOMs}_{530} &=& \frac{65}{4} \zeta_5 + 6 \zeta_3 ~~,~~
c^{\MOMs}_{531} ~=~ \frac{3}{2} \zeta_3 ~~,~~
c^{\MOMs}_{532} ~=~ \frac{9}{8} - \frac{1}{2} \zeta_3 ~~,~~
c^{\MOMs}_{533} ~=~ \frac{9}{4} - \frac{5}{4} \zeta_3 \nonumber \\
c^{\MOMs}_{534} &=& \frac{9}{4} - \frac{5}{4} \zeta_3 ~~,~~
c^{\MOMs}_{535} ~=~ \frac{3}{2} - \frac{3}{4} \zeta_3 ~~,~~
c^{\MOMs}_{536} ~=~ \frac{9}{4} - \zeta_3 ~~,~~
c^{\MOMs}_{537} ~=~ \frac{7}{2} - \frac{3}{2} \zeta_3 \nonumber \\
c^{\MOMs}_{538} &=& \frac{7}{2} ~~,~~
c^{\MOMs}_{539} ~=~ 10 \zeta_5 ~~,~~
c^{\MOMs}_{540} ~=~ 3 \zeta_3 ~~,~~
c^{\MOMs}_{541} ~=~ 3 \zeta_3 ~~,~~
c^{\MOMs}_{542} ~=~ 6 \zeta_3 \nonumber \\
c^{\MOMs}_{543} &=& 3 - \frac{1}{2} \zeta_3 ~~,~~
c^{\MOMs}_{544} ~=~ \frac{9}{2} - \frac{1}{2} \zeta_3 ~~,~~
c^{\MOMs}_{545} ~=~ \frac{3}{2} \zeta_3 ~~,~~
c^{\MOMs}_{546} ~=~ \frac{3}{2} \zeta_3 \nonumber \\
c^{\MOMs}_{547} &=& \frac{9}{8} ~~,~~
c^{\MOMs}_{548} ~=~ \frac{9}{8} ~~,~~
c^{\MOMs}_{549} ~=~ \frac{9}{4} ~~,~~
c^{\MOMs}_{550} ~=~ \frac{9}{4} ~~,~~
c^{\MOMs}_{551} ~=~ \frac{9}{4} \nonumber \\
c^{\MOMs}_{552} &=& \frac{9}{4} ~~,~~
c^{\MOMs}_{553} ~=~ \frac{7}{2} ~~,~~
c^{\MOMs}_{554} ~=~ \frac{7}{2} ~~,~~
c^{\MOMs}_{555} ~=~ \frac{3}{2} ~~,~~
c^{\MOMs}_{556} ~=~ \frac{9}{4} \nonumber \\
c^{\MOMs}_{557} &=& 3 \zeta_3 ~~,~~
c^{\MOMs}_{558} ~=~ \frac{9}{4} ~~,~~
c^{\MOMs}_{559} ~=~ \frac{9}{2} ~~,~~
c^{\MOMs}_{560} ~=~ \frac{9}{2} ~~,~~
c^{\MOMs}_{561} ~=~ 3 \nonumber \\
c^{\MOMs}_{562} &=& \frac{9}{2} ~~,~~
c^{\MOMs}_{563} ~=~ 7 ~.
\end{eqnarray}
To assist with the derivation of both sets of coefficients from the value of
$Z_\Phi^{ij}$ in each scheme we have recorded the explicit expression in
Appendix B. Indeed by providing them for each specific tensor means the 
divergence structure of all the individual diagrams are provided to five loops.
More tensors appear in $Z_\Phi^{ij}$ than $\gamma_\Phi^{ij}$. The extra ones
arise in terms with poles in $\epsilon$ higher than the simple one. They 
correspond to connected one-particle reducible Feynman graphs of the $\Phi$ 
$2$-point function. Such topologies and hence tensors clearly cannot appear in 
the final expression for $\gamma_\Phi^{ij}$ in either scheme which is a 
non-trivial check on the overall expression. This is because it is the 
generalization of the observation that in a conventional coupling constant 
renormalization the coefficients of the non-simple poles in $\epsilon$ are 
determined by the lower order renormalization constants.

\sect{XYZ model.}

As an application of the general tensor renormalization we consider a 
particular theory that is connected to the Wess-Zumino model which was examined
in \cite{40,103}. It was investigated in \cite{40} due to its connection with a
one dimensional conformal manifold. In particular several theories are of 
interest for the case when the Wess-Zumino model has three chiral superfields 
as they lie on the manifold. These are the XYZ model and a version of the model
itself with three copies. First we recall the relevant properties of the more
general model in order to extend the four loop analysis of \cite{91} to five 
loops here. As indicated in \cite{40} the model involves three chiral 
superfields and their anti-chiral counterparts with superpotential
\begin{equation}
W(\Phi_i) ~=~ g_1 \Phi_1 \Phi_2 \Phi_3 ~+~ 
\frac{g_2}{6} \left( \Phi_1^3 + \Phi_2^3 + \Phi_3^3 \right)
\label{pot3d}
\end{equation}
and its conjugate where $g_1$ and $g_2$ are complex coupling constants. 
Therefore the non-zero tensor coupling entries are 
\begin{equation}
d^{123} ~=~ g_1 ~~,~~
d^{111} ~=~ d^{222} ~=~ d^{333} ~=~ g_2 ~~,~~
\bar{d}^{123} ~=~ \bar{g}_1 ~~,~~
\bar{d}^{111} ~=~ \bar{d}^{222} ~=~ \bar{d}^{333} ~=~ \bar{g}_2 ~.
\label{cc3d}
\end{equation}
These variables were mapped to others which are similar to polar coordinates in
geometry through, \cite{40,104}, 
\begin{equation}
r^2 ~=~ 2 g_1 \bar{g}_1 ~+~ g_2 \bar{g}_2 ~~~,~~~
\tau ~=~ \frac{g_2}{g_1} ~~~,~~~
\bar{\tau} ~=~ \frac{\bar{g}_2}{\bar{g}_1} 
\label{parm3d}
\end{equation}
where the parameter $\tau$ takes values in $\Cc \Pp(1)$, \cite{104}. Using 
these combinations certain values of $\tau$ and $\bar{\tau}$ allow one to 
define various different theories with the justification recorded in \cite{40}.
We have provided these in Table \ref{modeldef} where the first three were given
in \cite{40} and cWZ${}^3$ is used as shorthand to denote the three copy 
Wess-Zumino model. This is also equivalent to the parameter choice of the final
row of Table \ref{modeldef} which was not noted in \cite{40} and will be 
another useful limit for checking results. For the $\Zz_2$~$\times$~$\Zz_2$ 
symmetric model the complex number $\omega$ and its conjugate appear are
\begin{equation}
\omega ~=~ -~ \frac{1}{2} ~+~ \frac{\sqrt{3}}{2} i ~~~,~~~
\bar{\omega} ~=~ -~ \frac{1}{2} ~-~ \frac{\sqrt{3}}{2} i ~.
\end{equation}

{\begin{table}[ht]
\begin{center}
\begin{tabular}{|c|c||c|}
\hline
$\tau$ & $\bar{\tau}$ & Theory \\
\hline
$0$ & $0$ & XYZ model \\
$1$ & $1$ & cWZ${}^3$ \\
$(1-\sqrt{3})\omega^2$ & $(1-\sqrt{3})\bar{\omega}^2$ & 
$\Zz_2$~$\times$~$\Zz_2$ symmetric \\
$\infty$ & $\infty$ & Wess-Zumino model (\ref{lagwz}) \\
\hline
\end{tabular}
\end{center}
\begin{center}
\caption{Definition of various models from the values of $\tau$ and
$\bar{\tau}$.}
\label{modeldef}
\end{center}
\end{table}}

With (\ref{parm3d}) the anomalous dimension is formally written as
\begin{equation}
\gamma_\Phi(r,\tau,\bar{\tau}) ~=~ \sum_{i=1}^\infty f_i(\tau,\bar{\tau}) r^{2i}
\label{gam3d}
\end{equation}
where the coefficients are given by
\begin{eqnarray}
f_1(\tau,\bar{\tau}) &=& \frac{1}{2} ~~,~~
f_2(\tau,\bar{\tau}) ~=~ -~ \frac{1}{2} ~~,~~ 
f_3(\tau,\bar{\tau}) ~=~ \frac{5}{8} ~+~ 
\frac{3}{2} \frac{[(\tau^3+2) (\bar{\tau}^3+2) + 18 \tau \bar{\tau} ]}
{[2+\tau\bar{\tau}]^3} \zeta_3 \nonumber \\
f_4(\tau,\bar{\tau}) &=& -~ \frac{9}{8} ~+~ 
\left[ \frac{9}{4} \zeta_4 - \frac{15}{2} \zeta_3 \right]
\frac{[(\tau^3+2) (\bar{\tau}^3+2) + 18 \tau \bar{\tau} ]}
{[2+\tau\bar{\tau}]^3} \nonumber \\
&& -~ 10 \frac{[ (2+\tau\bar{\tau})^4 - 8 (1-\tau^3)(1-\bar{\tau}^3)]}
{[2+\tau\bar{\tau}]^4} \zeta_5 \nonumber \\
f_5(\tau,\bar{\tau}) &=& \frac{79}{32} ~+~ 
\left[ \frac{3}{8}
+ \frac{423}{16} \frac{[(\tau^3+2) (\bar{\tau}^3+2) + 18 \tau \bar{\tau} ]}
{[2+\tau\bar{\tau}]^3} \right] \zeta_3 
-~ \frac{441}{32} 
\frac{[(\tau^3+2) (\bar{\tau}^3+2) + 18 \tau \bar{\tau} ]}
{[2+\tau\bar{\tau}]^3} \zeta_4 \nonumber \\
&& +~ \left[ 
\frac{305}{4} \frac{[ (2+\tau\bar{\tau})^4 - 8 (1-\tau^3)(1-\bar{\tau}^3)]}
{[2+\tau\bar{\tau}]^4} 
- \frac{153}{8} \frac{[(\tau^3+2) (\bar{\tau}^3+2) + 18 \tau \bar{\tau} ]}
{[2+\tau\bar{\tau}]^3} \right] \zeta_5 \nonumber \\
&& -~ \frac{225}{8} 
\frac{[ (2+\tau\bar{\tau})^4 - 8 (1-\tau^3)(1-\bar{\tau}^3)]}
{[2+\tau\bar{\tau}]^4} \zeta_6 \nonumber \\
&& -~ \left[ \frac{45}{4} 
- \frac{9}{2} \frac{[ (2+\tau\bar{\tau})^4 - 8 (1-\tau^3)(1-\bar{\tau}^3)]}
{[2+\tau\bar{\tau}]^4} 
- \frac{45}{2} \frac{[(\tau^3+2) (\bar{\tau}^3+2) + 18 \tau \bar{\tau} ]}
{[2+\tau\bar{\tau}]^3} \right] \zeta_3^2 \nonumber \\
&& +~ \frac{1323}{16} 
\frac{[ (2+\tau\bar{\tau})^4 - 10 (1-\tau^3)(1-\bar{\tau}^3)]}
{[2+\tau\bar{\tau}]^4} \zeta_7
\label{gam3dc}
\end{eqnarray}
with $f_1$ to $f_4$ in accord with \cite{40}. It is straightforward to check 
that $f_i(1,1)$~$=$~$f_i(\infty,\infty)$ for $i$~$=$~$1$ to $5$. Moreover  the
$f_i(1,1)$ correspond to the respective coefficients of (\ref{gam5}). While we 
have checked the values $f_i(\tau,\bar{\tau})$ to four loops and found
$f_5(\tau,\bar{\tau})$ using (\ref{cc3d}) and (\ref{parm3d}) they could also
have been derived from (\ref{gam5T}) from the simple identifications
\begin{eqnarray} 
T_2 &=& 1 ~~,~~
T_5 ~=~ \frac{[(\tau^3+2) (\bar{\tau}^3+2) + 18 \tau \bar{\tau} ]}
{[2+\tau\bar{\tau}]^3} \nonumber \\
T_{71} &=& \frac{[ (2+\tau\bar{\tau})^4 - 8 (1-\tau^3)(1-\bar{\tau}^3)]}
{[2+\tau\bar{\tau}]^4} ~~,~~
T_{94} ~=~ \frac{[ (2+\tau\bar{\tau})^4 - 10 (1-\tau^3)(1-\bar{\tau}^3)]}
{[2+\tau\bar{\tau}]^4} ~~~~
\end{eqnarray} 
thereby making the connection with the primitive graphs for the conformal
manifold case. It is worth remarking that given this relation between the $T_i$
invariants one could in principle repeat the analysis of \cite{40} and that 
which follows here for non-supersymmetric scalar $\phi^3$ theory. While that 
theory is renormalizable in six dimensions the four loop renormalization group 
functions have been expressed in terms of the four $T_i$ that appear here for 
chiral $\phi^3$ theory. 

The main topic of study in \cite{40} was the evaluation of the critical
exponents of the dimension two bilinear operators denoted by $\Delta_i$ where 
$i$~$\in$~$\{ \mathbf{1}, \mathbf{2}, \mathbf{2^\prime},
\mathbf{2^{\prime\prime}}, \mathbf{2^{\prime\prime\prime}} \}$ correspond to
the different representations of the
$\mathbf{3}$~$\otimes$~$\bar{\mathbf{3}}$ decomposition of the $9$ operators.
These operator dimensions were determined in three dimensions using conformal 
bootstrap methods as well as resumming four dimensional perturbation theory. 
For the latter the matrix of operator anomalous dimensions was computed to four
loops prior to being evaluated at the Wilson-Fisher fixed point. The critical 
point eigenvalues of this matrix then corresponded to the critical exponents 
$\Delta_i$, \cite{40}. We are now in a position to extend the four loop 
analysis of \cite{40} to five loops in order to compare with the bootstrap 
exponent estimates. First, the location of the Wilson-Fisher fixed point has to
be found. Since the $\beta$-function is synonymous with 
$\gamma_\Phi(r,\tau,\bar{\tau})$ in this model then the $\epsilon$ 
expansion of the critical value of $r$, denoted by $r_\ast$, is given by
solving $\gamma_\Phi(r_\ast,\tau,\bar{\tau})$~$=$~$\third \epsilon$. From 
(\ref{gam3d}) and defining 
\begin{equation}
r_\ast^2 ~=~ \sum_{i=1}^\infty h_i \epsilon^i
\end{equation}
the various coefficients of the critical coupling are
\begin{eqnarray}
h_1 &=& \frac{1}{3f_1} ~~,~~
h_2 ~=~ -~ \frac{f_2}{9f_1^3} ~~,~~
h_3 ~=~ \frac{2f_2^2}{27f_1^5} ~-~ \frac{f_3}{27f_1^4} ~~,~~
h_4 ~=~ \frac{5f_2f_3}{81f_1^6} ~-~ \frac{5f_2^3}{81f_1^7} ~-~ 
\frac{f_4}{81f_1^5} \nonumber \\ 
h_5 &=& \frac{14f_2^4}{243f_1^9} ~-~ \frac{7f_2^2f_3}{81f_1^8} ~+~ 
\frac{f_3^2}{81f_1^7} ~+~ \frac{2f_2f_4}{81f_1^7} ~-~ \frac{f_5}{243f_1^6} ~.
\end{eqnarray}

The $3$~$\times$~$3$ matrix of mass anomalous dimensions that was constructed
in \cite{40} is defined by 
\begin{equation}
\gamma_M^{ij} ~=~ \mu \frac{d M^{ij}}{d \mu}
\end{equation}
where the matrix $M^{ij}$ corresponds to the mass dimension $2$ matrix 
$(m^2)^{ij}$ of \cite{40} which is computed from $\gamma_\phi^{ij}$ using 
\begin{eqnarray}
\gamma_M^{ij} &=& -~ 2 M^{ij} ~+~
\left[ M^{ps} d^{sqr} + M^{qs} d^{psr} + M^{rs} d^{pqs} \right]
\frac{\partial \gamma_\Phi^{ij}}{d^{pqr}} \nonumber \\
&& +~ \left[ M^{ps} \bar{d}^{sqr} + M^{qs} \bar{d}^{psr} +
M^{rs} \bar{d}^{pqs} \right] \frac{\partial \gamma_\Phi^{ij}}{\bar{d}^{pqr}} ~.
\end{eqnarray}
The next stage is to construct the $9$~$\times$~$9$ matrix, $\Delta^{ijkl}$,
the eigenvalues of which produce the scaling dimensions of the bilinear
operators. It has $81$ elements since the matrix is labelled by the pairs of
indices $(ij)$ and $(kl)$ and defined by
\begin{equation}
\Delta^{ijkl} ~=~ d \delta^{ik} \delta^{jl} ~+~ 
\frac{\partial \gamma_M^{ij}}{\partial M^{kl}} ~.
\end{equation}
Following the prescription given in \cite{40} we have extended the four loop
expressions for the five critical exponents $\Delta_i$ to the next order. In
particular we found 
\begin{eqnarray}
\Delta_{\mathbf 1} &=& 2 ~-~ \frac{4}{3} \epsilon^2 ~+~ 
\left[ \frac{4}{9}
+ \frac{16}{3} \frac{[(\tau^3+2) (\bar{\tau}^3+2) + 18 \tau \bar{\tau}]}
{[2+\tau \bar{\tau}]^3} \right] \zeta_3 \epsilon^3
\nonumber \\
&& 
-~ \left[ \frac{28}{27}
+ \frac{112}{9} \frac{[(\tau^3+2) (\bar{\tau}^3+2) + 18 \tau \bar{\tau}]}
{[2+\tau \bar{\tau}]^3} \zeta_3 
- 8 \frac{[(\tau^3+2) (\bar{\tau}^3+2) + 18 \tau \bar{\tau}]}
{[2+\tau \bar{\tau}]^3} \zeta_4 
\right. \nonumber \\
&& \left. ~~~~
+~ \frac{320}{9} \frac{[(\tau \bar{\tau}+2)^4 - 8 (1-\tau^3) (1-\bar{\tau}^3)]}
{[2+\tau \bar{\tau}]^4} \zeta_5 \right] \epsilon^4
\nonumber \\
&& 
+~ \left[ \frac{76}{81}
+ \left[ \frac{496}{27}
\frac{[(\tau^3+2) (\bar{\tau}^3+2) + 18 \tau \bar{\tau}]}
{[2+\tau \bar{\tau}]^3} 
+ \frac{32}{27} \right] \zeta_3
- \frac{56}{3} \frac{[(\tau^3+2) (\bar{\tau}^3+2) + 18 \tau \bar{\tau}]}
{[2+\tau \bar{\tau}]^3} \zeta_4 
\right. \nonumber \\
&& \left. ~~~~
+~ \left[ \frac{3520}{27} 
\frac{[(\tau \bar{\tau}+2)^4 - 8 (1-\tau^3) (1-\bar{\tau}^3)]}
{[2+\tau \bar{\tau}]^4}
- \frac{544}{9} \frac{[(\tau^3+2) (\bar{\tau}^3+2) + 18 \tau \bar{\tau}]}
{[2+\tau \bar{\tau}]^3} \right] \zeta_5
\right. \nonumber \\
&& \left. ~~~~
-~ \frac{800}{9} 
\frac{[(\tau \bar{\tau}+2)^4 - 8 (1-\tau^3) (1-\bar{\tau}^3)]}
{[2+\tau \bar{\tau}]^4} \zeta_6 
\right. \nonumber \\
&& \left. ~~~~
+ \left[ \frac{256}{81}
\left[ \frac{9}{2} \frac{[(2+\tau \bar{\tau})^4 - 8(1-\tau^3)(1-\bar{\tau}^3)]}
{[2+\tau \bar{\tau}]^4}
- \frac{45}{4}
+ \frac{45}{2} \frac{[(\tau^3+2)(\bar{\tau}^3+2) + 18 \tau \bar{\tau}]}
{[2+\tau \bar{\tau}]^3} \right]
\right. \right. \nonumber \\
&& \left. \left. ~~~~~~~~
- \frac{64}{3} \frac{[(\tau^3+2) (\bar{\tau}^3+2) + 18 \tau \bar{\tau} ]^2}
{[2+\tau \bar{\tau}]^6} \right] \zeta_3^2
\right. \nonumber \\
&& \left. ~~~~
+~ \frac{784}{3} \frac{[(2+\tau \bar{\tau})^4 - 10 (1-\tau^3) (1-\bar{\tau}^3)]}
{[2+\tau \bar{\tau}]^4} \zeta_7 \right] \epsilon^5 ~+~ O(\epsilon^6)
\label{op31}
\end{eqnarray}
for the singlet operator as well as
\begin{eqnarray}
\Delta_{\mathbf 2} &=& 2 ~-~ \frac{4}{[2+\tau \bar{\tau}]} \epsilon
+ \frac{4}{3} \tau \bar{\tau} 
\frac{[1 - \tau \bar{\tau}]}{[2+\tau \bar{\tau}]^2} \epsilon^2
\nonumber \\
&&
+~ \tau \bar{\tau}
\left[ \frac{4}{9} \frac{[1 - \tau \bar{\tau}][10 - \tau \bar{\tau}]}
{[2+\tau \bar{\tau}]^3} 
+ \frac{16}{3} \frac{[3 (1 - \tau \bar{\tau})^2
+ (1 - \tau^3)(1 - \bar{\tau}^3)]}{[2+\tau \bar{\tau}]^4} \zeta_3 \right] 
\epsilon^3
\nonumber \\
&&
+~ \tau \bar{\tau}
\left[ \frac{4}{27} \frac{[ 7 \tau^2 \bar{\tau}^2 - 26 \tau \bar{\tau} + 100 ]
[1 - \tau \bar{\tau}]}{[2+\tau \bar{\tau}]^4} 
\right. \nonumber \\
&& \left. ~~~~~~~~
-~ \frac{16}{9} \frac{[ 2 (1-\tau \bar{\tau}) (2+\tau \bar{\tau})^2
+ [3 (1-\tau \bar{\tau})^2 + (1-\tau^3) (1-\bar{\tau}^3) ]
(2+7 \tau \bar{\tau}) ]}{[2+\tau \bar{\tau}]^5} \zeta_3 
\right. \nonumber \\
&& \left. ~~~~~~~~
+~ 8 \frac{[ 3 (1-\tau \bar{\tau})^2 + (1-\tau^3) (1-\bar{\tau}^3) ]}
{[2+\tau \bar{\tau}]^4} \zeta_4 
\right. \nonumber \\
&& \left. ~~~~~~~~
-~ \frac{320}{27} 
\frac{[3 \tau \bar{\tau} (2+\tau \bar{\tau}) (1-\tau \bar{\tau})^2 
+ 8 (1-\tau^3) (1-\bar{\tau}^3) ]}
{[2+\tau \bar{\tau}]^5} \zeta_5 \right] \epsilon^4
\nonumber \\
&& 
+~ \tau \bar{\tau}
\left[ \frac{4}{81} \frac{[19 \tau^2 \bar{\tau}^2 - 38 \tau \bar{\tau} + 100 ]
[1-\tau \bar{\tau}] [10-\tau \bar{\tau}]}{[2+\tau \bar{\tau}]^5}
\right. \nonumber \\
&& \left. ~~~~~~~~
+~ \frac{16}{27}
\left[ \frac{}{} 31 \tau^2 \bar{\tau}^2 (1-\tau^3) (1-\bar{\tau}^3) 
- 32 \tau \bar{\tau} (1-\tau^3) (1-\bar{\tau}^3) 
+ 28 (1-\tau^3) (1-\bar{\tau}^3) 
\right. \right. \nonumber \\
&& \left. \left. ~~~~~~~~~~~~~~~~~~
-~ (2 \tau^4 \bar{\tau}^4+85 \tau^3 \bar{\tau}^3-237 \tau^2 \bar{\tau}^2
+ 148 \tau \bar{\tau}-52) (1-\tau \bar{\tau}) \right]
\frac{\zeta_3}{[2+\tau \bar{\tau}]^6}
\right. \nonumber \\
&& \left. ~~~~~~~~
-~ \frac{8}{3} \left[ \frac{}{}
7 (1-\tau^3) (1-\bar{\tau}^3) \tau \bar{\tau} 
+ 21 \tau \bar{\tau} (1-\tau \bar{\tau})^2 + 12 (1-\tau \bar{\tau})
\right. \right. \nonumber \\
&& \left. \left. ~~~~~~~~~~~~~~~~
+ 2 (1-\tau^3) + 2 (1-\bar{\tau}^3) \right]
\frac{\zeta_4}{[2+\tau \bar{\tau}]^5}
\right. \nonumber \\
&& \left. ~~~~~~~~
-~ \frac{16}{81} \left[ \frac{}{}
3 (70 \tau^3 \bar{\tau}^3-386 \tau^2 \bar{\tau}^2
- 566 \tau \bar{\tau}+207) (2+\tau \bar{\tau}) (1-\tau \bar{\tau})
\right. \right. \nonumber \\
&& \left. \left. ~~~~~~~~~~~~~~~~~~
- 144 \tau^2 \bar{\tau}^2 (1-\tau^3) (1-\bar{\tau}^3)
+ 1916 \tau \bar{\tau} (1-\tau^3) (1-\bar{\tau}^3)
\right. \right. \nonumber \\
&& \left. \left. ~~~~~~~~~~~~~~~~~~
+ 1336 (1-\tau^3) (1-\bar{\tau}^3) \right] 
\frac{\zeta_5}{[2+\tau \bar{\tau}]^6}
\right. \nonumber \\
&& \left. ~~~~~~~~
-~ \frac{800}{27}
\frac{[ 8 (1-\tau) (1-\bar{\tau}) (1+\tau+\tau^2) (1+\bar{\tau}+\bar{\tau}^2)
+ 3 \tau \bar{\tau} (2+\tau \bar{\tau}) (1-\tau \bar{\tau})^2 ]}
{[2+\tau \bar{\tau}]^5} \zeta_6
\right. \nonumber \\
&& \left. ~~~~~~~~
+ \frac{128}{27} \left[ \frac{}{}
(18 \tau^4 \bar{\tau}^4-72 \tau^3 \bar{\tau}^3+137 \tau^2 \bar{\tau}^2
- 205 \tau \bar{\tau}-34) (1-\tau \bar{\tau}) 
\right. \right. \nonumber \\
&& \left. \left. ~~~~~~~~~~~~~~~~~~
+ 6 \tau^3 \bar{\tau}^3 (1-\tau^3) (1-\bar{\tau}^3)
+ 46 \tau^2 \bar{\tau}^2 (1-\tau^3) (1-\bar{\tau}^3)
\right. \right. \nonumber \\
&& \left. \left. ~~~~~~~~~~~~~~~~~~
- 77 \tau \bar{\tau} (1-\tau^3) (1-\bar{\tau}^3)
+ 9 (1-\tau^3)^2 + 9 (1-\bar{\tau}^3)^2 
\right. \right. \nonumber \\
&& \left. \left. ~~~~~~~~~~~~~~~~~~
+ 52 (1-\tau^3) + 52 (1-\bar{\tau}^3)
\right]
\frac{\zeta_3^2}{[2+\tau \bar{\tau}]^7}
\right. \nonumber \\
&& \left. ~~~~~~~~
+~ \frac{392}{3} 
\frac{[ \tau \bar{\tau} (1+2 \tau \bar{\tau}) (1-\tau \bar{\tau})^2 
+ 4 (1-\tau^3) (1-\bar{\tau}^3)]}{[2+\tau \bar{\tau}]^5} \zeta_7 \right] 
\epsilon^5
~+~ O(\epsilon^6) ~.
\label{op32}
\end{eqnarray}
Electronic expressions for these are included in the attached data file. While 
we have also calculated expressions for $\Delta_\mathbf{2^\prime}$, 
$\Delta_\mathbf{2^{\prime\prime}}$ and $\Delta_\mathbf{2^{\prime\prime\prime}}$
explicitly they can also be deduced from the following mappings given in 
\cite{40}, 
\begin{eqnarray}
&& \Delta_{\mathbf 2} ~\to~ \Delta_{\mathbf 2^\prime}: ~~~~~~~
\tau ~\to~ \frac{[\tau+2]}{[\tau-1]} ~~~,~~~
\bar{\tau} ~\to~ \frac{[\bar{\tau}+2]}{[\bar{\tau}-1]}
\nonumber \\
&& \Delta_{\mathbf 2} ~\to~ \Delta_{\mathbf 2^{\prime\prime}}: ~~~~~~~
\tau ~\to~ \frac{[\omega\tau+2]}{[\omega\tau-1]} ~~~,~~~
\bar{\tau} ~\to~ 
\frac{[\bar{\omega}\bar{\tau}+2]}{[\bar{\omega}\bar{\tau}-1]}
\nonumber \\
&& \Delta_{\mathbf 2} ~\to~ \Delta_{\mathbf 2^{\prime\prime\prime}}: ~~~~~~~
\tau ~\to~ \frac{[\omega^2\tau+2]}{[\omega^2\tau-1]} ~~~,~~~
\bar{\tau} ~\to~ 
\frac{[\bar{\omega}^2\bar{\tau}+2]}{[\bar{\omega}^2\bar{\tau}-1]} ~.
\end{eqnarray}
We note that each expression resulting from applying the mappings to 
$\Delta_{\mathbf 2}$ is consistent with the direct five loop evaluation which
provides a useful check on the critical exponents. Another consistency check
is that setting both $\tau$ and $\bar{\tau}$ to be equal to $1$ or $\infty$ in 
$\Delta_{\mathbf 1}$ reproduces the coefficients of $\epsilon$ in 
(\ref{omeps}). The discrepancy in the $O(\epsilon)$ term is due to the 
canonical part of $\Delta^{ijkl}$. 

Having determined the five loop corrections to $\Delta_i$ we can now extract
estimates for them in three dimensions. First we record the explicit
expressions for the $\epsilon$ expansion of the various exponents for each of
the three theories. We have
\begin{eqnarray}
\Delta_\mathbf{1}^{\mbox{\footnotesize{XYZ}}} &=& 2 ~-~ \frac{4}{3} \epsilon^2 
~+~ 4 (6 \zeta_3 + 1 ) \frac{\epsilon^3}{9}
~+~ 4 ( 27 \zeta_4 - 42 \zeta_3 - 120 \zeta_5 - 7 ) \frac{\epsilon^4}{27} 
\nonumber \\
&& +~ 2 (72 \zeta_3^2 + 420 \zeta_3 - 378 \zeta_4 + 1416 \zeta_5 - 1800 \zeta_6
+ 3969 \zeta_7 + 38 ) \frac{\epsilon^5}{81} ~+~ O(\epsilon^6)
\nonumber \\
\Delta_\mathbf{2}^{\mbox{\footnotesize{XYZ}}} &=& 2 ~-~ 2 \epsilon
\nonumber \\
\Delta_\mathbf{2^\prime}^{\mbox{\footnotesize{XYZ}}} &=& 
\Delta_\mathbf{2^{\prime\prime}}^{\mbox{\footnotesize{XYZ}}} ~=~
\Delta_\mathbf{2^{\prime\prime\prime}}^{\mbox{\footnotesize{XYZ}}} \nonumber \\
&=& 2 ~-~ \frac{2}{3} \epsilon ~-~ \frac{4}{9} \epsilon^2 
~+~ 4 (12 \zeta_3 - 1) \frac{\epsilon^3}{27} 
~+~ 4 ( 54 \zeta_4 - 56 \zeta_3 - 160 \zeta_5 - 3 ) \frac{\epsilon^4}{81} 
\nonumber \\
&& +~ 2 (528 \zeta_3^2 + 248 \zeta_3 - 504 \zeta_4 + 1467 \zeta_5 
- 2400 \zeta_6 + 5292 \zeta_7 - 14 ) \frac{\epsilon^5}{243} \nonumber \\
&& +~ O(\epsilon^6)
\label{expxyz}
\end{eqnarray}
\begin{eqnarray}
\Delta_\mathbf{1}^{\mbox{\footnotesize{cWZ${}^3$}}} &=& 
\Delta_\mathbf{2^\prime}^{\mbox{\footnotesize{cWZ${}^3$}}} \nonumber \\
&=& 2 ~-~ \frac{4}{3} \epsilon^2 ~+~  4 (12 \zeta_3 + 1 ) \frac{\epsilon^3}{9} 
~+~ 4 ( 54 \zeta_4 - 84 \zeta_3 - 240 \zeta_5 - 7 ) \frac{\epsilon^4}{27} 
\nonumber \\
&& +~ 4 (576 \zeta_3^2 + 396 \zeta_3 - 378 \zeta_4 + 1416 \zeta_5
- 1800 \zeta_6 + 5292 \zeta_7 + 19 ) \frac{\epsilon^5}{81} ~+~ O(\epsilon^6)
\nonumber \\
\Delta_\mathbf{2}^{\mbox{\footnotesize{cWZ${}^3$}}} &=& 
\Delta_\mathbf{2^{\prime\prime}}^{\mbox{\footnotesize{cWZ${}^3$}}} ~=~ 
\Delta_\mathbf{2^{\prime\prime\prime}}^{\mbox{\footnotesize{cWZ${}^3$}}} ~=~ 
2 ~-~ \frac{4}{3} \epsilon
\label{expcwz3}
\end{eqnarray}
and
\begin{eqnarray}
\Delta_\mathbf{1}^{\Zz_2\times\Zz_2} &=& 2 ~-~ \frac{4}{3} \epsilon^2
~+~ 4 (9 \zeta_3 + 1 ) \frac{\epsilon^3}{9} 
~+~ 2 ( 81 \zeta_4 - 126 \zeta_3 - 300 \zeta_5 - 14 ) \frac{\epsilon^4}{27}
\nonumber \\
&& +~ (2376 \zeta_3^2 + 2424 \zeta_3 - 2268 \zeta_4 + 5856 \zeta_5 
- 9000 \zeta_6 + 22491 \zeta_7 + 152) \frac{\epsilon^5}{162} ~+~ O(\epsilon^6)
\nonumber \\
\Delta_\mathbf{2}^{\Zz_2\times\Zz_2} &=& 
\Delta_\mathbf{2^{\prime\prime}}^{\Zz_2\times\Zz_2} \nonumber \\
&=& 2 ~+~ 2 \frac{( 26 - 15 \sqrt{3} )}{(71 \sqrt{3} - 123)} \epsilon 
~+~ \frac{( 265 - 153 \sqrt{3} )}{3 (71 \sqrt{3} - 123)} \epsilon^2 
\nonumber \\
&& +~ (1590 \sqrt{3} \zeta_3 - 41 \sqrt{3} - 2754 \zeta_3 + 71 )
\frac{\epsilon^3}{9 (71 \sqrt{3} - 123) } 
\nonumber \\
&& +~ ( 14310 \sqrt{3} \zeta_4 - 17452 \sqrt{3} \zeta_3 
- 53000 \sqrt{3} \zeta_5 - 1011 \sqrt{3} + 30228 \zeta_3 - 24786 \zeta_4
\nonumber \\
&& ~~~~
+ 91800 \zeta_5 + 1751) \frac{\epsilon^4}{54 (71 \sqrt{3} - 123)} 
\nonumber \\
&& +~ (177624 \sqrt{3} \zeta_3^2 + 107664 \sqrt{3} \zeta_3
- 157068 \sqrt{3} \zeta_4 + 451070 \sqrt{3} \zeta_5 - 795000 \sqrt{3} \zeta_6
\nonumber \\
&& ~~~~
+ 1912617 \sqrt{3} \zeta_7 - 1602 \sqrt{3} - 307656 \zeta_3^2 - 186480 \zeta_3
+ 272052 \zeta_4 - 781293 \zeta_5 
\nonumber \\
&& ~~~~
+ 1377000 \zeta_6 - 3312792 \zeta_7 + 2774)
\frac{\epsilon^5}{324 (71 \sqrt{3} - 123)} ~+~ O(\epsilon^6)
\nonumber \\
\Delta_\mathbf{2^\prime}^{\Zz_2\times\Zz_2} &=& 
\Delta_\mathbf{2^{\prime\prime\prime}}^{\Zz_2\times\Zz_2} \nonumber \\
&=& 2 ~+~ 2 \frac{( 97 - 56 \sqrt{3} )}{(71 \sqrt{3} - 123)} \epsilon 
~+~ \frac{(11 \sqrt{3} - 19)}{3 (71 \sqrt{3} - 123)} \epsilon^2 
\nonumber \\
&& +~ (114 \sqrt{3} \zeta_3 + 41 \sqrt{3} - 198 \zeta_3 - 71)
\frac{\epsilon^3}{9 (71 \sqrt{3} - 123)} 
\nonumber \\
&& +~ ( 1026 \sqrt{3} \zeta_4 - 724 \sqrt{3} \zeta_3 - 3800 \sqrt{3} \zeta_5
+ 301 \sqrt{3} + 1260 \zeta_3 - 1782 \zeta_4
\nonumber \\
&& ~~~~
+ 6600 \zeta_5 - 521) \frac{\epsilon^4}{54 (71 \sqrt{3} - 123)} 
\nonumber \\
&& +~ (6408 \sqrt{3} \zeta_3^2 + 1392 \sqrt{3} \zeta_3 - 6516 \sqrt{3} \zeta_4
+ 48983 \sqrt{3} \zeta_5 - 57000 \sqrt{3} \zeta_6 
\nonumber \\
&& ~~~~
+ 122598 \sqrt{3} \zeta_7 + 2170 \sqrt{3} - 11160 \zeta_3^2 - 2448 \zeta_3
+ 11340 \zeta_4 - 84996 \zeta_5 
\nonumber \\
&& ~~~~
+ 99000 \zeta_6 - 213003 \zeta_7 - 3758) 
\frac{\epsilon^5}{324 (71 \sqrt{3} - 123)} ~+~ O(\epsilon^6) ~.
\label{expz2sq}
\end{eqnarray}
We note that both $\Delta_\mathbf{1}^{\mbox{\footnotesize{cWZ${}^3$}}}$ and
$\Delta_\mathbf{2}^{\mbox{\footnotesize{cWZ${}^3$}}}$ are indeed consistent 
with (\ref{omeps}) as expected after allowance is made for the canonical 
dimension contribution of $2$~$-$~$2\epsilon$. For several exponents the series
truncates at $O(\epsilon)$ and no order symbol is included. This is because 
these are exact to all orders in $\epsilon$ and their three dimensional values 
tally precisely with those of \cite{40}. In deriving (\ref{expxyz}), 
(\ref{expcwz3}) and (\ref{expz2sq}) we have encoded (\ref{op31}) and 
(\ref{op32}) together with the $\tau$ and $\bar{\tau}$ dependent expressions 
for $\Delta_{\mathbf 2^\prime}$, $\Delta_{\mathbf 2^{\prime\prime}}$ and 
$\Delta_{\mathbf 2^{\prime\prime\prime}}$ in one programme and then evaluated
each explicitly. For the XYZ and the cWZ${}^3$ cases we find that several
non-exact exponents are equal and this agrees with \cite{40}. However in the
$\Zz_2$~$\times$~$\Zz_2$ case we disagree with the equivalences recorded in
Table 2 of \cite{40} for the $\mathbf{2^{\prime\prime}}$ and 
$\mathbf{2^{\prime\prime\prime}}$ dimensions. Instead we found
$\Delta^{\Zz_2\times\Zz_2}_{\mathbf{2}}$~$=$~$\Delta^{\Zz_2\times\Zz_2}_{\mathbf{2^{\prime\prime}}}$ and
$\Delta^{\Zz_2\times\Zz_2}_{\mathbf{2^{\prime}}}$~$=$~$\Delta^{\Zz_2\times\Zz_2}_{\mathbf{2^{\prime\prime\prime}}}$.
To see the alternating sign pattern and the magnitude of the coefficients the 
numerical values of the non-exact exponents are 
\begin{eqnarray}
\Delta_\mathbf{1}^{\mbox{\footnotesize{XYZ}}} &=& 
2 ~-~ 1.333333 \epsilon^2 ~+~ 3.649929 \epsilon^3 ~-~ 22.621480 \epsilon^4 
~+~ 95.728196 \epsilon^5 ~+~ O(\epsilon^6)
\nonumber \\
\Delta_\mathbf{2^\prime}^{\mbox{\footnotesize{XYZ}}} &=& 
\Delta_\mathbf{2^{\prime\prime}}^{\mbox{\footnotesize{XYZ}}} ~=~
\Delta_\mathbf{2^{\prime\prime\prime}}^{\mbox{\footnotesize{XYZ}}} \nonumber \\
&=& 2 ~-~ 0.666667 \epsilon ~-~ 0.444444 \epsilon^2 ~+~ 1.988842 \epsilon^3 
~-~ 8.779169 \epsilon^4 ~+~ 40.471457 \epsilon^5 \nonumber \\
&& +~ O(\epsilon^6)
\nonumber \\
\Delta_\mathbf{1}^{\mbox{\footnotesize{cWZ${}^3$}}} &=& 
\Delta_\mathbf{2^\prime}^{\mbox{\footnotesize{cWZ${}^3$}}} \nonumber \\
&=& 2 ~-~ 1.333333 \epsilon^2 ~+~ 6.855415 \epsilon^3 ~-~ 44.205924 \epsilon^4 
~+~ 290.935250 \epsilon^5 ~+~ O(\epsilon^6)
\nonumber \\
\Delta_\mathbf{1}^{\Zz_2\times\Zz_2} &=& 
2 ~-~ 1.333333 \epsilon^2 ~+~ 5.252672 \epsilon^3 ~-~ 28.805134 \epsilon^4 
~+~ 145.920995 \epsilon^5 ~+~ O(\epsilon^6)
\nonumber \\
\Delta_\mathbf{2}^{\Zz_2\times\Zz_2} &=& 
\Delta_\mathbf{2^{\prime\prime}}^{\Zz_2\times\Zz_2} \nonumber \\
&=& 2 ~-~ 1.577350 \epsilon ~+~ 0.051567 \epsilon^2 ~+~ 0.278877 \epsilon^3 
~-~ 0.888082 \epsilon^4 ~+~ 5.331310 \epsilon^5 \nonumber \\
&& +~ O(\epsilon^6)
\nonumber \\
\Delta_\mathbf{2^\prime}^{\Zz_2\times\Zz_2} &=& 
\Delta_\mathbf{2^{\prime\prime\prime}}^{\Zz_2\times\Zz_2} \nonumber \\
&=& 2 ~-~ 0.422650 \epsilon ~-~ 0.718233 \epsilon^2 ~+~ 2.926608 \epsilon^3 
~-~ 16.028343 \epsilon^4 ~+~ 78.326933 \epsilon^5
\nonumber \\
&& +~ O(\epsilon^6) ~.
\end{eqnarray}
For the exponents which have an $O(\epsilon)$ term the series are alternating
when the canonical value of $(2$~$-$~$2\epsilon)$ is allowed for.

{\begin{table}[ht]
\begin{center}
\begin{tabular}{|c||r|r|}
\hline
Pad\'{e} & $\Delta_{\mathbf 1}~~~$ & $\Delta_{\mathbf 2^\prime}~~~$ \\
\hline
$[2,1]$ & 1.859277 & 1.632346 \\
$[1,2]$ & 1.868528 & 1.660704 \\
\hline
$[3,1]$ & 1.777975 & 1.633073 \\
$[2,2]$ & ----------- & 1.633070 \\
$[1,3]$ & 1.797562 & 1.639170 \\
\hline
$[4,1]$ & 1.669152 & 1.638139 \\
$[3,2]$ & ----------- & 1.632229 \\
$[2,3]$ & ----------- & 1.637434 \\
$[1,4]$ & 1.705650 & 1.637537 \\
\hline
\end{tabular}
\end{center}
\begin{center}
\caption{Pad\'{e} approximants at three, four and five loops for non-exact
operator dimensions in the XYZ model.}
\label{tabxyz}
\end{center}
\end{table}}

{\begin{table}[hb]
\begin{center}
\begin{tabular}{|c||r|r|}
\hline
Pad\'{e} & $\Delta_{\mathbf 1}~~~$ \\
\hline
$[2,1]$ & 1.906650 \\
$[1,2]$ & 1.910813 \\
\hline
$[3,1]$ & 1.869530 \\
$[2,2]$ & ----------- \\
$[1,3]$ & 1.874821 \\
\hline
$[4,1]$ & 1.879670 \\
$[3,2]$ & 1.877593 \\
$[2,3]$ & 1.879319 \\
$[1,4]$ & 1.879929 \\
\hline
\end{tabular}
\end{center}
\begin{center}
\caption{Pad\'{e} approximants at three, four and five loops for the non-exact
operator dimensions in cWZ${}^3$ model.}
\label{tabcwz3}
\end{center}
\end{table}}
In \cite{44} the perturbative expansion was used to estimate the exponents in
three dimensions in order to compare them with the conformal bootstrap
calculation. Therefore we have extended that study here using the same method.
This was to construct the Pad\'{e} approximants for the five loop non-exact 
exponents. The results for each of the three theories are given in Tables 
\ref{tabxyz}, \ref{tabcwz3} and \ref{tabz2sq} where the Pad\'{e} approximants 
for three and four loops are also given. The $[L,0]$ and $[0,L]$ approximants 
at each loop order $L$ are excluded as they either do not converge or are 
singular in $2$~$<$~$d$~$<$~$4$. There are no entries in each table for some 
operator dimensions. This is because for those cases the Pad\'{e} approximant 
is also singular above three dimensions. So because there is no continuous 
connection down from four dimensions to three in these cases any evaluation at 
the latter dimension is unreliable. What is generally evident for each of the
theories is that the five loop Pad\'{e} approximants are similar especially in 
the cases where there are no singularities. Table \ref{tabsum} summarizes the 
situation at three, four and five loops for each of the three theories and also
records the conformal bootstrap results of \cite{40}. Each loop estimate is the
average of the Pad\'{e} approximants in the individual table of each theory.
In \cite{40} the three loop Pad\'{e} approximants were used to compare with the
bootstrap. By providing the same data for the next two loop orders gives an 
overall indication of the trend of including higher order loops. For the XYZ 
model the $\Delta_\mathbf{1}$ estimates are decreasing towards the bootstrap 
value and is a significant improvement on the three loop estimate. The 
estimates for the other exponents are slowly decreasing away from the value 
given in \cite{40}. It might be tempting to surmise that the operator 
dimensions in the XYZ model have been interchanged since swapping them would 
give agreement to a few percent. However this is not the case from analysing 
(\ref{expxyz}). A similar feature occurs for the non-exact exponent of the 
cWZ${}^3$ theory although the five loop value is within $2\%$ of the bootstrap 
value. The situation for the three non-exact dimensions for the 
$\Zz_2$~$\times$~$\Zz_2$ case is somewhat mixed. Clearly the estimate for 
$\Delta_{\mathbf 2}^{\Zz_2\times\Zz_2}$ is within less than a percentage of the
value of \cite{40} and is stable at each loop order. For the other operators 
the tolerance is around $5\%$ but the trend with loop order is not as settled.

{\begin{table}[ht]
\begin{center}
\begin{tabular}{|c||r|r|r|}
\hline
Pad\'{e} & $\Delta_{\mathbf 1}~~~$ & $\Delta_{\mathbf 2}~~~$ &
$\Delta_{\mathbf 2^\prime}~~~$ \\
\hline
$[2,1]$ & 1.887757 & ----------- & 1.729559 \\
$[1,2]$ & 1.893722 & 1.253242 & 1.747789 \\
\hline
$[3,1]$ & 1.842132 & 1.237664 & 1.706973 \\
$[2,2]$ & ----------- & 1.237098 & 1.702425 \\
$[1,3]$ & 1.850355 & ----------- & 1.716789 \\
\hline
$[4,1]$ & 1.813663 & 1.245205 & 1.684017 \\
$[3,2]$ & ----------- & 1.255392 & ----------- \\
$[2,3]$ & ----------- & 1.243920 & ----------- \\
$[1,4]$ & 1.821597 & 1.253878 & 1.692667 \\
\hline
\end{tabular}
\end{center}
\begin{center}
\caption{Pad\'{e} approximants at three, four and five loops for the non-exact
operator dimensions in $\Zz_2$~$\times$~$\Zz_2$ model.}
\label{tabz2sq}
\end{center}
\end{table}}

{\begin{table}[hb]
\begin{center}
\begin{tabular}{|c|c||r|r|r||r|}
\hline
Model & Dimension & $3$ loop & $4$ loop & $5$ loop & \cite{40} \\
\hline
XYZ & $\Delta_{\mathbf 1}$ & 1.863902 & 1.787768 & 1.687401 & 1.639 \\
      & $\Delta_{\mathbf 2^\prime}$ & 1.646525 & 1.635104 & 1.636335 & 1.681 \\
\hline
\rule{0pt}{12pt}
cWZ${}^3$ & $\Delta_{\mathbf 1}$ & 1.908732 & 1.872175 & 1.879128 & 1.910 \\
\hline
$\Zz_2$~$\times$~$\Zz_2$ & $\Delta_{\mathbf 1}$ & 1.890740 & 1.846243 & 1.817630 & 1.898 \\
                         & $\Delta_{\mathbf 2}$ & 1.253242 & 1.237381 & 1.249599 & 1.259 \\
                         & $\Delta_{\mathbf 2^\prime}$ & 1.738674 & 1.708729 & 1.688342 & 1.727 \\
\hline
\end{tabular}
\end{center}
\begin{center}
\caption{Averages of three, four and five loop Pad\'{e} approximants for 
non-exact operator dimensions compared with conformal bootstrap results.}
\label{tabsum}
\end{center}
\end{table}}

{\begin{figure}[ht]
\begin{center}
\includegraphics[width=15.0cm,height=8.7cm]{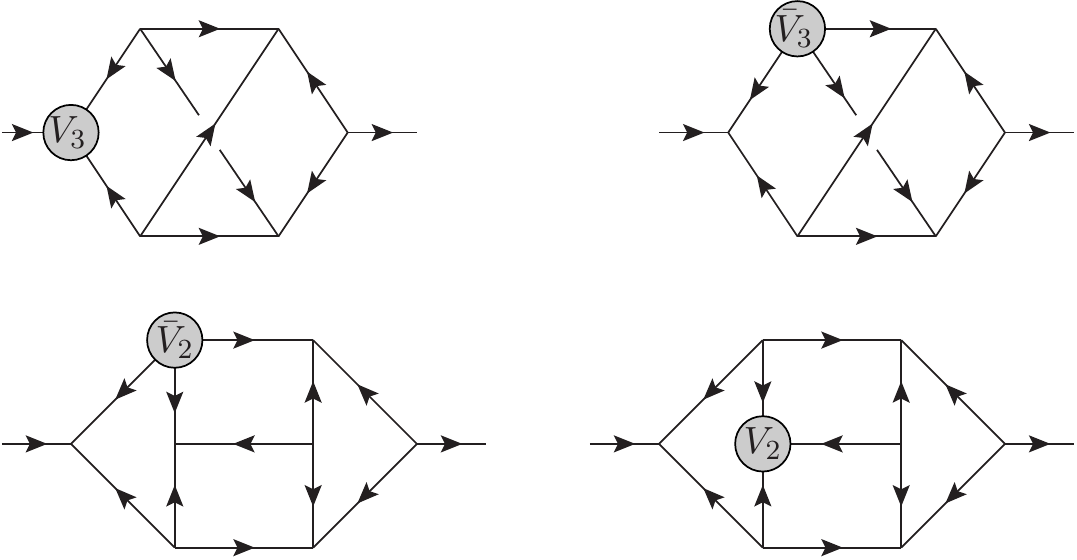}
\end{center}
\caption{Six loop product primitive graphs.}
\label{figtw61pr}
\end{figure}}

\sect{Beyond five loops.}

While our focus to this point has been on the five loop renormalization group
functions, the next stage in studying (\ref{lagwz}) would be to extend this to 
six loops. Given what we have established here it is worth giving guidance on 
what would be required for that as several common features emerged. First, at
six loops there are $324$ Feynman graphs contributing to the $\Phi$ $2$-point
function. The content of $\gamma_\Phi(a)$ at that order will involve rationals
as well as what we term irrationals. The majority of these will be $\zeta_n$ 
for $n$~$=$~$3$ to $9$. In addition their products such as $\zeta_3 \zeta_5$ 
and $\zeta_3^3$, which are both present in the six loop $\phi^4$ 
$\beta$-function \cite{32,34}, should appear if the structure of the 
renormalization group functions of this non-supersymmetric paradigm theory is 
valid. That would therefore imply the potential additional presence of the 
multiple zeta $\zeta_{3,5}$. As noted earlier the $O(1/N^3)$ expression for the
exponent $\eta$, \cite{49}, may indicate that such an irrational is actually
absent. However if it were present it would have to arise in a primitive graph
whose $O(N)$ group theory factor is beyond $O(1/N^3)$. Alternatively candidate 
primitive graphs from $\phi^4$ theory may be excluded because of the 
restriction the chiral symmetry places on the graph topologies. 

{\begin{figure}[ht]
\begin{center}
\includegraphics[width=9.95cm,height=4.0cm]{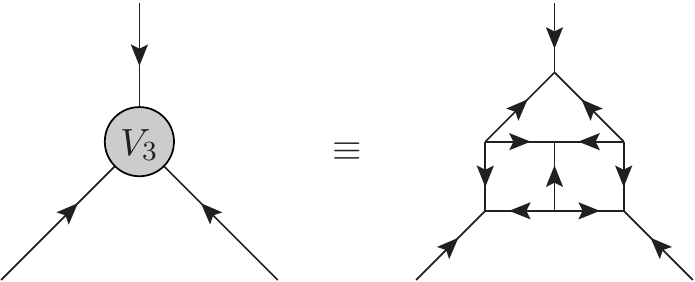}
\end{center}
\caption{Three loop planar vertex correction.}
\label{figv3}
\end{figure}}

Of the $324$ graphs it turns out that $17$ of these are primitive. One feature 
to emerge from the five loop evaluation of the Feynman graphs was the 
appearance of what was termed the product primitives. These are $2$-point 
graphs with vertex subgraphs. As the vertex function is finite, we noted that 
the simple pole can be deduced from the finite value of the vertex itself. At 
six loops we have illustrated the $8$ graphs of the total primitives that are 
product primitives in Figure \ref{figtw61pr} where the vertex $V_3$ is defined 
in Figure \ref{figv3}. The residue of the simple pole in $\epsilon$ of each of 
the graphs will be proportional to $\zeta_3 \zeta_5$ and have a group factor of
$T_2 T_5 T_{71}$ for (\ref{lagwzt}). The explicit coefficient of this residue
requires the implementation of the $D$-algebra. This is also an issue for the
remaining non-product primitives especially as the power of the irreducible 
scalar products increases with loop order. The remaining graphs intermediate to
those with rational contributions and the primitives correspond to the 
decoration of the lower loop primitives with an extra one loop bubble. A subset
of these should be calculable with the use of subtractions and {\sc Forcer}. 
The remainder of this type, similar to the non-product primitives, could only 
be reliably evaluated with a five loop version of {\sc Forcer}. 

{\begin{figure}[ht]
\begin{center}
\includegraphics[width=15.0cm,height=8.70cm]{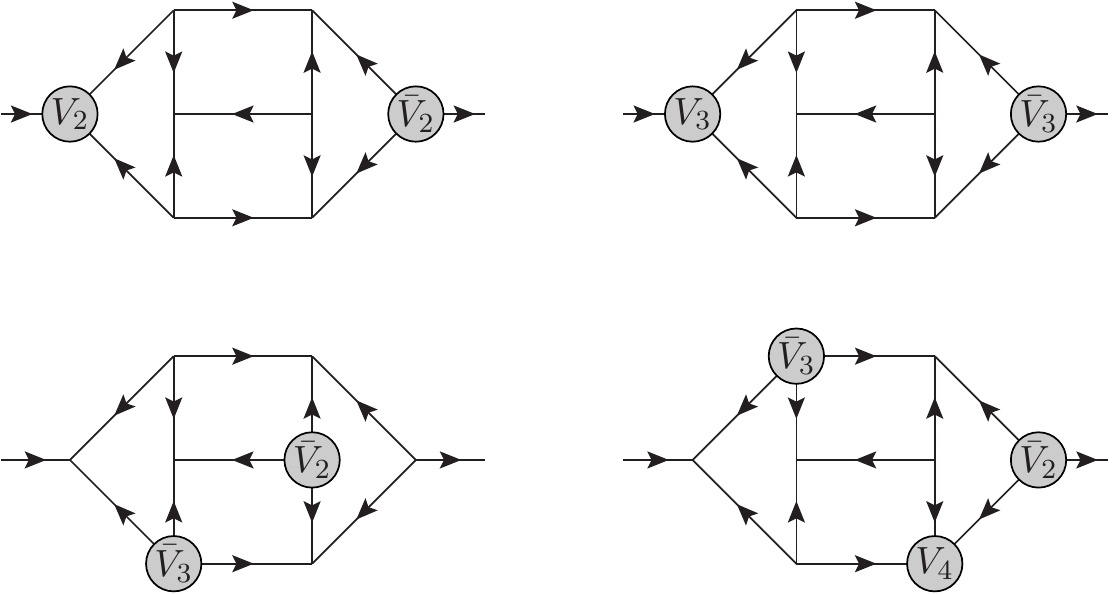}
\end{center}
\caption{Higher order product primitive graphs.}
\label{fighopr}
\end{figure}}

Next we note that the concept of product primitives naturally continues at 
higher loop order. We have provided several examples in Figure \ref{fighopr} to
illustrate the point. A new vertex function $V_4$ has been defined in Figure
\ref{figv4} where the actual $3$-point function is isolated by amputating the
right external vertex. In Figure \ref{fighopr} the graphs are $8$, $10$, $9$ 
and $13$ loops respectively from top left to bottom right. The simple pole 
residue of each would be $\zeta_3^2 \zeta_5$, $\zeta_5^3$, $\zeta_3 \zeta_5^2$ 
and $\zeta_3 \zeta_5^2 \zeta_7$ in the same respective order with the equally 
associated group factors of $T_2 T_5^2 T_{71}$, $T_2 T_{71}^3$, 
$T_2 T_5 T_{71}^2$ and $T_2 T_5 T_{71}^2 T_{94}$. So there is a clear 
association of each group factor with a specific $\zeta_n$. 

{\begin{figure}[ht]
\begin{center}
\includegraphics[width=14.0cm,height=4.0cm]{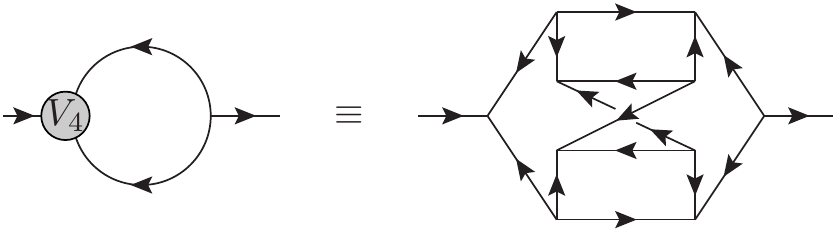}
\end{center}
\caption{Definition of four loop primitive vertex.}
\label{figv4}
\end{figure}}

Finally we return to the rational part of $\gamma_\Phi(a)$ and note that it is
possible to deduce the contribution in the $\MSbar$ scheme purely from the five
loop computation. This is because the rational part of the six loop MOM scheme
is known from the Hopf algebra solution of the Dyson-Schwinger equation given
in \cite{81}. As we showed earlier the five loop MOM expression for
$\gamma_\Phi(a)$ could be deduced from the $\MSbar$ expression by using the
coupling constant map (\ref{ccmommap}) and the formalism of (\ref{betamomms}) 
and (\ref{gammamomms}). To extract the rational part at six loops requires one 
ingredient which is the finite part of the $\Phi$ $2$-point function at five 
loops. This is because the coupling constant mapping at $L$ loops determines 
the $(L+1)$ loop renormalization group functions from (\ref{betamomms}) and 
(\ref{gammamomms}) once they are available at $L$ loops in one specific scheme.
Previously the MOM five loop $\beta$-function was deduced from the $\MSbar$ 
one. Here we reverse the process given the result of \cite{81}. So all that is 
required is the rational part of the $\Phi$ $2$-point function at five loops. 
As these are the bubble graphs which are simple to evaluate to the finite part
we have applied the formalism to find the rational piece of the six loop 
$\MSbar$ $\beta$-function which is
\begin{eqnarray}
\beta(a) &=& \frac{3}{2} a^2 ~-~ \frac{3}{2} a^3 ~+~ 
\left[ 36 \zeta_3 + 15 \right] \frac{a^4}{8} ~+~
\left[ 54 \zeta_4 - 180 \zeta_3 - 240 \zeta_5 - 27 \right] \frac{a^5}{8}
\nonumber \\
&& +~ \left[ 1512 \zeta_3^2 + 2574 \zeta_3 - 1323 \zeta_4 + 5484 \zeta_5
- 2700 \zeta_6 + 7938 \zeta_7 + 237 \right] \frac{a^6}{32} \nonumber \\
&& +~ \left[ -~ \frac{369}{20} ~+~ \mbox{non-rational contribution}
\right] a^7 ~+~ O(a^8) 
\label{bet6rat}
\end{eqnarray}
and we note that the alternating sign pattern of the rationals is maintained. 
To determine the non-rational contribution of (\ref{bet6rat}) is of course a
more strenuous exercise.

\sect{Discussion.}

We have completed a comprehensive study of the Wess-Zumino model at five loops.
This has proceeded in two phases with the initial one outlining the algorithm
for carrying out the computation of the five loop Feynman graphs that are
required for the $\beta$-function of the original model of \cite{1}. Once
established the second part addressed applications to various extensions of the
core theory by allowing the fields to lie in various symmetry groups or take
the couplings to be general tensors. One consequence was to extend the 
precision of the $\epsilon$ expansion of critical exponents to a new order. 
This is important in the context of other methods such as the conformal
bootstrap and the functional renormalization group techniques. These have been
applied to several problems like the emergent supersymmetric fixed point that 
is present in Gross-Neveu-Yukawa systems which relate to materials in Nature 
and could be the first manifestation of supersymmetry in reality. As a 
corollary the five loop Wess-Zumino renormalization could be a useful 
independent check on any future higher order renormalization of that system. 
However, to effect such a calculation in the Gross-Neveu-Yukawa model in four 
dimensions at five loops would be a massive undertaking especially given the 
number of graphs that would need to be evaluated. At four loops either $7384$ 
or $188531$ Feynman graphs were determined in \cite{15} where the two totals 
depend on whether real or complex scalars were used together with their
respective Dirac or left and right handed Weyl fermions. These are 
substantially larger numbers than the four loop ones given in Table 
\ref{feynnum}. This is primarily due to the fact that unlike the component 
Wess-Zumino model each interaction of the Gross-Neveu-Yukawa system has an 
independent coupling constant. Consequently all the $3$- and $4$-point vertices
have to be renormalized separately in the absence of any Ward identities. One 
interesting aspect of the $\epsilon$ expansion analysis was the close agreement
of the five loop estimates with other methods for the Gross-Neveu-Yukawa system
as is evident from Table \ref{expd3sum}. While the five loop results appear 
competitive with the latest bootstrap estimates there is still not precise 
agreement. Whether this is an indication of some discrepancy or not, such as 
non-perturbative contributions outside the scope of perturbation theory, is 
worth pursuing. If so it should not violate the underlying supersymmetry in the
extension from four to three dimensions in an $\epsilon$ expansion approach. 
The other case where we produced exponent estimates to compare with bootstrap 
methods, which concerned the one dimensional conformal manifold connected to 
the XYZ model, we found values that in some instances were close to the values 
quoted in \cite{40}. This suggests that perhaps higher orders in $\epsilon$ 
would be necessary to produce a more accurate comparison. While we have 
sketched out some basic ideas as to how a six loop computation could proceed 
again such a task is not trivial. Perhaps the graphical function methods of 
\cite{31,32,33} offers the best direction to follow especially if the method 
could be adapted to superspace in the first instance rather than have to use a 
component Lagrangian. Such a six loop renormalization would give insight into 
whether there are multiple zetas in the $\beta$-function of the Wess-Zumino 
model. This is the order where $\zeta_{3,5}$ first appears in its 
non-supersymmetric cousin $\phi^4$ theory which also has no chiral symmetry. If
it was present at this order in (\ref{lagwz}) then there would be no more 
debate.

\vspace{1cm}
\noindent
{\bf Acknowledgements.} We thank I. Jack for useful discussions at various 
stages of this work, H. Osborn for pointing out reference \cite{40}, D. Poland 
for discussions on the conformal bootstrap formalism as well as G. Dunne for 
posing several questions. It is also a pleasure to thank D.J. Broadhurst for 
enlightening discussions over many years concerning multiple zetas in primitive
Feynman integrals at very high loop order. This work was supported by a DFG 
Mercator Fellowship and in part through the STFC Consolidated Grants 
ST/J000493/1 and ST/T000988/1. The graphs were drawn with the {\sc Axodraw} 
package \cite{105}.

\appendix

\sect{Tensor definitions.}

In this appendix we define the tensors $T_{Lr}^{ij}$ that appear in the 
anomalous dimension of the general action (\ref{lagwzgend}). Each of these 
tensors depends on the tensor couplings $d^{ijk}$ and $\bar{d}^{ijk}$. The
subscript of each dummy index $j_n$ in each of the definitions is in direct 
correspondence to the label used in the {\sc Qgraf} electronic output that 
defines the underlying graph. In particular the bridge between $T_{Lr}^{ij}$ 
and the settings of the {\tt qgraf.dat} file in partnership with the 
{\tt form.sty} style file is to use the {\tt notadp} and {\tt onepi} options. 
To three loops the first set of tensors is
\begin{eqnarray}
&& d^{i j_1 j_2} \bar{d}^{j j_1 j_2} ~=~ T_{11}^{i j} \nonumber \\
&& d^{i j_1 j_2} d^{j_3 j_4 j_5} \bar{d}^{j j_1 j_3} \bar{d}^{j_2 j_4 j_5} ~=~ T_{21}^{i j} \nonumber \\
&& d^{i j_1 j_2} d^{j_3 j_5 j_7} d^{j_4 j_6 j_8} \bar{d}^{j j_3 j_4} \bar{d}^{j_1 j_5 j_6} \bar{d}^{j_2 j_7 j_8} ~=~ T_{31}^{i j} \nonumber \\
&& d^{i j_1 j_2} d^{j_3 j_5 j_6} d^{j_4 j_7 j_8} \bar{d}^{j j_3 j_4} \bar{d}^{j_1 j_5 j_6} \bar{d}^{j_2 j_7 j_8} ~=~ T_{32}^{i j} \nonumber \\
&& d^{i j_1 j_2} d^{j_3 j_6 j_7} d^{j_4 j_5 j_8} \bar{d}^{j j_1 j_3} \bar{d}^{j_2 j_4 j_5} \bar{d}^{j_8 j_6 j_7} ~=~ T_{33}^{i j} \nonumber \\
&& d^{i j_1 j_2} d^{j_3 j_4 j_6} d^{j_5 j_7 j_8} \bar{d}^{j j_1 j_3} \bar{d}^{j_2 j_4 j_5} \bar{d}^{j_6 j_7 j_8} ~=~ T_{34}^{i j} ~.
\end{eqnarray}
For orientation $T_{11}^{ij}$ and $T_{21}^{ij}$ correspond to the graphs of
Figures \ref{figtw1} and \ref{figtw2} respectively while $T_{31}^{ij}$ is the
non-planar graph of Figure \ref{figtw3}. We note that in \cite{91} a factor of
$\half$ was included in the definition of the tensor corresponding to
$T_{11}^{ij}$. At four loops the $13$ tensors are
\begin{eqnarray}
&& d^{i j_1 j_2} d^{j_3 j_5 j_9} d^{j_4 j_7 j_{10}} d^{j_6 j_8 j_{11}} \bar{d}^{j j_3 j_4} \bar{d}^{j_1 j_5 j_6} \bar{d}^{j_2 j_7 j_8} \bar{d}^{j_9 j_{10} j_{11}} ~=~ T_{41}^{i j} \nonumber \\
&& d^{i j_1 j_2} d^{j_3 j_5 j_9} d^{j_4 j_6 j_{10}} d^{j_{11} j_7 j_8} \bar{d}^{j j_3 j_4} \bar{d}^{j_1 j_5 j_6} \bar{d}^{j_2 j_7 j_8} \bar{d}^{j_9 j_{10} j_{11}} ~=~ T_{42}^{i j} \nonumber \\
&& d^{i j_3 j_4} d^{j_1 j_5 j_6} d^{j_2 j_7 j_8} d^{j_9 j_{10} j_{11}} \bar{d}^{j j_1 j_2} \bar{d}^{j_3 j_5 j_9} \bar{d}^{j_4 j_6 j_{10}} \bar{d}^{j_{11} j_7 j_8} ~=~ T_{43}^{i j} \nonumber \\
&& d^{i j_1 j_2} d^{j_3 j_5 j_7} d^{j_4 j_6 j_9} d^{j_8 j_{10} j_{11}} \bar{d}^{j j_3 j_4} \bar{d}^{j_1 j_5 j_6} \bar{d}^{j_2 j_7 j_8} \bar{d}^{j_9 j_{10} j_{11}} ~=~ T_{44}^{i j} \nonumber \\
&& d^{i j_1 j_2} d^{j_3 j_5 j_6} d^{j_4 j_9 j_{10}} d^{j_7 j_8 j_{11}} \bar{d}^{j j_3 j_4} \bar{d}^{j_1 j_5 j_6} \bar{d}^{j_2 j_7 j_8} \bar{d}^{j_{11} j_9 j_{10}} ~=~ T_{45}^{i j} \nonumber \\
&& d^{i j_1 j_2} d^{j_3 j_5 j_6} d^{j_4 j_7 j_9} d^{j_8 j_{10} j_{11}} \bar{d}^{j j_3 j_4} \bar{d}^{j_1 j_5 j_6} \bar{d}^{j_2 j_7 j_8} \bar{d}^{j_9 j_{10} j_{11}} ~=~ T_{46}^{i j} \nonumber \\
&& d^{i j_1 j_2} d^{j_3 j_6 j_7} d^{j_4 j_8 j_9} d^{j_5 j_{10} j_{11}} \bar{d}^{j j_1 j_3} \bar{d}^{j_2 j_4 j_5} \bar{d}^{j_6 j_8 j_{10}} \bar{d}^{j_7 j_9 j_{11}} ~=~ T_{47}^{i j} \nonumber \\
&& d^{i j_1 j_2} d^{j_3 j_6 j_7} d^{j_4 j_8 j_9} d^{j_5 j_{10} j_{11}} \bar{d}^{j j_1 j_3} \bar{d}^{j_2 j_4 j_5} \bar{d}^{j_6 j_8 j_9} \bar{d}^{j_7 j_{10} j_{11}} ~=~ T_{48}^{i j} \nonumber \\
&& d^{i j_1 j_2} d^{j_3 j_6 j_7} d^{j_4 j_5 j_8} d^{j_9 j_{10} j_{11}} \bar{d}^{j j_1 j_3} \bar{d}^{j_2 j_4 j_5} \bar{d}^{j_7 j_{10} j_{11}} \bar{d}^{j_8 j_6 j_9} ~=~ T_{49}^{i j} \nonumber \\
&& d^{i j_1 j_3} d^{j_2 j_4 j_5} d^{j_7 j_{10} j_{11}} d^{j_8 j_6 j_9} \bar{d}^{j j_1 j_2} \bar{d}^{j_3 j_6 j_7} \bar{d}^{j_4 j_5 j_8} \bar{d}^{j_9 j_{10} j_{11}} ~=~ T_{410}^{i j} \nonumber \\
&& d^{i j_1 j_2} d^{j_3 j_6 j_7} d^{j_4 j_5 j_8} d^{j_9 j_{10} j_{11}} \bar{d}^{j j_1 j_3} \bar{d}^{j_2 j_4 j_5} \bar{d}^{j_6 j_7 j_9} \bar{d}^{j_8 j_{10} j_{11}} ~=~ T_{411}^{i j} \nonumber \\
&& d^{i j_1 j_2} d^{j_3 j_4 j_6} d^{j_5 j_7 j_8} d^{j_{11} j_9 j_{10}} \bar{d}^{j j_1 j_3} \bar{d}^{j_2 j_4 j_5} \bar{d}^{j_6 j_9 j_{10}} \bar{d}^{j_7 j_8 j_{11}} ~=~ T_{412}^{i j} \nonumber \\
&& d^{i j_1 j_2} d^{j_3 j_4 j_6} d^{j_5 j_7 j_8} d^{j_9 j_{10} j_{11}} \bar{d}^{j j_1 j_3} \bar{d}^{j_2 j_4 j_5} \bar{d}^{j_6 j_7 j_9} \bar{d}^{j_8 j_{10} j_{11}} ~=~ T_{413}^{i j} 
\end{eqnarray}
where $T_{44}^{ij}$ and $T_{41}^{ij}$ respectively correspond to the graphs in
the bottom row of Figure \ref{figtw4}. 

At five loops the $63$ different tensors are
\begin{eqnarray}
&& d^{i j_1 j_2} d^{j_3 j_9 j_{10}} d^{j_4 j_{11} j_{12}} d^{j_5 j_7 j_{13}} d^{j_6 j_8 j_{14}} \bar{d}^{j j_3 j_4} \bar{d}^{j_1 j_5 j_6} \bar{d}^{j_2 j_7 j_8} \bar{d}^{j_{13} j_9 j_{11}} \bar{d}^{j_{14} j_{10} j_{12}} ~=~ T_{51}^{i j} \nonumber \\
&& d^{i j_1 j_2} d^{j_3 j_9 j_{10}} d^{j_4 j_{11} j_{12}} d^{j_5 j_6 j_{13}} d^{j_7 j_8 j_{14}} \bar{d}^{j j_3 j_4} \bar{d}^{j_1 j_5 j_6} \bar{d}^{j_2 j_7 j_8} \bar{d}^{j_{13} j_9 j_{11}} \bar{d}^{j_{14} j_{10} j_{12}} ~=~ T_{52}^{i j} \nonumber \\
&& d^{i j_3 j_4} d^{j_1 j_5 j_6} d^{j_2 j_7 j_8} d^{j_{13} j_9 j_{11}} d^{j_{14} j_{10} j_{12}} \bar{d}^{j j_1 j_2} \bar{d}^{j_3 j_9 j_{10}} \bar{d}^{j_4 j_{11} j_{12}} \bar{d}^{j_5 j_6 j_{13}} \bar{d}^{j_7 j_8 j_{14}} ~=~ T_{53}^{i j} \nonumber \\
&& d^{i j_1 j_2} d^{j_3 j_9 j_{10}} d^{j_4 j_{11} j_{12}} d^{j_5 j_6 j_{13}} d^{j_7 j_8 j_{14}} \bar{d}^{j j_3 j_4} \bar{d}^{j_1 j_5 j_6} \bar{d}^{j_2 j_7 j_8} \bar{d}^{j_{13} j_9 j_{10}} \bar{d}^{j_{14} j_{11} j_{12}} ~=~ T_{54}^{i j} \nonumber \\
&& d^{i j_1 j_2} d^{j_3 j_5 j_9} d^{j_4 j_{10} j_{11}} d^{j_6 j_{12} j_{13}} d^{j_7 j_8 j_{14}} \bar{d}^{j j_3 j_4} \bar{d}^{j_1 j_5 j_6} \bar{d}^{j_2 j_7 j_8} \bar{d}^{j_9 j_{14} j_{12}} \bar{d}^{j_{13} j_{10} j_{11}} ~=~ T_{55}^{i j} \nonumber \\
&& d^{i j_1 j_2} d^{j_3 j_5 j_9} d^{j_4 j_{10} j_{11}} d^{j_6 j_{12} j_{13}} d^{j_7 j_8 j_{14}} \bar{d}^{j j_3 j_4} \bar{d}^{j_1 j_5 j_6} \bar{d}^{j_2 j_7 j_8} \bar{d}^{j_9 j_{12} j_{13}} \bar{d}^{j_{14} j_{10} j_{11}} ~=~ T_{56}^{i j} \nonumber \\
&& d^{i j_1 j_2} d^{j_3 j_5 j_9} d^{j_4 j_{10} j_{11}} d^{j_6 j_7 j_{12}} d^{j_8 j_{13} j_{14}} \bar{d}^{j j_3 j_4} \bar{d}^{j_1 j_5 j_6} \bar{d}^{j_2 j_7 j_8} \bar{d}^{j_9 j_{13} j_{14}} \bar{d}^{j_{12} j_{10} j_{11}} ~=~ T_{57}^{i j} \nonumber \\
&& d^{i j_3 j_4} d^{j_1 j_5 j_6} d^{j_2 j_7 j_8} d^{j_9 j_{13} j_{14}} d^{j_{12} j_{10} j_{11}} \bar{d}^{j j_1 j_2} \bar{d}^{j_3 j_5 j_9} \bar{d}^{j_4 j_{10} j_{11}} \bar{d}^{j_6 j_7 j_{12}} \bar{d}^{j_8 j_{13} j_{14}} ~=~ T_{58}^{i j} \nonumber \\
&& d^{i j_1 j_2} d^{j_3 j_5 j_9} d^{j_4 j_{10} j_{11}} d^{j_6 j_7 j_{12}} d^{j_8 j_{13} j_{14}} \bar{d}^{j j_3 j_4} \bar{d}^{j_1 j_5 j_6} \bar{d}^{j_2 j_7 j_8} \bar{d}^{j_9 j_{12} j_{13}} \bar{d}^{j_{14} j_{10} j_{11}} ~=~ T_{59}^{i j} \nonumber \\
&& d^{i j_3 j_4} d^{j_1 j_5 j_6} d^{j_2 j_7 j_8} d^{j_9 j_{12} j_{13}} d^{j_{14} j_{10} j_{11}} \bar{d}^{j j_1 j_2} \bar{d}^{j_3 j_5 j_9} \bar{d}^{j_4 j_{10} j_{11}} \bar{d}^{j_6 j_7 j_{12}} \bar{d}^{j_8 j_{13} j_{14}} ~=~ T_{510}^{i j} \nonumber \\
&& d^{i j_1 j_2} d^{j_3 j_5 j_9} d^{j_4 j_{10} j_{11}} d^{j_6 j_7 j_{12}} d^{j_8 j_{13} j_{14}} \bar{d}^{j j_3 j_4} \bar{d}^{j_1 j_5 j_6} \bar{d}^{j_2 j_7 j_8} \bar{d}^{j_9 j_{10} j_{13}} \bar{d}^{j_{12} j_{11} j_{14}} ~=~ T_{511}^{i j} \nonumber \\
&& d^{i j_1 j_2} d^{j_3 j_5 j_9} d^{j_4 j_{10} j_{11}} d^{j_6 j_7 j_{12}} d^{j_8 j_{13} j_{14}} \bar{d}^{j j_3 j_4} \bar{d}^{j_1 j_5 j_6} \bar{d}^{j_2 j_7 j_8} \bar{d}^{j_9 j_{12} j_{10}} \bar{d}^{j_{11} j_{13} j_{14}} ~=~ T_{512}^{i j} \nonumber \\
&& d^{i j_1 j_2} d^{j_3 j_5 j_9} d^{j_4 j_7 j_{10}} d^{j_6 j_{11} j_{12}} d^{j_8 j_{13} j_{14}} \bar{d}^{j j_3 j_4} \bar{d}^{j_1 j_5 j_6} \bar{d}^{j_2 j_7 j_8} \bar{d}^{j_9 j_{13} j_{14}} \bar{d}^{j_{10} j_{11} j_{12}} ~=~ T_{513}^{i j} \nonumber \\
&& d^{i j_1 j_2} d^{j_3 j_5 j_9} d^{j_4 j_7 j_{10}} d^{j_6 j_{11} j_{12}} d^{j_8 j_{13} j_{14}} \bar{d}^{j j_3 j_4} \bar{d}^{j_1 j_5 j_6} \bar{d}^{j_2 j_7 j_8} \bar{d}^{j_9 j_{11} j_{13}} \bar{d}^{j_{10} j_{12} j_{14}} ~=~ T_{514}^{i j} \nonumber \\
&& d^{i j_1 j_2} d^{j_3 j_5 j_9} d^{j_4 j_7 j_{10}} d^{j_6 j_{11} j_{12}} d^{j_8 j_{13} j_{14}} \bar{d}^{j j_3 j_4} \bar{d}^{j_1 j_5 j_6} \bar{d}^{j_2 j_7 j_8} \bar{d}^{j_9 j_{11} j_{12}} \bar{d}^{j_{10} j_{13} j_{14}} ~=~ T_{515}^{i j} \nonumber \\
&& d^{i j_1 j_2} d^{j_3 j_5 j_9} d^{j_4 j_7 j_{10}} d^{j_6 j_8 j_{11}} d^{j_{12} j_{13} j_{14}} \bar{d}^{j j_3 j_4} \bar{d}^{j_1 j_5 j_6} \bar{d}^{j_2 j_7 j_8} \bar{d}^{j_9 j_{11} j_{12}} \bar{d}^{j_{10} j_{13} j_{14}} ~=~ T_{516}^{i j} \nonumber \\
&& d^{i j_3 j_4} d^{j_1 j_5 j_6} d^{j_2 j_7 j_8} d^{j_9 j_{11} j_{12}} d^{j_{10} j_{13} j_{14}} \bar{d}^{j j_1 j_2} \bar{d}^{j_3 j_5 j_9} \bar{d}^{j_4 j_7 j_{10}} \bar{d}^{j_6 j_8 j_{11}} \bar{d}^{j_{12} j_{13} j_{14}} ~=~ T_{517}^{i j} \nonumber \\
&& d^{i j_1 j_2} d^{j_3 j_5 j_9} d^{j_4 j_7 j_{10}} d^{j_6 j_8 j_{11}} d^{j_{12} j_{13} j_{14}} \bar{d}^{j j_3 j_4} \bar{d}^{j_1 j_5 j_6} \bar{d}^{j_2 j_7 j_8} \bar{d}^{j_9 j_{10} j_{12}} \bar{d}^{j_{11} j_{13} j_{14}} ~=~ T_{518}^{i j} \nonumber \\
&& d^{i j_1 j_2} d^{j_3 j_5 j_9} d^{j_4 j_6 j_{10}} d^{j_7 j_{11} j_{12}} d^{j_8 j_{13} j_{14}} \bar{d}^{j j_3 j_4} \bar{d}^{j_1 j_5 j_6} \bar{d}^{j_2 j_7 j_8} \bar{d}^{j_9 j_{11} j_{13}} \bar{d}^{j_{10} j_{12} j_{14}} ~=~ T_{519}^{i j} \nonumber \\
&& d^{i j_3 j_4} d^{j_1 j_5 j_6} d^{j_2 j_7 j_8} d^{j_9 j_{11} j_{13}} d^{j_{10} j_{12} j_{14}} \bar{d}^{j j_1 j_2} \bar{d}^{j_3 j_5 j_9} \bar{d}^{j_4 j_6 j_{10}} \bar{d}^{j_7 j_{11} j_{12}} \bar{d}^{j_8 j_{13} j_{14}} ~=~ T_{520}^{i j} \nonumber \\
&& d^{i j_1 j_2} d^{j_3 j_5 j_9} d^{j_4 j_6 j_{10}} d^{j_7 j_{11} j_{12}} d^{j_8 j_{13} j_{14}} \bar{d}^{j j_3 j_4} \bar{d}^{j_1 j_5 j_6} \bar{d}^{j_2 j_7 j_8} \bar{d}^{j_9 j_{11} j_{12}} \bar{d}^{j_{10} j_{13} j_{14}} ~=~ T_{521}^{i j} \nonumber \\
&& d^{i j_3 j_4} d^{j_1 j_5 j_6} d^{j_2 j_7 j_8} d^{j_9 j_{11} j_{12}} d^{j_{10} j_{13} j_{14}} \bar{d}^{j j_1 j_2} \bar{d}^{j_3 j_5 j_9} \bar{d}^{j_4 j_6 j_{10}} \bar{d}^{j_7 j_{11} j_{12}} \bar{d}^{j_8 j_{13} j_{14}} ~=~ T_{522}^{i j} \nonumber \\
&& d^{i j_1 j_2} d^{j_3 j_5 j_9} d^{j_4 j_6 j_{10}} d^{j_8 j_{13} j_{14}} d^{j_{11} j_7 j_{12}} \bar{d}^{j j_3 j_4} \bar{d}^{j_1 j_5 j_6} \bar{d}^{j_2 j_7 j_8} \bar{d}^{j_9 j_{10} j_{11}} \bar{d}^{j_{12} j_{13} j_{14}} ~=~ T_{523}^{i j} \nonumber \\
&& d^{i j_3 j_4} d^{j_1 j_5 j_6} d^{j_2 j_7 j_8} d^{j_9 j_{10} j_{11}} d^{j_{12} j_{13} j_{14}} \bar{d}^{j j_1 j_2} \bar{d}^{j_3 j_5 j_9} \bar{d}^{j_4 j_6 j_{10}} \bar{d}^{j_8 j_{13} j_{14}} \bar{d}^{j_{11} j_7 j_{12}} ~=~ T_{524}^{i j} \nonumber \\
&& d^{i j_1 j_2} d^{j_3 j_5 j_9} d^{j_4 j_6 j_{10}} d^{j_7 j_8 j_{11}} d^{j_{12} j_{13} j_{14}} \bar{d}^{j j_3 j_4} \bar{d}^{j_1 j_5 j_6} \bar{d}^{j_2 j_7 j_8} \bar{d}^{j_9 j_{11} j_{12}} \bar{d}^{j_{10} j_{13} j_{14}} ~=~ T_{525}^{i j} \nonumber \\
&& d^{i j_3 j_4} d^{j_1 j_5 j_6} d^{j_2 j_7 j_8} d^{j_9 j_{11} j_{12}} d^{j_{10} j_{13} j_{14}} \bar{d}^{j j_1 j_2} \bar{d}^{j_3 j_5 j_9} \bar{d}^{j_4 j_6 j_{10}} \bar{d}^{j_7 j_8 j_{11}} \bar{d}^{j_{12} j_{13} j_{14}} ~=~ T_{526}^{i j} \nonumber \\
&& d^{i j_1 j_2} d^{j_3 j_5 j_9} d^{j_4 j_6 j_{10}} d^{j_7 j_8 j_{11}} d^{j_{12} j_{13} j_{14}} \bar{d}^{j j_3 j_4} \bar{d}^{j_1 j_5 j_6} \bar{d}^{j_2 j_7 j_8} \bar{d}^{j_9 j_{10} j_{12}} \bar{d}^{j_{11} j_{13} j_{14}} ~=~ T_{527}^{i j} \nonumber \\
&& d^{i j_3 j_4} d^{j_1 j_5 j_6} d^{j_2 j_7 j_8} d^{j_9 j_{10} j_{12}} d^{j_{11} j_{13} j_{14}} \bar{d}^{j j_1 j_2} \bar{d}^{j_3 j_5 j_9} \bar{d}^{j_4 j_6 j_{10}} \bar{d}^{j_7 j_8 j_{11}} \bar{d}^{j_{12} j_{13} j_{14}} ~=~ T_{528}^{i j} \nonumber \\
&& d^{i j_1 j_2} d^{j_3 j_5 j_7} d^{j_4 j_6 j_9} d^{j_8 j_{10} j_{11}} d^{j_{14} j_{12} j_{13}} \bar{d}^{j j_3 j_4} \bar{d}^{j_1 j_5 j_6} \bar{d}^{j_2 j_7 j_8} \bar{d}^{j_9 j_{12} j_{13}} \bar{d}^{j_{10} j_{11} j_{14}} ~=~ T_{529}^{i j} \nonumber \\
&& d^{i j_1 j_2} d^{j_3 j_5 j_7} d^{j_4 j_6 j_9} d^{j_8 j_{10} j_{11}} d^{j_{12} j_{13} j_{14}} \bar{d}^{j j_3 j_4} \bar{d}^{j_1 j_5 j_6} \bar{d}^{j_2 j_7 j_8} \bar{d}^{j_9 j_{10} j_{12}} \bar{d}^{j_{11} j_{13} j_{14}} ~=~ T_{530}^{i j} \nonumber \\
&& d^{i j_1 j_2} d^{j_3 j_5 j_6} d^{j_4 j_9 j_{10}} d^{j_7 j_{11} j_{12}} d^{j_8 j_{13} j_{14}} \bar{d}^{j j_3 j_4} \bar{d}^{j_1 j_5 j_6} \bar{d}^{j_2 j_7 j_8} \bar{d}^{j_9 j_{11} j_{13}} \bar{d}^{j_{10} j_{12} j_{14}} ~=~ T_{531}^{i j} \nonumber \\
&& d^{i j_1 j_2} d^{j_3 j_5 j_6} d^{j_4 j_9 j_{10}} d^{j_7 j_{11} j_{12}} d^{j_8 j_{13} j_{14}} \bar{d}^{j j_3 j_4} \bar{d}^{j_1 j_5 j_6} \bar{d}^{j_2 j_7 j_8} \bar{d}^{j_9 j_{11} j_{12}} \bar{d}^{j_{10} j_{13} j_{14}} ~=~ T_{532}^{i j} \nonumber \\
&& d^{i j_1 j_2} d^{j_3 j_5 j_6} d^{j_4 j_9 j_{10}} d^{j_7 j_8 j_{11}} d^{j_{12} j_{13} j_{14}} \bar{d}^{j j_3 j_4} \bar{d}^{j_1 j_5 j_6} \bar{d}^{j_2 j_7 j_8} \bar{d}^{j_{10} j_{13} j_{14}} \bar{d}^{j_{11} j_9 j_{12}} ~=~ T_{533}^{i j} \nonumber \\
&& d^{i j_3 j_4} d^{j_1 j_5 j_6} d^{j_2 j_7 j_8} d^{j_{10} j_{13} j_{14}} d^{j_{11} j_9 j_{12}} \bar{d}^{j j_1 j_2} \bar{d}^{j_3 j_5 j_6} \bar{d}^{j_4 j_9 j_{10}} \bar{d}^{j_7 j_8 j_{11}} \bar{d}^{j_{12} j_{13} j_{14}} ~=~ T_{534}^{i j} \nonumber \\
&& d^{i j_1 j_2} d^{j_3 j_5 j_6} d^{j_4 j_9 j_{10}} d^{j_7 j_8 j_{11}} d^{j_{12} j_{13} j_{14}} \bar{d}^{j j_3 j_4} \bar{d}^{j_1 j_5 j_6} \bar{d}^{j_2 j_7 j_8} \bar{d}^{j_9 j_{10} j_{12}} \bar{d}^{j_{11} j_{13} j_{14}} ~=~ T_{535}^{i j} \nonumber \\
&& d^{i j_1 j_2} d^{j_3 j_5 j_6} d^{j_4 j_7 j_9} d^{j_8 j_{10} j_{11}} d^{j_{14} j_{12} j_{13}} \bar{d}^{j j_3 j_4} \bar{d}^{j_1 j_5 j_6} \bar{d}^{j_2 j_7 j_8} \bar{d}^{j_9 j_{12} j_{13}} \bar{d}^{j_{10} j_{11} j_{14}} ~=~ T_{536}^{i j} \nonumber \\
&& d^{i j_1 j_2} d^{j_3 j_5 j_6} d^{j_4 j_7 j_9} d^{j_8 j_{10} j_{11}} d^{j_{12} j_{13} j_{14}} \bar{d}^{j j_3 j_4} \bar{d}^{j_1 j_5 j_6} \bar{d}^{j_2 j_7 j_8} \bar{d}^{j_9 j_{10} j_{12}} \bar{d}^{j_{11} j_{13} j_{14}} ~=~ T_{537}^{i j} \nonumber \\
&& d^{i j_1 j_2} d^{j_3 j_6 j_7} d^{j_4 j_8 j_9} d^{j_5 j_{10} j_{11}} d^{j_{12} j_{13} j_{14}} \bar{d}^{j j_1 j_3} \bar{d}^{j_2 j_4 j_5} \bar{d}^{j_6 j_8 j_{12}} \bar{d}^{j_7 j_{13} j_{14}} \bar{d}^{j_9 j_{10} j_{11}} ~=~ T_{538}^{i j} \nonumber \\
&& d^{i j_1 j_2} d^{j_3 j_6 j_7} d^{j_4 j_8 j_9} d^{j_5 j_{10} j_{11}} d^{j_{12} j_{13} j_{14}} \bar{d}^{j j_1 j_3} \bar{d}^{j_2 j_4 j_5} \bar{d}^{j_6 j_8 j_{12}} \bar{d}^{j_7 j_{10} j_{13}} \bar{d}^{j_9 j_{11} j_{14}} ~=~ T_{539}^{i j} \nonumber \\
&& d^{i j_1 j_2} d^{j_3 j_6 j_7} d^{j_4 j_8 j_9} d^{j_5 j_{10} j_{11}} d^{j_{12} j_{13} j_{14}} \bar{d}^{j j_1 j_3} \bar{d}^{j_2 j_4 j_5} \bar{d}^{j_6 j_8 j_{12}} \bar{d}^{j_7 j_9 j_{13}} \bar{d}^{j_{14} j_{10} j_{11}} ~=~ T_{540}^{i j} \nonumber \\
&& d^{i j_1 j_3} d^{j_2 j_4 j_5} d^{j_6 j_8 j_{12}} d^{j_7 j_9 j_{13}} d^{j_{14} j_{10} j_{11}} \bar{d}^{j j_1 j_2} \bar{d}^{j_3 j_6 j_7} \bar{d}^{j_4 j_8 j_9} \bar{d}^{j_5 j_{10} j_{11}} \bar{d}^{j_{12} j_{13} j_{14}} ~=~ T_{541}^{i j} \nonumber \\
&& d^{i j_1 j_2} d^{j_3 j_6 j_7} d^{j_4 j_8 j_9} d^{j_5 j_{10} j_{11}} d^{j_{12} j_{13} j_{14}} \bar{d}^{j j_1 j_3} \bar{d}^{j_2 j_4 j_5} \bar{d}^{j_6 j_8 j_{10}} \bar{d}^{j_7 j_9 j_{12}} \bar{d}^{j_{11} j_{13} j_{14}} ~=~ T_{542}^{i j} \nonumber \\
&& d^{i j_1 j_2} d^{j_3 j_6 j_7} d^{j_4 j_8 j_9} d^{j_5 j_{10} j_{11}} d^{j_{14} j_{12} j_{13}} \bar{d}^{j j_1 j_3} \bar{d}^{j_2 j_4 j_5} \bar{d}^{j_6 j_8 j_9} \bar{d}^{j_7 j_{12} j_{13}} \bar{d}^{j_{10} j_{11} j_{14}} ~=~ T_{543}^{i j} \nonumber \\
&& d^{i j_1 j_2} d^{j_3 j_6 j_7} d^{j_4 j_8 j_9} d^{j_5 j_{10} j_{11}} d^{j_{12} j_{13} j_{14}} \bar{d}^{j j_1 j_3} \bar{d}^{j_2 j_4 j_5} \bar{d}^{j_6 j_8 j_9} \bar{d}^{j_7 j_{10} j_{12}} \bar{d}^{j_{11} j_{13} j_{14}} ~=~ T_{544}^{i j} \nonumber \\
&& d^{i j_1 j_2} d^{j_3 j_6 j_7} d^{j_4 j_5 j_8} d^{j_9 j_{11} j_{13}} d^{j_{10} j_{12} j_{14}} \bar{d}^{j j_1 j_3} \bar{d}^{j_2 j_4 j_5} \bar{d}^{j_6 j_9 j_{10}} \bar{d}^{j_7 j_{11} j_{12}} \bar{d}^{j_8 j_{13} j_{14}} ~=~ T_{545}^{i j} \nonumber \\
&& d^{i j_1 j_3} d^{j_2 j_4 j_5} d^{j_6 j_9 j_{10}} d^{j_7 j_{11} j_{12}} d^{j_8 j_{13} j_{14}} \bar{d}^{j j_1 j_2} \bar{d}^{j_3 j_6 j_7} \bar{d}^{j_4 j_5 j_8} \bar{d}^{j_9 j_{11} j_{13}} \bar{d}^{j_{10} j_{12} j_{14}} ~=~ T_{546}^{i j} \nonumber \\
&& d^{i j_1 j_2} d^{j_3 j_6 j_7} d^{j_4 j_5 j_8} d^{j_9 j_{10} j_{13}} d^{j_{14} j_{11} j_{12}} \bar{d}^{j j_1 j_3} \bar{d}^{j_2 j_4 j_5} \bar{d}^{j_6 j_9 j_{10}} \bar{d}^{j_7 j_{11} j_{12}} \bar{d}^{j_8 j_{13} j_{14}} ~=~ T_{547}^{i j} \nonumber \\
&& d^{i j_1 j_3} d^{j_2 j_4 j_5} d^{j_6 j_9 j_{10}} d^{j_7 j_{11} j_{12}} d^{j_8 j_{13} j_{14}} \bar{d}^{j j_1 j_2} \bar{d}^{j_3 j_6 j_7} \bar{d}^{j_4 j_5 j_8} \bar{d}^{j_9 j_{10} j_{13}} \bar{d}^{j_{14} j_{11} j_{12}} ~=~ T_{548}^{i j} \nonumber \\
&& d^{i j_1 j_2} d^{j_3 j_6 j_7} d^{j_4 j_5 j_8} d^{j_9 j_{10} j_{11}} d^{j_{12} j_{13} j_{14}} \bar{d}^{j j_1 j_3} \bar{d}^{j_2 j_4 j_5} \bar{d}^{j_6 j_9 j_{10}} \bar{d}^{j_7 j_{11} j_{12}} \bar{d}^{j_8 j_{13} j_{14}} ~=~ T_{549}^{i j} \nonumber \\
&& d^{i j_1 j_3} d^{j_2 j_4 j_5} d^{j_6 j_9 j_{10}} d^{j_7 j_{11} j_{12}} d^{j_8 j_{13} j_{14}} \bar{d}^{j j_1 j_2} \bar{d}^{j_3 j_6 j_7} \bar{d}^{j_4 j_5 j_8} \bar{d}^{j_9 j_{10} j_{11}} \bar{d}^{j_{12} j_{13} j_{14}} ~=~ T_{550}^{i j} \nonumber \\
&& d^{i j_1 j_2} d^{j_3 j_6 j_7} d^{j_4 j_5 j_8} d^{j_9 j_{12} j_{13}} d^{j_{10} j_{11} j_{14}} \bar{d}^{j j_1 j_3} \bar{d}^{j_2 j_4 j_5} \bar{d}^{j_7 j_{10} j_{11}} \bar{d}^{j_8 j_6 j_9} \bar{d}^{j_{14} j_{12} j_{13}} ~=~ T_{551}^{i j} \nonumber \\
&& d^{i j_1 j_3} d^{j_2 j_4 j_5} d^{j_7 j_{10} j_{11}} d^{j_8 j_6 j_9} d^{j_{14} j_{12} j_{13}} \bar{d}^{j j_1 j_2} \bar{d}^{j_3 j_6 j_7} \bar{d}^{j_4 j_5 j_8} \bar{d}^{j_9 j_{12} j_{13}} \bar{d}^{j_{10} j_{11} j_{14}} ~=~ T_{552}^{i j} \nonumber \\
&& d^{i j_1 j_2} d^{j_3 j_6 j_7} d^{j_4 j_5 j_8} d^{j_9 j_{10} j_{12}} d^{j_{11} j_{13} j_{14}} \bar{d}^{j j_1 j_3} \bar{d}^{j_2 j_4 j_5} \bar{d}^{j_7 j_{10} j_{11}} \bar{d}^{j_8 j_6 j_9} \bar{d}^{j_{12} j_{13} j_{14}} ~=~ T_{553}^{i j} \nonumber \\
&& d^{i j_1 j_3} d^{j_2 j_4 j_5} d^{j_7 j_{10} j_{11}} d^{j_8 j_6 j_9} d^{j_{12} j_{13} j_{14}} \bar{d}^{j j_1 j_2} \bar{d}^{j_3 j_6 j_7} \bar{d}^{j_4 j_5 j_8} \bar{d}^{j_9 j_{10} j_{12}} \bar{d}^{j_{11} j_{13} j_{14}} ~=~ T_{554}^{i j} \nonumber \\
&& d^{i j_1 j_2} d^{j_3 j_6 j_7} d^{j_4 j_5 j_8} d^{j_9 j_{12} j_{13}} d^{j_{10} j_{11} j_{14}} \bar{d}^{j j_1 j_3} \bar{d}^{j_2 j_4 j_5} \bar{d}^{j_6 j_7 j_9} \bar{d}^{j_8 j_{10} j_{11}} \bar{d}^{j_{14} j_{12} j_{13}} ~=~ T_{555}^{i j} \nonumber \\
&& d^{i j_1 j_2} d^{j_3 j_6 j_7} d^{j_4 j_5 j_8} d^{j_9 j_{10} j_{12}} d^{j_{11} j_{13} j_{14}} \bar{d}^{j j_1 j_3} \bar{d}^{j_2 j_4 j_5} \bar{d}^{j_6 j_7 j_9} \bar{d}^{j_8 j_{10} j_{11}} \bar{d}^{j_{12} j_{13} j_{14}} ~=~ T_{556}^{i j} \nonumber \\
&& d^{i j_1 j_2} d^{j_3 j_4 j_6} d^{j_5 j_7 j_8} d^{j_9 j_{11} j_{13}} d^{j_{10} j_{12} j_{14}} \bar{d}^{j j_1 j_3} \bar{d}^{j_2 j_4 j_5} \bar{d}^{j_6 j_9 j_{10}} \bar{d}^{j_7 j_{11} j_{12}} \bar{d}^{j_8 j_{13} j_{14}} ~=~ T_{557}^{i j} \nonumber \\
&& d^{i j_1 j_2} d^{j_3 j_4 j_6} d^{j_5 j_7 j_8} d^{j_9 j_{11} j_{12}} d^{j_{10} j_{13} j_{14}} \bar{d}^{j j_1 j_3} \bar{d}^{j_2 j_4 j_5} \bar{d}^{j_6 j_9 j_{10}} \bar{d}^{j_7 j_{11} j_{12}} \bar{d}^{j_8 j_{13} j_{14}} ~=~ T_{558}^{i j} \nonumber \\
&& d^{i j_1 j_2} d^{j_3 j_4 j_6} d^{j_5 j_7 j_8} d^{j_{10} j_{13} j_{14}} d^{j_{11} j_9 j_{12}} \bar{d}^{j j_1 j_3} \bar{d}^{j_2 j_4 j_5} \bar{d}^{j_6 j_9 j_{10}} \bar{d}^{j_7 j_8 j_{11}} \bar{d}^{j_{12} j_{13} j_{14}} ~=~ T_{559}^{i j} \nonumber \\
&& d^{i j_1 j_3} d^{j_2 j_4 j_5} d^{j_6 j_9 j_{10}} d^{j_7 j_8 j_{11}} d^{j_{12} j_{13} j_{14}} \bar{d}^{j j_1 j_2} \bar{d}^{j_3 j_4 j_6} \bar{d}^{j_5 j_7 j_8} \bar{d}^{j_{10} j_{13} j_{14}} \bar{d}^{j_{11} j_9 j_{12}} ~=~ T_{560}^{i j} \nonumber \\
&& d^{i j_1 j_2} d^{j_3 j_4 j_6} d^{j_5 j_7 j_8} d^{j_9 j_{10} j_{12}} d^{j_{11} j_{13} j_{14}} \bar{d}^{j j_1 j_3} \bar{d}^{j_2 j_4 j_5} \bar{d}^{j_6 j_9 j_{10}} \bar{d}^{j_7 j_8 j_{11}} \bar{d}^{j_{12} j_{13} j_{14}} ~=~ T_{561}^{i j} \nonumber \\
&& d^{i j_1 j_2} d^{j_3 j_4 j_6} d^{j_5 j_7 j_8} d^{j_9 j_{12} j_{13}} d^{j_{10} j_{11} j_{14}} \bar{d}^{j j_1 j_3} \bar{d}^{j_2 j_4 j_5} \bar{d}^{j_6 j_7 j_9} \bar{d}^{j_8 j_{10} j_{11}} \bar{d}^{j_{14} j_{12} j_{13}} ~=~ T_{562}^{i j} \nonumber \\
&& d^{i j_1 j_2} d^{j_3 j_4 j_6} d^{j_5 j_7 j_8} d^{j_9 j_{10} j_{12}} d^{j_{11} j_{13} j_{14}} \bar{d}^{j j_1 j_3} \bar{d}^{j_2 j_4 j_5} \bar{d}^{j_6 j_7 j_9} \bar{d}^{j_8 j_{10} j_{11}} \bar{d}^{j_{12} j_{13} j_{14}} ~=~ T_{563}^{i j} ~.
\end{eqnarray}
Again to assist with orientation the graphs in the top row of Figure 
\ref{figpr5} are respectively $T_{511}^{ij}$ and $T_{514}^{ij}$. Those of the 
lower row correspond to the tensors $T_{51}^{ij}$ and $T_{519}^{ij}$.

\sect{Renormalization constants.}

In this appendix we record the explicit form of the wave function 
renormalization constant for the action with the general tensor couplings 
(\ref{lagwzgend}). This is primarily to illustrate the structure of such a 
tensor renormalization constant as well as to provide the numerical value of 
each pole in $\epsilon$ for each tensor. To record the result in a compact way 
we decompose the renormalization constant $Z^{ij}_\Phi$ into a basis of tensors
as well as the residues of the respective poles giving 
\begin{equation}
Z^{ij}_\Phi ~=~ \delta^{ij} ~+~ 
\sum_{L=1}^5 \sum_{q=0}^L \sum_{r=1}^{k_L} a^{\cal S}_{Lq|Lr} T^{ij}_{Lr} 
\frac{1}{\epsilon^q} ~+~
\sum_{L=2}^5 \sum_{q=0}^L \sum_{r=1}^{d_L} b^{\cal S}_{Lq|Lr} D^{ij}_{Lr}
\frac{1}{\epsilon^q}
\end{equation}
where $k_L$ is defined in the last column of Table \ref{feynnum}. The 
coefficients $a^{\cal S}_{Lq|Lr}$ and $b^{\cal S}_{Lq|Lr}$ have pairs of 
labels. The first pair identifies the loop order and the power of the 
$\epsilon$ pole while the second pair relates to the relevant tensor. The label
${\cal S}$ denotes either the $\MSbar$ or MOM scheme. Clearly 
$a^{\MSbars}_{L0|Lr}$~$=$~$0$ and $b^{\MSbars}_{L0|Lr}$~$=$~$0$ as $q$~$=$~$0$
would indicate the finite part of the renormalization constant.
 
In addition to the tensors $T_{Lr}^{ij}$ that ultimately appear in the related
renormalization group functions, other ones arise for poles in $\epsilon$ of 
order higher than the simple one. These are denoted by $D_{Lr}^{ij}$ and those 
that arise to five loops are
\begin{eqnarray}
D_{21}^{ij} &=& \left( T_{11}^2 \right)^{ij} ~~,~~
D_{31}^{ij} ~=~ \left( T_{11}^3 \right)^{ij} ~~,~~
D_{32}^{ij} ~=~ \left( T_{21} T_{11} \right)^{ij} \nonumber \\
D_{41}^{ij} &=& \left( T_{11}^4 \right)^{ij} ~~,~~
D_{42}^{ij} ~=~ \left( T_{21}^2 \right)^{ij} ~~,~~
D_{43}^{ij} ~=~ \left( T_{21} T_{11}^2 \right)^{ij} ~~,~~
D_{44}^{ij} ~=~ \left( T_{31} T_{11} \right)^{ij} \nonumber \\
D_{45}^{ij} &=& \left( T_{32} T_{11} \right)^{ij} ~~,~~
D_{46}^{ij} ~=~ \left( T_{33} T_{11} \right)^{ij} ~~,~~
D_{47}^{ij} ~=~ \left( T_{34} T_{11} \right)^{ij} \nonumber \\
D_{51}^{ij} &=& \left( T_{11}^5 \right)^{ij} ~~,~~
D_{52}^{ij} ~=~ \left( T_{21}^2 \right)^{ij} ~~,~~
D_{53}^{ij} ~=~ \left( T_{21} T_{11}^3 \right)^{ij} ~~,~~
D_{54}^{ij} ~=~ \left( T_{21}^2 T_{11} \right)^{ij} \nonumber \\
D_{55}^{ij} &=& \left( T_{31} T_{21} \right)^{ij} ~~,~~
D_{56}^{ij} ~=~ \left( T_{32} T_{21} \right)^{ij} ~~,~~
D_{57}^{ij} ~=~ \left( T_{33} T_{21} \right)^{ij} ~~,~~
D_{58}^{ij} ~=~ \left( T_{34} T_{21} \right)^{ij} \nonumber \\
D_{59}^{ij} &=& \left( T_{31} T_{11}^2 \right)^{ij} ~~,~~
D_{510}^{ij} ~=~ \left( T_{32} T_{11}^2 \right)^{ij} ~~,~~
D_{511}^{ij} ~=~ \left( T_{33} T_{11}^2 \right)^{ij} ~~,~~
D_{512}^{ij} ~=~ \left( T_{34} T_{11}^2 \right)^{ij} \nonumber \\
D_{513}^{ij} &=& \left( T_{41} T_{11} \right)^{ij} ~~,~~
D_{514}^{ij} ~=~ \left( T_{42} T_{11} \right)^{ij} ~~,~~
D_{515}^{ij} ~=~ \left( T_{43} T_{11} \right)^{ij} ~~,~~
D_{516}^{ij} ~=~ \left( T_{44} T_{11} \right)^{ij} \nonumber \\
D_{517}^{ij} &=& \left( T_{45} T_{11} \right)^{ij} ~~,~~
D_{518}^{ij} ~=~ \left( T_{46} T_{11} \right)^{ij} ~~,~~
D_{519}^{ij} ~=~ \left( T_{47} T_{11} \right)^{ij} ~~,~~
D_{520}^{ij} ~=~ \left( T_{48} T_{11} \right)^{ij} \nonumber \\
D_{521}^{ij} &=& \left( T_{49} T_{11} \right)^{ij} ~~,~~
D_{522}^{ij} ~=~ \left( T_{410} T_{11} \right)^{ij} ~~,~~
D_{523}^{ij} ~=~ \left( T_{411} T_{11} \right)^{ij} ~~,~~
D_{524}^{ij} ~=~ \left( T_{412} T_{11} \right)^{ij} \nonumber \\
D_{525}^{ij} &=& \left( T_{413} T_{11} \right)^{ij} ~.
\end{eqnarray}
Graphically these correspond to the product of one-particle irreducible graphs.
Their coefficients in the Laurent expansion in $\epsilon$ are determined by
lower loop orders consistent with the renormalization group function.

For the $\MSbar$ scheme the residue of the poles to three loops are
\begin{eqnarray}
a^{\MSbars}_{11|11} &=& - \frac{1}{4} ~~,~~
a^{\MSbars}_{22|21} ~=~ - \frac{1}{8} ~~,~~
a^{\MSbars}_{21|21} ~=~ \frac{1}{8} ~~,~~
a^{\MSbars}_{33|32} ~=~ - \frac{1}{48} ~~,~~
a^{\MSbars}_{33|33} ~=~ - \frac{1}{24} \nonumber \\
a^{\MSbars}_{33|34} &=& - \frac{1}{24} ~~,~~
a^{\MSbars}_{32|32} ~=~ \frac{1}{48} ~~,~~
a^{\MSbars}_{32|33} ~=~ \frac{1}{24} ~~,~~
a^{\MSbars}_{32|34} ~=~ \frac{1}{8} ~~,~~
a^{\MSbars}_{31|31} ~=~ - \frac{1}{4} \zeta_3 \nonumber \\
a^{\MSbars}_{31|32} &=& \frac{1}{48} ~~,~~
a^{\MSbars}_{31|33} ~=~ \frac{1}{24} ~~,~~
a^{\MSbars}_{31|34} ~=~ - \frac{1}{6} 
\end{eqnarray}
with those at four loop being given by
\begin{eqnarray}
a^{\MSbars}_{44|45} &=& - \frac{1}{64} ~~,~~
a^{\MSbars}_{44|46} ~=~ - \frac{1}{64} ~~,~~
a^{\MSbars}_{44|48} ~=~ - \frac{1}{192} ~~,~~
a^{\MSbars}_{44|49} ~=~ - \frac{1}{64} \nonumber \\ 
a^{\MSbars}_{44|410} &=& - \frac{1}{64} ~~,~~
a^{\MSbars}_{44|411} ~=~ - \frac{1}{64} ~~,~~ 
a^{\MSbars}_{44|412} ~=~ - \frac{1}{96} ~~,~~
a^{\MSbars}_{44|413} ~=~ - \frac{1}{96} \nonumber \\
a^{\MSbars}_{43|45} &=& \frac{1}{64} ~~,~~
a^{\MSbars}_{43|46} ~=~ \frac{1}{24} ~~,~~
a^{\MSbars}_{43|48} ~=~ \frac{1}{64} ~~,~~
a^{\MSbars}_{43|49} ~=~ \frac{1}{24} ~~,~~
a^{\MSbars}_{43|410} ~=~ \frac{1}{24} \nonumber \\
a^{\MSbars}_{43|411} &=& \frac{1}{64} ~~,~~
a^{\MSbars}_{43|412} ~=~ \frac{1}{32} ~~,~~
a^{\MSbars}_{43|413} ~=~ \frac{1}{16} \nonumber \\
a^{\MSbars}_{42|42} &=& - \frac{1}{16} \zeta_3 ~~,~~
a^{\MSbars}_{42|43} ~=~ - \frac{1}{16} \zeta_3 ~~,~~
a^{\MSbars}_{42|44} ~=~ - \frac{1}{8} \zeta_3 ~~,~~
a^{\MSbars}_{42|45} ~=~ \frac{1}{64} \nonumber \\
a^{\MSbars}_{42|46} &=& - \frac{5}{192} ~~,~~
a^{\MSbars}_{42|47} ~=~ - \frac{3}{16} \zeta_3 ~~,~~
a^{\MSbars}_{42|48} ~=~ - \frac{1}{192} ~~,~~
a^{\MSbars}_{42|49} ~=~ - \frac{5}{192} \nonumber \\
a^{\MSbars}_{42|410} &=& - \frac{5}{192} ~~,~~
a^{\MSbars}_{42|411} ~=~ \frac{1}{64} ~~,~~ 
a^{\MSbars}_{42|412} ~=~ - \frac{1}{96} ~~,~~
a^{\MSbars}_{42|413} ~=~ - \frac{19}{96} \nonumber \\
a^{\MSbars}_{41|41} &=& \frac{5}{4} \zeta_5 ~~,~~
a^{\MSbars}_{41|42} ~=~ - \frac{3}{32} \zeta_4 + \frac{3}{16} \zeta_3 ~~,~~
a^{\MSbars}_{41|43} ~=~ - \frac{3}{32} \zeta_4 + \frac{3}{16} \zeta_3 \nonumber \\
a^{\MSbars}_{41|44} &=& - \frac{3}{16} \zeta_4 + \frac{3}{8} \zeta_3 ~~,~~
a^{\MSbars}_{41|45} ~=~ \frac{1}{64} - \frac{1}{32} \zeta_3 ~~,~~
a^{\MSbars}_{41|46} ~=~ - \frac{1}{24} \nonumber \\
a^{\MSbars}_{41|47} &=& \frac{3}{32} \zeta_4 + \frac{3}{16} \zeta_3 ~~,~~
a^{\MSbars}_{41|48} ~=~ - \frac{5}{192} ~~,~~
a^{\MSbars}_{41|49} ~=~ - \frac{1}{24} ~~,~~
a^{\MSbars}_{41|410} ~=~ - \frac{1}{24} \nonumber \\
a^{\MSbars}_{41|411} &=& \frac{1}{64} - \frac{1}{32} \zeta_3 ~~,~~
a^{\MSbars}_{41|412} ~=~ - \frac{5}{96} + \frac{1}{16} \zeta_3 ~~,~~
a^{\MSbars}_{41|413} ~=~ \frac{5}{16} ~.
\end{eqnarray}
The coefficients of the connected higher pole tensors are
\begin{eqnarray}
b^{\MSbars}_{22|21} &=& - \frac{1}{32} ~~,~~
b^{\MSbars}_{33|31} ~=~ - \frac{1}{128} ~~,~~
b^{\MSbars}_{33|32} ~=~ - \frac{1}{32} ~~,~~
b^{\MSbars}_{32|32} ~=~ \frac{1}{32} \nonumber \\
b^{\MSbars}_{44|41} &=& - \frac{5}{2048} ~~,~~
b^{\MSbars}_{44|42} ~=~ - \frac{1}{128} ~~,~~
b^{\MSbars}_{44|43} ~=~ - \frac{3}{256} ~~,~~
b^{\MSbars}_{44|45} ~=~ - \frac{1}{192} \nonumber \\
b^{\MSbars}_{44|46} &=& - \frac{1}{96} ~~,~~
b^{\MSbars}_{44|47} ~=~ - \frac{1}{96} \nonumber \\
b^{\MSbars}_{43|42} &=& \frac{1}{64} ~~,~~
b^{\MSbars}_{43|43} ~=~ \frac{3}{256} ~~,~~
b^{\MSbars}_{43|45} ~=~ \frac{1}{192} ~~,~~
b^{\MSbars}_{43|46} ~=~ \frac{1}{96} ~~,~~
b^{\MSbars}_{43|47} ~=~ \frac{1}{32} \nonumber \\
b^{\MSbars}_{42|42} &=& - \frac{1}{128} ~~,~~
b^{\MSbars}_{42|44} ~=~ - \frac{1}{16} \zeta_3 ~~,~~
b^{\MSbars}_{42|45} ~=~ \frac{1}{192} ~~,~~
b^{\MSbars}_{42|46} ~=~ \frac{1}{96} \nonumber \\
b^{\MSbars}_{42|47} &=& - \frac{1}{24} 
\end{eqnarray}
where obviously there can be no one loop coefficient.

Given that there are more tensors at five loops we record the data for this
part of $Z_\Phi^{ij}$ by the order of the pole. First, the leading pole 
coefficients are
\begin{eqnarray}
a^{\MSbars}_{55|54} &=& - \frac{1}{320} ~~,~~
a^{\MSbars}_{55|56} ~=~ - \frac{1}{160} ~~,~~
a^{\MSbars}_{55|515} ~=~ - \frac{1}{320} ~~,~~
a^{\MSbars}_{55|532} ~=~ - \frac{1}{480} \nonumber \\
a^{\MSbars}_{55|533} &=& - \frac{1}{160} ~~,~~
a^{\MSbars}_{55|534} ~=~ - \frac{1}{160} ~~,~~
a^{\MSbars}_{55|535} ~=~ - \frac{1}{160} ~~,~~
a^{\MSbars}_{55|536} ~=~ - \frac{1}{240} \nonumber \\
a^{\MSbars}_{55|537} &=& - \frac{1}{240} ~~,~~
a^{\MSbars}_{55|538} ~=~ - \frac{1}{160} ~~,~~
a^{\MSbars}_{55|543} ~=~ - \frac{1}{320} ~~,~~
a^{\MSbars}_{55|544} ~=~ - \frac{1}{320} \nonumber \\
a^{\MSbars}_{55|547} &=& - \frac{1}{480} ~~,~~
a^{\MSbars}_{55|548} ~=~ - \frac{1}{480} ~~,~~
a^{\MSbars}_{55|549} ~=~ - \frac{1}{160} ~~,~~
a^{\MSbars}_{55|550} ~=~ - \frac{1}{160} \nonumber \\
a^{\MSbars}_{55|551} &=& - \frac{1}{240} ~~,~~
a^{\MSbars}_{55|552} ~=~ - \frac{1}{240} ~~,~~
a^{\MSbars}_{55|553} ~=~ - \frac{1}{240} ~~,~~
a^{\MSbars}_{55|554} ~=~ - \frac{1}{240} \nonumber \\
a^{\MSbars}_{55|555} &=& - \frac{1}{160} ~~,~~
a^{\MSbars}_{55|556} ~=~ - \frac{1}{160} ~~,~~
a^{\MSbars}_{55|558} ~=~ - \frac{1}{960} ~~,~~
a^{\MSbars}_{55|559} ~=~ - \frac{1}{320} \nonumber \\
a^{\MSbars}_{55|560} &=& - \frac{1}{320} ~~,~~
a^{\MSbars}_{55|561} ~=~ - \frac{1}{320} ~~,~~
a^{\MSbars}_{55|562} ~=~ - \frac{1}{480} ~~,~~
a^{\MSbars}_{55|563} ~=~ - \frac{1}{480} 
\end{eqnarray}
then 
\begin{eqnarray}
a^{\MSbars}_{54|54} &=& \frac{1}{320} ~~,~~
a^{\MSbars}_{54|56} ~=~ \frac{1}{64} ~~,~~
a^{\MSbars}_{54|515} ~=~ \frac{1}{80} ~~,~~
a^{\MSbars}_{54|532} ~=~ \frac{11}{1920} \nonumber \\
a^{\MSbars}_{54|533} &=& \frac{1}{64} ~~,~~
a^{\MSbars}_{54|534} ~=~ \frac{1}{64} ~~,~~
a^{\MSbars}_{54|535} ~=~ \frac{1}{160} ~~,~~
a^{\MSbars}_{54|536} ~=~ \frac{11}{960} \nonumber \\
a^{\MSbars}_{54|537} &=& \frac{11}{480} ~~,~~
a^{\MSbars}_{54|538} ~=~ \frac{1}{40} ~~,~~
a^{\MSbars}_{54|543} ~=~ \frac{3}{320} ~~,~~
a^{\MSbars}_{54|544} ~=~ \frac{1}{60} \nonumber \\
a^{\MSbars}_{54|547} &=& \frac{11}{1920} ~~,~~
a^{\MSbars}_{54|548} ~=~ \frac{11}{1920} ~~,~~
a^{\MSbars}_{54|549} ~=~ \frac{1}{64} ~~,~~
a^{\MSbars}_{54|550} ~=~ \frac{1}{64} \nonumber \\
a^{\MSbars}_{54|551} &=& \frac{11}{960} ~~,~~
a^{\MSbars}_{54|552} ~=~ \frac{11}{960} ~~,~~
a^{\MSbars}_{54|553} ~=~ \frac{11}{480} ~~,~~
a^{\MSbars}_{54|554} ~=~ \frac{11}{480} \nonumber \\
a^{\MSbars}_{54|555} &=& \frac{1}{160} ~~,~~
a^{\MSbars}_{54|556} ~=~ \frac{1}{64} ~~,~~
a^{\MSbars}_{54|558} ~=~ \frac{1}{160} ~~,~~
a^{\MSbars}_{54|559} ~=~ \frac{1}{60} \nonumber \\
a^{\MSbars}_{54|560} &=& \frac{1}{60} ~~,~~
a^{\MSbars}_{54|561} ~=~ \frac{3}{320} ~~,~~
a^{\MSbars}_{54|562} ~=~ \frac{1}{80} ~~,~~
a^{\MSbars}_{54|563} ~=~ \frac{1}{48} 
\end{eqnarray}
are the quartic pole ones. Continuing the triple pole coefficients are
\begin{eqnarray}
a^{\MSbars}_{53|563} &=& - \frac{11}{96} ~~,~~
a^{\MSbars}_{53|562} ~=~ - \frac{13}{480} ~~,~~
a^{\MSbars}_{53|561} ~=~ - \frac{1}{320} ~~,~~
a^{\MSbars}_{53|560} ~=~ - \frac{11}{320} \nonumber \\
a^{\MSbars}_{53|559} &=& - \frac{11}{320} ~~,~~
a^{\MSbars}_{53|558} ~=~ - \frac{13}{960} ~~,~~
a^{\MSbars}_{53|556} ~=~ - \frac{1}{160} ~~,~~
a^{\MSbars}_{53|555} ~=~ \frac{1}{160} \nonumber \\
a^{\MSbars}_{53|554} &=& - \frac{7}{120} ~~,~~
a^{\MSbars}_{53|553} ~=~ - \frac{7}{120} ~~,~~
a^{\MSbars}_{53|550} ~=~ - \frac{1}{160} ~~,~~
a^{\MSbars}_{53|549} ~=~ - \frac{1}{160} \nonumber \\
a^{\MSbars}_{53|544} &=& - \frac{11}{320} ~~,~~
a^{\MSbars}_{53|543} ~=~ - \frac{1}{320} ~~,~~
a^{\MSbars}_{53|538} ~=~ - \frac{17}{480} ~~,~~
a^{\MSbars}_{53|537} ~=~ - \frac{7}{120} \nonumber \\
a^{\MSbars}_{53|535} &=& \frac{1}{160} ~~,~~
a^{\MSbars}_{53|534} ~=~ - \frac{1}{160} ~~,~~
a^{\MSbars}_{53|533} ~=~ - \frac{1}{160} ~~,~~
a^{\MSbars}_{53|515} ~=~ - \frac{17}{960} \nonumber \\
a^{\MSbars}_{53|56} &=& - \frac{1}{160} ~~,~~
a^{\MSbars}_{53|54} ~=~ \frac{1}{320} ~~,~~
a^{\MSbars}_{53|557} ~=~ - \frac{3}{40} \zeta_3 ~~,~~
a^{\MSbars}_{53|546} ~=~ - \frac{9}{160} \zeta_3 \nonumber \\
a^{\MSbars}_{53|545} &=& - \frac{9}{160} \zeta_3 ~~,~~
a^{\MSbars}_{53|542} ~=~ - \frac{3}{40} \zeta_3 ~~,~~
a^{\MSbars}_{53|541} ~=~ - \frac{3}{80} \zeta_3 ~~,~~
a^{\MSbars}_{53|540} ~=~ - \frac{3}{80} \zeta_3 \nonumber \\
a^{\MSbars}_{53|531} &=& - \frac{9}{160} \zeta_3 ~~,~~
a^{\MSbars}_{53|530} ~=~ - \frac{1}{40} \zeta_3 ~~,~~
a^{\MSbars}_{53|529} ~=~ - \frac{1}{40} \zeta_3 ~~,~~
a^{\MSbars}_{53|528} ~=~ - \frac{1}{80} \zeta_3 \nonumber \\
a^{\MSbars}_{53|527} &=& - \frac{1}{80} \zeta_3 ~~,~~
a^{\MSbars}_{53|526} ~=~ - \frac{1}{40} \zeta_3 ~~,~~
a^{\MSbars}_{53|525} ~=~ - \frac{1}{40} \zeta_3 ~~,~~
a^{\MSbars}_{53|524} ~=~ - \frac{1}{80} \zeta_3 \nonumber \\
a^{\MSbars}_{53|523} &=& - \frac{1}{80} \zeta_3 ~~,~~
a^{\MSbars}_{53|522} ~=~ - \frac{1}{80} \zeta_3 ~~,~~
a^{\MSbars}_{53|521} ~=~ - \frac{1}{80} \zeta_3 ~~,~~
a^{\MSbars}_{53|513} ~=~ - \frac{1}{80} \zeta_3 \nonumber \\
a^{\MSbars}_{53|58} &=& - \frac{1}{40} \zeta_3 ~~,~~
a^{\MSbars}_{53|57} ~=~ - \frac{1}{40} \zeta_3 ~~,~~
a^{\MSbars}_{53|55} ~=~ - \frac{1}{40} \zeta_3 ~~,~~
a^{\MSbars}_{53|53} ~=~ - \frac{1}{160} \zeta_3 \nonumber \\
a^{\MSbars}_{53|52} &=& - \frac{1}{160} \zeta_3
\end{eqnarray}
with 
\begin{eqnarray}
a^{\MSbars}_{52|52} &=& - \frac{3}{320} \zeta_4 + \frac{3}{160} \zeta_3 ~~,~~
a^{\MSbars}_{52|53} ~=~ - \frac{3}{320} \zeta_4 + \frac{3}{160} \zeta_3 ~~,~~
a^{\MSbars}_{52|54} ~=~ \frac{1}{320} - \frac{1}{160} \zeta_3 \nonumber \\
a^{\MSbars}_{52|55} &=& - \frac{3}{80} \zeta_4 + \frac{3}{40} \zeta_3 ~~,~~
a^{\MSbars}_{52|56} ~=~ - \frac{3}{320} - \frac{1}{160} \zeta_3 ~~,~~
a^{\MSbars}_{52|57} ~=~ - \frac{3}{80} \zeta_4 + \frac{3}{40} \zeta_3 \nonumber \\
a^{\MSbars}_{52|58} &=& - \frac{3}{80} \zeta_4 + \frac{3}{40} \zeta_3 ~~,~~
a^{\MSbars}_{52|59} ~=~ \frac{1}{4} \zeta_5 ~~,~~
a^{\MSbars}_{52|510} ~=~ \frac{1}{4} \zeta_5 ~~,~~
a^{\MSbars}_{52|512} ~=~ \frac{1}{4} \zeta_5 \nonumber \\
a^{\MSbars}_{52|513} &=& - \frac{3}{160} \zeta_4 + \frac{3}{80} \zeta_3 ~~,~~
a^{\MSbars}_{52|515} ~=~ \frac{1}{120} ~~,~~
a^{\MSbars}_{52|516} ~=~ \frac{1}{4} \zeta_5 ~~,~~
a^{\MSbars}_{52|517} ~=~ \frac{1}{4} \zeta_5 \nonumber \\
a^{\MSbars}_{52|518} &=& \frac{1}{8} \zeta_5 ~~,~~
a^{\MSbars}_{52|521} ~=~ - \frac{3}{160} \zeta_4 + \frac{3}{80} \zeta_3 ~~,~~
a^{\MSbars}_{52|522} ~=~ - \frac{3}{160} \zeta_4 + \frac{3}{80} \zeta_3 \nonumber \\
a^{\MSbars}_{52|523} &=& - \frac{3}{160} \zeta_4 + \frac{7}{80} \zeta_3 ~~,~~
a^{\MSbars}_{52|524} ~=~ - \frac{3}{160} \zeta_4 + \frac{7}{80} \zeta_3 ~~,~~
a^{\MSbars}_{52|525} ~=~ - \frac{3}{80} \zeta_4 + \frac{3}{40} \zeta_3 \nonumber \\
a^{\MSbars}_{52|526} &=& - \frac{3}{80} \zeta_4 + \frac{3}{40} \zeta_3 ~~,~~
a^{\MSbars}_{52|527} ~=~ - \frac{3}{160} \zeta_4 + \frac{3}{80} \zeta_3 ~~,~~
a^{\MSbars}_{52|528} ~=~ - \frac{3}{160} \zeta_4 + \frac{3}{80} \zeta_3 \nonumber \\
a^{\MSbars}_{52|529} &=& - \frac{3}{80} \zeta_4 + \frac{3}{40} \zeta_3 ~~,~~
a^{\MSbars}_{52|530} ~=~ - \frac{3}{80} \zeta_4 + \frac{7}{40} \zeta_3 ~~,~~
a^{\MSbars}_{52|531} ~=~ \frac{3}{320} \zeta_4 + \frac{9}{160} \zeta_3 \nonumber \\
a^{\MSbars}_{52|532} &=& - \frac{19}{1920} ~~,~~
a^{\MSbars}_{52|533} ~=~ - \frac{3}{320} - \frac{1}{160} \zeta_3 ~~,~~
a^{\MSbars}_{52|534} ~=~ - \frac{3}{320} - \frac{1}{160} \zeta_3 \nonumber \\
a^{\MSbars}_{52|535} &=& \frac{1}{160} - \frac{1}{80} \zeta_3 ~~,~~
a^{\MSbars}_{52|536} ~=~ - \frac{19}{960} + \frac{1}{160} \zeta_3 ~~,~~
a^{\MSbars}_{52|537} ~=~ \frac{19}{480} ~~,~~
a^{\MSbars}_{52|538} ~=~ \frac{1}{60} \nonumber \\
a^{\MSbars}_{52|539} &=& \zeta_5 ~~,~~
a^{\MSbars}_{52|540} ~=~ - \frac{9}{160} \zeta_4 + \frac{3}{16} \zeta_3 ~~,~~
a^{\MSbars}_{52|541} ~=~ - \frac{9}{160} \zeta_4 + \frac{3}{16} \zeta_3 \nonumber \\
a^{\MSbars}_{52|542} &=& -\frac{ 9}{80} \zeta_4 + \frac{3}{8} \zeta_3 ~~,~~
a^{\MSbars}_{52|543} ~=~ - \frac{1}{320} - \frac{3}{160} \zeta_3 ~~,~~
a^{\MSbars}_{52|544} ~=~ \frac{1}{240} \nonumber \\
a^{\MSbars}_{52|545} &=& \frac{3}{320} \zeta_4 + \frac{9}{160} \zeta_3 ~~,~~
a^{\MSbars}_{52|546} ~=~ \frac{3}{320} \zeta_4 + \frac{9}{160} \zeta_3 ~~,~~
a^{\MSbars}_{52|547} ~=~ - \frac{19}{1920} \nonumber \\
a^{\MSbars}_{52|548} &=& - \frac{19}{1920} ~~,~~
a^{\MSbars}_{52|549} ~=~ - \frac{3}{320} - \frac{1}{160} \zeta_3 ~~,~~
a^{\MSbars}_{52|550} ~=~ - \frac{3}{320} - \frac{1}{160} \zeta_3 \nonumber \\
a^{\MSbars}_{52|551} &=& - \frac{19}{960} + \frac{1}{160} \zeta_3 ~~,~~
a^{\MSbars}_{52|552} ~=~ - \frac{19}{960} + \frac{1}{160} \zeta_3 ~~,~~
a^{\MSbars}_{52|553} ~=~ \frac{19}{480} \nonumber \\
a^{\MSbars}_{52|554} &=& \frac{19}{480} ~~,~~
a^{\MSbars}_{52|555} ~=~ \frac{1}{160} - \frac{1}{80} \zeta_3 ~~,~~
a^{\MSbars}_{52|556} ~=~ - \frac{3}{320} - \frac{1}{160} \zeta_3 \nonumber \\
a^{\MSbars}_{52|557} &=& \frac{3}{40} \zeta_4 + \frac{9}{40} \zeta_3 ~~,~~
a^{\MSbars}_{52|558} ~=~ \frac{1}{240} ~~,~~
a^{\MSbars}_{52|559} ~=~ \frac{1}{240} + \frac{1}{80} \zeta_3 \nonumber \\
a^{\MSbars}_{52|560} &=& \frac{1}{240} + \frac{1}{80} \zeta_3 ~~,~~
a^{\MSbars}_{52|561} ~=~ - \frac{1}{320} - \frac{1}{160} \zeta_3 ~~,~~
a^{\MSbars}_{52|562} ~=~ \frac{1}{120} + \frac{1}{20} \zeta_3 \nonumber \\
a^{\MSbars}_{52|563} &=& \frac{19}{48}
\end{eqnarray}
as the coefficients for the double pole. Finally the simple poles that lead to
$\gamma_\Phi^{ij}$ are
\begin{eqnarray}
a^{\MSbars}_{51|51} &=& - \frac{9}{20} \zeta_3^2 ~~,~~
a^{\MSbars}_{51|52} ~=~ \frac{143}{160} \zeta_5 + \frac{9}{320} \zeta_4 - \frac{1}{160} \zeta_3 ~~,~~
a^{\MSbars}_{51|53} ~=~ \frac{143}{160} \zeta_5 + \frac{9}{320} \zeta_4 - \frac{1}{160} \zeta_3 \nonumber \\
a^{\MSbars}_{51|54} &=& \frac{1}{320} - \frac{3}{320} \zeta_4 + \frac{1}{160} \zeta_3 ~~,~~
a^{\MSbars}_{51|55} ~=~ - \frac{67}{40} \zeta_5 + \frac{9}{80} \zeta_4 - \frac{1}{40} \zeta_3 \nonumber \\
a^{\MSbars}_{51|56} &=& - \frac{3}{160} - \frac{3}{320} \zeta_4 + \frac{1}{32} \zeta_3 ~~,~~
a^{\MSbars}_{51|57} ~=~ - \frac{9}{5} \zeta_5 + \frac{9}{80} \zeta_4 - \frac{1}{40} \zeta_3 \nonumber \\
a^{\MSbars}_{51|58} &=& - \frac{9}{5} \zeta_5 + \frac{9}{80} \zeta_4 - \frac{1}{40} \zeta_3 ~~,~~
a^{\MSbars}_{51|59} ~=~ \frac{5}{8} \zeta_6 - \frac{5}{4} \zeta_5 - \frac{1}{20} \zeta_3^2 \nonumber \\
a^{\MSbars}_{51|510} &=& \frac{5}{8} \zeta_6 - \frac{5}{4} \zeta_5 - \frac{1}{20} \zeta_3^2 ~~,~~
a^{\MSbars}_{51|511} ~=~ - \frac{441}{80} \zeta_7 ~~,~~
a^{\MSbars}_{51|512} ~=~ \frac{5}{8} \zeta_6 - \frac{5}{4} \zeta_5 - \frac{1}{20} \zeta_3^2 \nonumber \\
a^{\MSbars}_{51|513} &=& \frac{79}{40} \zeta_5 + \frac{9}{160} \zeta_4 - \frac{1}{80} \zeta_3 ~~,~~
a^{\MSbars}_{51|514} ~=~ - \frac{441}{160} \zeta_7 ~~,~~
a^{\MSbars}_{51|515} ~=~ - \frac{1}{60} \nonumber \\
a^{\MSbars}_{51|516} &=& \frac{5}{8} \zeta_6 - \frac{5}{4} \zeta_5 + \frac{1}{4} \zeta_3^2 ~~,~~
a^{\MSbars}_{51|517} ~=~ \frac{5}{8} \zeta_6 - \frac{5}{4} \zeta_5 + \frac{1}{4} \zeta_3^2 \nonumber \\
a^{\MSbars}_{51|518} &=& \frac{5}{16} \zeta_6 - \frac{5}{8} \zeta_5 + \frac{11}{40} \zeta_3^2 ~~,~~
a^{\MSbars}_{51|519} ~=~ - \frac{9}{10} \zeta_3^2 ~~,~~
a^{\MSbars}_{51|520} ~=~ - \frac{9}{10} \zeta_3^2 \nonumber \\
a^{\MSbars}_{51|521} &=& \frac{153}{80} \zeta_5 + \frac{9}{160} \zeta_4 - \frac{1}{80} \zeta_3 ~~,~~
a^{\MSbars}_{51|522} ~=~ \frac{153}{80} \zeta_5 + \frac{9}{160} \zeta_4 - \frac{1}{80} \zeta_3 \nonumber \\
a^{\MSbars}_{51|523} &=& \frac{143}{80} \zeta_5 + \frac{21}{160} \zeta_4 - \frac{31}{80} \zeta_3 ~~,~~
a^{\MSbars}_{51|524} ~=~ \frac{143}{80} \zeta_5 + \frac{21}{160} \zeta_4 - \frac{31}{80} \zeta_3 \nonumber \\
a^{\MSbars}_{51|525} &=& - \frac{9}{5} \zeta_5 + \frac{9}{80} \zeta_4 - \frac{1}{40} \zeta_3 ~~,~~
a^{\MSbars}_{51|526} ~=~ - \frac{9}{5} \zeta_5 + \frac{9}{80} \zeta_4 - \frac{1}{40} \zeta_3 \nonumber \\
a^{\MSbars}_{51|527} &=& \frac{143}{80} \zeta_5 + \frac{9}{160} \zeta_4 - \frac{1}{80} \zeta_3 ~~,~~
a^{\MSbars}_{51|528} ~=~ \frac{143}{80} \zeta_5 + \frac{9}{160} \zeta_4 - \frac{1}{80} \zeta_3 \nonumber \\
a^{\MSbars}_{51|529} &=& - \frac{41}{20} \zeta_5 + \frac{9}{80} \zeta_4 - \frac{1}{40} \zeta_3 ~~,~~
a^{\MSbars}_{51|530} ~=~ - \frac{41}{20} \zeta_5 + \frac{21}{80} \zeta_4 - \frac{31}{40} \zeta_3 \nonumber \\
a^{\MSbars}_{51|531} &=& - \frac{1}{20} \zeta_5 - \frac{3}{320} \zeta_4 + \frac{13}{160} \zeta_3 ~~,~~
a^{\MSbars}_{51|533} ~=~ - \frac{3}{160} - \frac{3}{320} \zeta_4 + \frac{1}{32} \zeta_3 \nonumber \\
a^{\MSbars}_{51|534} &=& - \frac{3}{160} - \frac{3}{320} \zeta_4 + \frac{1}{32} \zeta_3 ~~,~~
a^{\MSbars}_{51|535} ~=~ \frac{1}{160} - \frac{3}{160} \zeta_4 + \frac{1}{80} \zeta_3 \nonumber \\
a^{\MSbars}_{51|536} &=& \frac{3}{320} \zeta_4 + \frac{1}{160} \zeta_3 ~~,~~
a^{\MSbars}_{51|537} ~=~ \frac{7}{60} ~~,~~
a^{\MSbars}_{51|538} ~=~ - \frac{1}{30} \nonumber \\
a^{\MSbars}_{51|539} &=& - \frac{5}{8} \zeta_6 - \frac{3}{4} \zeta_5 + \frac{1}{20} \zeta_3^2 ~~,~~
a^{\MSbars}_{51|540} ~=~ - \frac{1}{80} \zeta_5 - \frac{1}{5} \zeta_3 ~~,~~
a^{\MSbars}_{51|541} ~=~ - \frac{1}{80} \zeta_5 - \frac{1}{5} \zeta_3 \nonumber \\
a^{\MSbars}_{51|542} &=& \frac{9}{40} \zeta_5 - \frac{2}{5} \zeta_3 ~~,~~ 
a^{\MSbars}_{51|543} ~=~ - \frac{3}{160} + \frac{3}{160} \zeta_4 + \frac{1}{40} \zeta_3 ~~,~~
a^{\MSbars}_{51|544} ~=~ \frac{1}{12} \nonumber \\
a^{\MSbars}_{51|545} &=& - \frac{1}{20} \zeta_5 - \frac{3}{320} \zeta_4 + \frac{13}{160} \zeta_3 ~~,~~
a^{\MSbars}_{51|546} ~=~ - \frac{1}{20} \zeta_5 - \frac{3}{320} \zeta_4 + \frac{13}{160} \zeta_3 \nonumber \\
a^{\MSbars}_{51|549} &=& - \frac{3}{160} - \frac{3}{320} \zeta_4 + \frac{1}{32} \zeta_3 ~~,~~ 
a^{\MSbars}_{51|550} ~=~ - \frac{3}{160} - \frac{3}{320} \zeta_4 + \frac{1}{32} \zeta_3 \nonumber \\
a^{\MSbars}_{51|551} &=& \frac{3}{320} \zeta_4 + \frac{1}{160} \zeta_3 ~~,~~
a^{\MSbars}_{51|552} ~=~ \frac{3}{320} \zeta_4 + \frac{1}{160} \zeta_3 ~~,~~
a^{\MSbars}_{51|553} ~=~ \frac{7}{60} ~~,~~
a^{\MSbars}_{51|554} ~=~ \frac{7}{60} \nonumber \\
a^{\MSbars}_{51|555} &=& \frac{1}{160} - \frac{3}{160} \zeta_4 + \frac{1}{80} \zeta_3 ~~,~~
a^{\MSbars}_{51|556} ~=~ - \frac{3}{160} - \frac{3}{320} \zeta_4 + \frac{1}{32} \zeta_3 \nonumber \\
a^{\MSbars}_{51|557} &=& \frac{1}{10} \zeta_5 - \frac{21}{160} \zeta_4 - \frac{5}{16} \zeta_3 ~~,~~
a^{\MSbars}_{51|558} ~=~ \frac{1}{80} ~~,~~
a^{\MSbars}_{51|559} ~=~ \frac{1}{12} + \frac{3}{160} \zeta_4 - \frac{7}{80} \zeta_3 \nonumber \\
a^{\MSbars}_{51|560} &=& \frac{1}{12} + \frac{3}{160} \zeta_4 - \frac{7}{80} \zeta_3 ~~,~~
a^{\MSbars}_{51|561} ~=~ - \frac{3}{160} + \frac{3}{80} \zeta_4 - \frac{1}{80} \zeta_3 \nonumber \\
a^{\MSbars}_{51|562} &=& \frac{1}{40} - \frac{3}{160} \zeta_4 - \frac{9}{80} \zeta_3 ~~,~~
a^{\MSbars}_{51|563} ~=~ - \frac{7}{10} 
\end{eqnarray}
from which it is straightforward to see the connection with $c^{\MSbars}_{5r}$.

For the coefficients of $D_{Lr}^{ij}$ we have
\begin{eqnarray}
b^{\MSbars}_{55|51} &=& - \frac{7}{8192} ~~,~~
b^{\MSbars}_{55|53} ~=~ - \frac{5}{1024} ~~,~~
b^{\MSbars}_{55|54} ~=~ - \frac{3}{512} ~~,~~
b^{\MSbars}_{55|56} ~=~ - \frac{1}{384} \nonumber \\
b^{\MSbars}_{55|57} &=& - \frac{1}{192} ~~,~~
b^{\MSbars}_{55|58} ~=~ - \frac{1}{192} ~~,~~
b^{\MSbars}_{55|510} ~=~ - \frac{1}{512} ~~,~~
b^{\MSbars}_{55|511} ~=~ - \frac{1}{256} \nonumber \\
b^{\MSbars}_{55|512} &=& - \frac{1}{256} ~~,~~
b^{\MSbars}_{55|517} ~=~ - \frac{1}{256} ~~,~~
b^{\MSbars}_{55|518} ~=~ - \frac{1}{256} ~~,~~
b^{\MSbars}_{55|520} ~=~ - \frac{1}{768} \nonumber \\
b^{\MSbars}_{55|521} &=& - \frac{1}{256} ~~,~~
b^{\MSbars}_{55|522} ~=~ - \frac{1}{256} ~~,~~
b^{\MSbars}_{55|523} ~=~ - \frac{1}{256} ~~,~~
b^{\MSbars}_{55|524} ~=~ - \frac{1}{384} \nonumber \\
b^{\MSbars}_{55|525} &=& - \frac{1}{384} 
\end{eqnarray}
for the $\frac{1}{\epsilon^5}$ coefficients and
\begin{eqnarray}
b^{\MSbars}_{54|53} &=& \frac{5}{1024} ~~,~~
b^{\MSbars}_{54|54} ~=~ \frac{3}{256} ~~,~~
b^{\MSbars}_{54|56} ~=~ \frac{1}{192} ~~,~~
b^{\MSbars}_{54|57} ~=~ \frac{1}{96} \nonumber \\
b^{\MSbars}_{54|58} &=& \frac{1}{48} ~~,~~
b^{\MSbars}_{54|510} ~=~ \frac{1}{512} ~~,~~
b^{\MSbars}_{54|511} ~=~ \frac{1}{256} ~~,~~
b^{\MSbars}_{54|512} ~=~ \frac{3}{256} \nonumber \\
b^{\MSbars}_{54|517} &=& \frac{1}{256} ~~,~~
b^{\MSbars}_{54|518} ~=~ \frac{1}{96} ~~,~~
b^{\MSbars}_{54|520} ~=~ \frac{1}{256} ~~,~~
b^{\MSbars}_{54|521} ~=~ \frac{1}{96} \nonumber \\
b^{\MSbars}_{54|522} &=& \frac{1}{96} ~~,~~
b^{\MSbars}_{54|523} ~=~ \frac{1}{256} ~~,~~
b^{\MSbars}_{54|524} ~=~ \frac{1}{128} ~~,~~
b^{\MSbars}_{54|525} ~=~ \frac{1}{64} 
\end{eqnarray}
for the next order. The remaining two sets of coefficients are
\begin{eqnarray}
b^{\MSbars}_{53|54} &=& - \frac{3}{512} ~~,~~
b^{\MSbars}_{53|55} ~=~ - \frac{1}{32} \zeta_3 ~~,~~
b^{\MSbars}_{53|58} ~=~ - \frac{7}{192} ~~,~~
b^{\MSbars}_{53|59} ~=~ - \frac{3}{128} \zeta_3 \nonumber \\
b^{\MSbars}_{53|510} &=& \frac{1}{512} ~~,~~
b^{\MSbars}_{53|511} ~=~ \frac{1}{256} ~~,~~
b^{\MSbars}_{53|512} ~=~ - \frac{1}{64} ~~,~~
b^{\MSbars}_{53|514} ~=~ - \frac{1}{64} \zeta_3 \nonumber \\
b^{\MSbars}_{53|515} &=& - \frac{1}{64} \zeta_3 ~~,~~
b^{\MSbars}_{53|516} ~=~ - \frac{1}{32} \zeta_3 ~~,~~
b^{\MSbars}_{53|517} ~=~ \frac{1}{256} ~~,~~
b^{\MSbars}_{53|518} ~=~ - \frac{5}{768} \nonumber \\
b^{\MSbars}_{53|519} &=& - \frac{3}{64} \zeta_3 ~~,~~
b^{\MSbars}_{53|520} ~=~ - \frac{1}{768} ~~,~~
b^{\MSbars}_{53|521} ~=~ - \frac{5}{768} ~~,~~
b^{\MSbars}_{53|522} ~=~ - \frac{5}{768} \nonumber \\
b^{\MSbars}_{53|523} &=& \frac{1}{256} ~~,~~
b^{\MSbars}_{53|524} ~=~ - \frac{1}{384} ~~,~~
b^{\MSbars}_{53|525} ~=~ - \frac{19}{384} 
\end{eqnarray}
and
\begin{eqnarray}
b^{\MSbars}_{52|55} &=& \frac{1}{32} \zeta_3 ~~,~~
b^{\MSbars}_{52|56} ~=~ - \frac{1}{384} ~~,~~
b^{\MSbars}_{52|57} ~=~ - \frac{1}{192} ~~,~~
b^{\MSbars}_{52|58} ~=~ \frac{1}{48} ~~,~~
b^{\MSbars}_{52|513} ~=~ \frac{5}{16} \zeta_5 \nonumber \\
b^{\MSbars}_{52|514} &=& - \frac{3}{128} \zeta_4 + \frac{3}{64} \zeta_3 ~~,~~
b^{\MSbars}_{52|515} ~=~ - \frac{3}{128} \zeta_4 + \frac{3}{64} \zeta_3 ~~,~~
b^{\MSbars}_{52|516} ~=~ - \frac{3}{64} \zeta_4 + \frac{3}{32} \zeta_3 \nonumber \\
b^{\MSbars}_{52|517} &=& \frac{1}{256} - \frac{1}{128} \zeta_3 ~~,~~
b^{\MSbars}_{52|518} ~=~ - \frac{1}{96} ~~,~~
b^{\MSbars}_{52|519} ~=~ \frac{3}{128} \zeta_4 + \frac{3}{64} \zeta_3  \nonumber \\
b^{\MSbars}_{52|520} &=& - \frac{5}{768} ~~,~~
b^{\MSbars}_{52|521} ~=~ - \frac{1}{96} ~~,~~
b^{\MSbars}_{52|522} ~=~ - \frac{1}{96} ~~,~~
b^{\MSbars}_{52|523} ~=~ \frac{1}{256} - \frac{1}{128} \zeta_3  \nonumber \\
b^{\MSbars}_{52|524} &=& - \frac{5}{384} + \frac{1}{64} \zeta_3 ~~,~~
b^{\MSbars}_{52|525} ~=~ \frac{5}{64} ~.
\end{eqnarray}
The analogous expressions for $a^{\MOMs}_{Lq|Lr}$ and $b^{\MOMs}_{Lq|Lr}$ are 
available in the attached data file.

\end{document}